\documentclass[
reprint, longbibliography,
superscriptaddress,
preprintnumbers,nofootinbib,
nobibnotes,
amsmath,amssymb,
aps, pra,
floatfix,
]{revtex4-2}

\usepackage[normalem]{ulem}
\usepackage{gensymb}
\usepackage{makebox}
\usepackage{upgreek}
\usepackage{mathrsfs}
\usepackage{comment}
\usepackage{graphicx}
\usepackage{dcolumn}
\usepackage{placeins}
\usepackage{bm}
\usepackage{float}
\usepackage{url}
\usepackage[]{hyperref}
\usepackage{xcolor}
\usepackage{textcomp}
\usepackage{amsmath}
\usepackage{varwidth}
\usepackage{placeins}
\usepackage{gensymb}
\usepackage{braket}
\usepackage{booktabs}
\usepackage{makecell}
\usepackage{tabularx}
\usepackage{bm}      
\usepackage{amsmath}
\usepackage{makecell}
\usepackage{array}
\usepackage{booktabs}
\usepackage{siunitx}
\hypersetup{
    colorlinks,
    linkcolor={red!50!black},
    citecolor={blue!50!black},
    urlcolor={blue!80!black}
}
\usepackage[center]{titlesec}
  \titleformat{\section}
   {\centering\normalfont\fontsize{10}{10}\bfseries}{\thesection}{1em}{}
  \titlespacing{\section}{0pt}{12pt plus 4pt minus 2pt}{6pt plus 2pt minus 2pt}
    \titleformat{\subsection}
   {\centering\normalfont\fontsize{10}{10}\bfseries}{\thesubsection}{1em}{}
  \titlespacing{\subsection}{0pt}{12pt plus 4pt minus 2pt}{6pt plus 2pt minus 2pt}
\usepackage{tabularx}
\newcolumntype{C}{>{\centering\arraybackslash}X} 
\usepackage{multirow}

\usepackage{amsfonts, amsmath, amssymb, graphicx, xcolor, verbatim, lipsum, dsfont, bm, hyperref, makecell, comment, upgreek, graphicx, dcolumn, array}
\usepackage[export]{adjustbox}
\usepackage[mathlines]{lineno}
\usepackage[T1]{fontenc}
\DeclareLanguageAlias{en}{english}
\DeclareLanguageAlias{eng}{english}

\begin{document}

\title{Optically detected nuclear magnetic resonance of carbon-13 in bulk diamond}

\author{Maxwell~D.~Aiello}
\affiliation{Center for High Technology Materials, 
University of New Mexico, Albuquerque, NM, USA}
\affiliation{Department of Physics and Astronomy,
University of New Mexico, Albuquerque, NM, USA}

\author{Janis~Smits}
\affiliation{Center for High Technology Materials, 
University of New Mexico, Albuquerque, NM, USA}

\author{Yaser~Silani}
\affiliation{Center for High Technology Materials, 
University of New Mexico, Albuquerque, NM, USA}

\author{Andris~Berzins}
\affiliation{Center for High Technology Materials, 
University of New Mexico, Albuquerque, NM, USA}

\author{David~Lidsky}
\affiliation{Center for High Technology Materials, 
University of New Mexico, Albuquerque, NM, USA}

\author{Bryan~A.~Richards}
\affiliation{Center for High Technology Materials, 
University of New Mexico, Albuquerque, NM, USA}
\affiliation{Department of Physics and Astronomy,
University of New Mexico, Albuquerque, NM, USA}

\author{Amilcar~Jeronimo~Perez}
\affiliation{Center for High Technology Materials, 
University of New Mexico, Albuquerque, NM, USA}
\affiliation{Department of Physics and Astronomy,
University of New Mexico, Albuquerque, NM, USA}

\author{Chandrasekhar~Ramanathan}
\affiliation{Department of Physics and Astronomy, Dartmouth College, Hanover, NH, USA}

\author{Sebasti\'an~C.~Carrasco}
\affiliation{DEVCOM Army Research Laboratory, Adelphi, MD, USA}

\author{Jabir Chathanathil}
\affiliation{DEVCOM Army Research Laboratory, Adelphi, MD, USA}

\author{Michael Goerz}
\affiliation{DEVCOM Army Research Laboratory, Adelphi, MD, USA}

\author{Vladimir~S.~Malinovsky}
\affiliation{DEVCOM Army Research Laboratory, Adelphi, MD, USA}

\author{Dmitry~Budker}
\affiliation{Helmholtz-Institut Mainz, 55128 Mainz, Germany}
\affiliation{GSI Helmholtzzentrum f{\"u}r Schwerionenforschung GmbH, 64291 Darmstadt, Germany}
\affiliation{QUANTUM, Institut für Physik, Johannes Gutenberg-Universit{\"a}t, 55128, Mainz, Germany}
\affiliation{Department of Physics, University of California, 94720-7300, Berkeley, CA, USA}

\author{Sean~Lourette}
\affiliation{Department of Physics, University of California, 94720-7300, Berkeley, CA, USA}
\affiliation{DEVCOM Army Research Laboratory, Adelphi, MD, USA}

\author{Andrey~Jarmola}
\email[Email address: ]{andrey.jarmola@gmail.com}
\affiliation{Department of Physics, University of California, 94720-7300, Berkeley, CA, USA}
\affiliation{ODMR Technologies Inc., El Cerrito, CA, USA}

\author{Victor~M.~Acosta}
\email[Email address: ]{victormarcelacosta@gmail.com}
\affiliation{Center for High Technology Materials, 
University of New Mexico, Albuquerque, NM, USA}
\affiliation{Department of Physics and Astronomy,
University of New Mexico, Albuquerque, NM, USA}

\date{\today}

\begin{abstract}
Precision measurements based on optically detected nuclear magnetic resonance offer exquisite sensitivity to absolute shifts in spin transition frequencies, with potential applications in fundamental physics experiments and inertial sensing. We investigate $^{13}$C nuclear spins in diamond as a candidate system for solid-state implementations, which hold the promise for high-fidelity readout of large numbers of coherent nuclear spins in millitesla or lower magnetic fields. We demonstrate a technique that allows for both optical polarization and readout of large ensembles of ${\sim}10^{16}$ polarized nuclear spins. Our method takes advantage of state-selective Landau-Zener transitions under microwave frequency sweeping, which bidirectionally transfer spin polarization between Nitrogen-Vacancy (NV) electron spins and remote $^{13}$C nuclear spins. Using natural isotopic abundance diamonds with nitrogen densities of ${\sim}0.5\mbox{--}10~{\rm ppm}$, we perform optically-detected $^{13}$C Ramsey spectroscopy and realize a nuclear-spin-dependent fluorescence contrast exceeding $0.5\%~{\rm peak{\mbox{-}}to{\mbox{-}}peak}$. We observe nuclear spin dephasing times $T_2^{\ast}\approx2~{\rm ms}$ that only modestly improve with homonuclear dipolar decoupling, indicating that they are limited by the longitudinal spin relaxation of nearby NV electron spins. We study the magnetic field dependence of the optical readout and find comparable contrast and dephasing times for magnetic fields in the range $8\mbox{--}20~{\rm mT}$. Our method can be interpreted as a type of repetitive readout, where each NV center optically reads out the spin state of ${\sim}10^2$ nuclei before nuclear spins depolarize.
\end{abstract}

\maketitle

\section{\label{sec:level1}Introduction}
Precision measurements of non-magnetic spin interactions are widely used in tabletop searches for new physics~\cite{SAF2018,DOB2006,CON2025} and the development of highly stable gyroscopes~\cite{DON2010,MEY2014,WAL2016}. These applications typically require high sensitivity to absolute shifts in the spin precession frequency. For an uncorrelated ensemble of $N_p$ polarized spins, the minimum detectable change in precession frequency is given by $\delta f\approx1/(2\pi\mathcal{F}\sqrt{N_p T_2^{\ast} t})$, where $T_2^*$ is the spin-dephasing time, $t$ is the total measurement time, and $\mathcal{F}$ is the spin readout fidelity that relates $\delta f$ to the spin-projection noise limit ($\mathcal{F}=1$)~\cite{LED2012,AJO2012,DEG2017}. In practice, the precision is often limited by environmental instabilities, such as fluctuations in magnetic field and temperature, which lead to uncompensated shifts in the spin precession frequency~\cite{MEY2014}. Thus, when probing non-magnetic spin interactions, nuclear spins possessing a small magnetic moment are preferable due to their relatively weak coupling to environmental fluctuations and relatively long $T_2^{\ast}$ dephasing times~\cite{JAC1995}. 

Hyperpolarized solid-state spins are attractive for compact precision measurement systems, as their high polarized-spin density offers the largest $N_p$ values and their $T_2^{\ast}$ times can be tuned via dipolar decoupling methods~\cite{PAR2019,JOS2025}. However, the conventional nuclear magnetic resonance (NMR) readout via radio-frequency (RF) induction is fundamentally limited by Johnson noise~\cite{HOU1976,SIL2023} (\ref{app:johnson}). At room temperature and low magnetic fields ($B_0\lesssim10~{\rm mT}$, corresponding to NMR frequencies of order $\lesssim100~{\rm kHz}$) this restricts the readout fidelity to $\mathcal{F}\ll10^{-3}$ (\ref{app:johnsonexpt}), limiting the attainable sensitivity.

Optically-detected NMR (ODNMR) is an intriguing alternative to inductive detection that can offer superior readout fidelity at low $B_0$ and ambient temperatures. Typically, an optically detected electron-spin system is used to transfer polarization to and from neighboring, hyperfine-coupled nuclear spins \cite{WAL1997}. The electron spin system can simultaneously be used as a secondary sensor to suppress the impact of environmental fluctuations in a ``co-magnetometer'' configuration \cite{WAL2016,KLI2023}. In noble gas/alkali-metal vapor mixtures, these techniques have matured to the point where sub-nanohertz frequency precision has been realized \cite{ZHA2025}. However, for small vapor cells (volume $V\lesssim10~{\rm mm^3}$), the $N_p T_2^{\ast}$ product is fundamentally limited by spin-altering collisions of noble gas nuclei with alkali atoms and cell walls \cite{LIM2025}, which raises challenges for miniature sensor applications~\cite{EKL2008,NOO2019,CHE2021,WAN2022}. 

To increase spin density and circumvent the impact of collisional dephasing, liquid-state co-magnetometers have been explored~\cite{LED2012L,WU2018}. However, the ODNMR liquid systems with the most favorable $N_p T_2^{\ast}$ products require careful thermal cycling in a near-cryogenic environment \cite{SAU1997,ROM2001}, which can be impractical. 

Precision ODNMR measurements based on Nitrogen-Vacancy (NV) centers (ground-state spin $S=1$) in diamond have been proposed as a compact, solid-state alternative to alkali-metal vapor~\cite{LED2012,AJO2012}. Initial demonstrations used ODNMR of the intrinsic $^{14}$N nuclear spins associated with NV centers~\cite{JAR2021,SOS2021}. This method provides a spin readout fidelity, $\mathcal{F}\approx10^{-3}$, but the spin density is limited, as there is only one intrinsic $^{14}$N spin for each NV electron spin. 

We hypothesized that ODNMR measurements of the abundant weakly-coupled $^{13}$C nuclei (spin $I=1/2$) in diamond could provide a larger polarized spin density. Numerous experiments have demonstrated room-temperature hyperpolarization of bulk $^{13}$C spins by transferring the near-unity optical polarization of NV electron spins to nuclei via either the solid-effect~\cite{REY1998,KIN2015,ALV2015}, pulsed~\cite{SCH2018} and microwave-swept~\cite{HEN1988,CHE2015,AJO2018,ZAN2019,KAV2025,MIY2021,BLI2025} Landau-Zener dynamics, or microwave~\cite{LON2013} and magnetic-field~\cite{JAC2009,KIN2010,SCO2016,PAG2018} tuned spin energy degeneracy. Long-lived $^{13}$C bulk polarization exceeding $5\%$ has been reported for mm-sized diamonds~\cite{KIN2010,ALV2015,KIN2015,KAV2025}. However, bulk detection has largely relied on post-polarization shuttling of diamonds into a superconducting NMR magnet for inductive coil-based readout. This has two primary drawbacks: i) the large magnetic field is impractical for some applications and is difficult to stabilize for absolute NMR frequency measurements, and ii) the readout fidelity of inductive detection is relatively poor, limiting sensitivity (\ref{app:johnsonexpt}). This motivates the use of optical $^{13}$C NMR detection techniques, but prior studies have focused on the low $N_p$ regime~\cite{JAC2009,SME2009,TAM2012,KOL2012,DRE2012,LAR2013,FIS2013B,FIS2013L,FOR2021,REN2023,MEI2023,CHA2025}. 

Here, we demonstrate a method for low-field ODNMR spectroscopy of $^{13}$C nuclei that are weakly coupled to NV centers. Our technique uses state-selective Landau-Zener transitions under microwave frequency sweeping, which bidirectionally transfer spin polarization between NV electron spins and remote $^{13}$C nuclear spins. This enables a repetitive readout, where each NV center is used to optically read out the spin state of ${\sim}10^2$ nuclei before nuclear spins depolarize. We perform optically-detected $^{13}$C Ramsey spectroscopy and realize a nuclear-spin-dependent fluorescence contrast exceeding $0.5\%~{\rm peak{\mbox{-}}to{\mbox{-}}peak}$. We study the $^{13}$C spin coherence times of four diamonds, with nitrogen density in the ${\sim}0.5\mbox{--}10~{\rm ppm}$ range, and find they are likely limited by the ${\sim}3~{\rm ms}$ longitudinal spin relaxation of neighboring NV electron spins. The results indicate the potential for compact ODNMR devices, featuring relatively high spin readout fidelity of large ensembles of polarized diamond $^{13}$C nuclear spins at room temperature and in magnetic fields down to ${\sim}8~{\rm mT}$.

\begin{figure*}[hbt]
    \centering
    \includegraphics[width=0.98\linewidth]{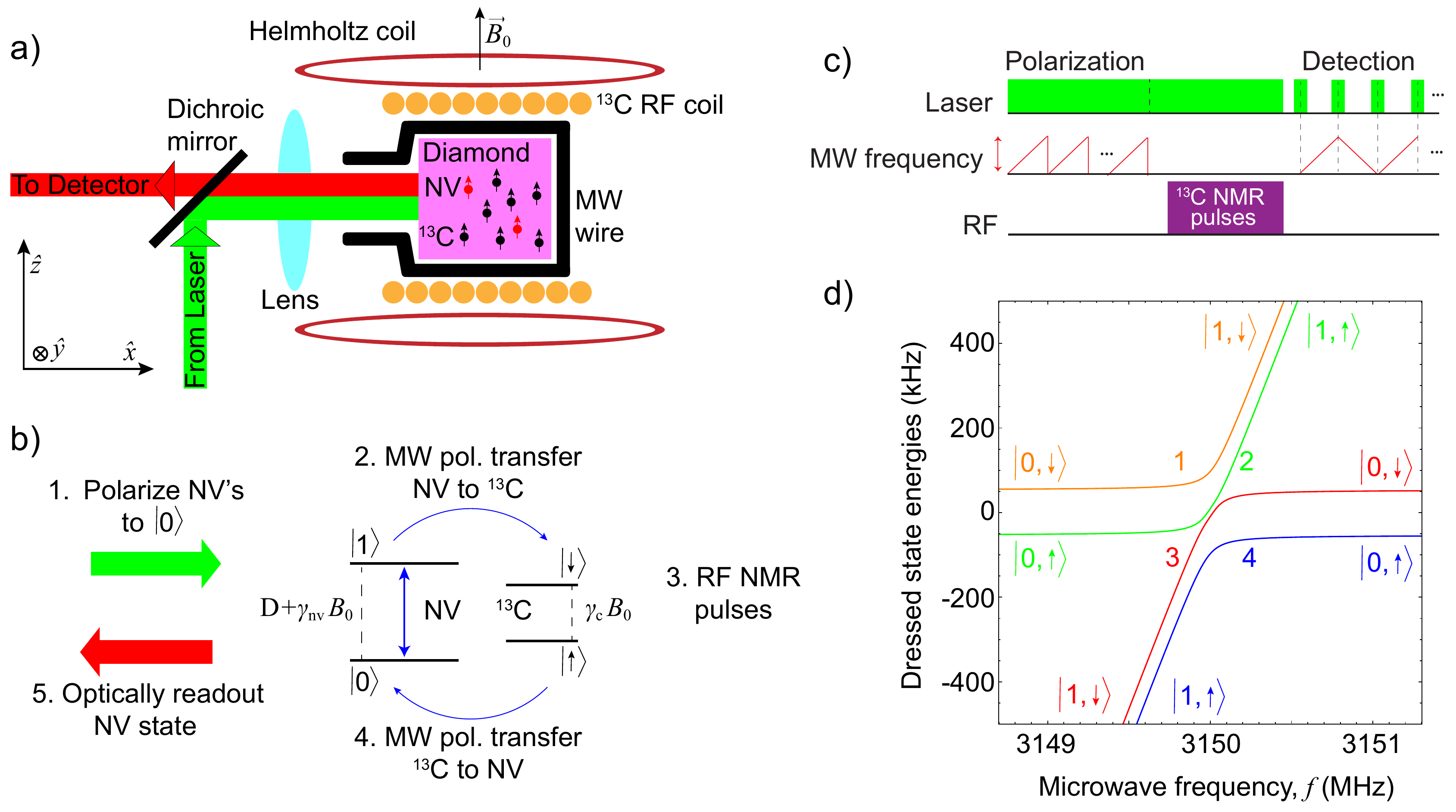}
    \caption{\textbf{$^{13}$C ODNMR apparatus and detection principle.} (a) Experimental apparatus. Magnetic fields of $B_0=2\mbox{--}12~{\rm mT}$ are applied along the NV axis ($\hat{z}$) using Helmholtz coils. A loop of wire wrapped around the diamond delivers microwaves (MW) that address the NV $f_+$ transition. A 20-turn copper-wire radio frequency (RF) coil is wrapped around the diamond to drive the $^{13}$C NMR transition. (b) Schematic of the ODNMR principle, as described in the main text. (c) ODNMR pulse sequence, consisting of three phases: polarization, RF NMR encoding, and detection. The MW frequency is swept about the NV $f_+$ transition in a unidirectional manner (low to high) for the polarization phase, and bidirectionally for the detection phase. (d) Example plot of the dressed state energy levels versus frequency, $\nu_{1,2,3,4}(f)$, of a NV-$^{13}$C spin pair, calculated by diagonalizing the rotating-frame Hamiltonian in Eq.~\eqref{eq:H}. When the microwave frequency is swept, nuclear spin flips primarily occur for the narrowest avoided crossing, between states $\ket{\psi_2}$ and $\ket{\psi_3}$. The following representative values are used: $B_0=10~{\rm mT}$, $\Omega=100~{\rm kHz}$, $f_+=3150~{\rm MHz}$, $f_n=107~{\rm kHz}$, and $A_{zz}=A_{zx}=30~{\rm kHz}$, using the notation of Ref.~\cite{ZAN2019}. The $\ket{m_s,m_i}$ labels correspond to the states in the far-detuned limit.
    }
    \label{fig:fig1}
\end{figure*}

\section{Experimental setup}
\label{sec:expt}
The experimental configuration is depicted in Fig.~\ref{fig:fig1}(a). We focus our study on millimeter-sized diamonds, grown by chemical vapor deposition, with natural-abundance ($1.1\%$) $^{13}$C, that have an initial nitrogen density in the range ${\sim}0.5\mbox{--}10~{\rm ppm}$. Approximately $10\%$ of the initial nitrogen impurities were converted to NV centers via electron irradiation and annealing (\ref{sec:SI_diamond_tables}). 

A Helmholtz coil pair is used to supply a uniform external magnetic field, $B_0$, in the range $2\mbox{--}12~{\rm mT}$ along the $\hat{z}$ direction. The diamond is positioned such that one of the four NV crystallographic axes is aligned along $\hat{z}$. 

A copper-wire coil, wrapped around the diamond, is used to deliver $20\mbox{--}130~{\rm kHz}$ RF magnetic-field pulses (${\sim}50~{\rm \upmu s}$ duration), polarized along the $\hat{x}$ axis, that address the $^{13}$C nuclear spins. A second copper wire loop delivers $2.9\mbox{--}3.2~{\rm GHz}$ microwave magnetic fields polarized along the $\hat{y}$ axis that drive the $m_s=0\leftrightarrow m_s=1$ NV ground-state electron spin transition. The diamond is mounted on a water-cooled aluminum nitride substrate, and the entire assembly, excluding Helmholtz coils and optics, is housed in an RF-shielding aluminum box with holes for optical and electrical access. 

Laser light ($532~{\rm nm}$, $0.01\mbox{--}1.2~{\rm W}$) is amplitude modulated with an acousto-optic modulator and shaped with an anamorphic prism pair and lenses to a ${\sim}0.5\times2~{\rm mm^2}$ beam that is incident on a side facet of the diamond. An aspheric condenser lens (numerical aperture: ${\sim}0.6$) collects NV fluorescence, which is then spectrally filtered with a dichroic mirror and relayed by another lens onto an amplified photodetector. Typically, the incident peak laser power is ${\sim}0.4~{\rm W}$, the collected fluorescence power is ${\sim}1~{\rm mW}$, and the NV fluorescence contrast (relative difference in fluorescence rate between $m_s=0$ and $m_s=1$) is $(\Delta F/F)_{\rm pp,max}\approx3\%$. For a more detailed description of the experimental setup, see~\ref{sec:SI_exp_layout}.

Figure~\ref{fig:fig1}(b) depicts the ODNMR concept. Absorption of green laser light polarizes NV electron spins to the $m_s=0$ ground-state spin sublevel, denoted as $\ket{0}$. The frequency of a microwave field is swept about the NV $\ket{0}\leftrightarrow \ket{1}$ transition frequency, $f_+$. Here $f_\pm=D\pm\gamma_{\rm nv} B_0$, where $D=2.87~{\rm GHz}$ is the axial zero-field splitting parameter and $\gamma_{\rm nv}=28.03~{\rm GHz/T}$ is the NV gyromagnetic ratio. This process transfers NV electron spin polarization to $^{13}$C nuclear spins, mediated by transverse hyperfine coupling. RF pulses encode the $^{13}$C NMR frequency, $f_n=\gamma_{\rm nuc} B_0$, where $\gamma_{\rm nuc}=10.7~{\rm MHz/T}$, into a change in the $^{13}$C polarization. The resulting nuclear spin polarization is transferred back to NV electron spins via a similar frequency-swept microwave transfer process. Finally, the NV electron spin polarization is read out optically via the spin-dependent fluorescence rate.

\subsection{Optical hyperpolarization of $^{13}$C spins}
Figure~\ref{fig:fig1}(c) shows a typical ODNMR pulse sequence. During the polarization phase of the sequence, laser light is on continuously to repolarize NV centers to $m_s=0$, while the microwave frequency is repeatedly swept across the $f_+$ transition unidirectionally, from low to high frequency, to transfer polarization to $^{13}$C nuclear spins. This polarization protocol has previously been demonstrated with NV-doped diamonds~\cite{CHE2015,AJO2018,ZAN2019,KAV2025,MIY2021,BLI2025} and other solid-state systems~\cite{HEN1988,MIY2021,SAK2023,TAT2026}. The polarization process can be understood by first considering a single NV center coupled to a single $^{13}$C nuclear spin under continuous microwave driving of frequency $f$ and Rabi frequency $\Omega$. The $B_0$ field is taken to be large enough that we may consider a reduced two-level basis of the NV center, $\{\ket{1},\ket{0}\}$. In the NV frame rotating with frequency $f_+$, the Hamiltonian can be written approximately as~\cite{ZAN2019}:
\begin{equation}
\label{eq:H}
H_{\rm eff} = (f-f_+) S_z' + \Omega S_x' + A_{zz} S_z' I_z + A_{zx} S_z' I_x - f_n I_z,
\end{equation}
where  $S_z' = \ket{1}\!\bra{1}$ and $S_x'=(\ket{1}\!\bra{0}+\ket{0}\!\bra{1})/2$ are spin operators in the reduced NV basis, and $A_{zz}$ and $A_{zx}$ are axial and transverse hyperfine coupling parameters, respectively. Throughout, we neglect hyperfine coupling with the NV center's $^{14}$N nuclear spin, as it has little impact on the $^{13}$C spin dynamics; its main effect is to produce small shifts ($0,\pm2.2~{\rm MHz}$) in $f_+$, depending on the $^{14}$N spin state.

Diagonalization of $H_{\rm eff}$ in Eq.~\eqref{eq:H} yields four dressed states, denoted $\ket{\psi_{1,2,3,4}}$, with eigenfrequencies, $\nu_{1,2,3,4}$, that depend on the microwave drive frequency, $f$. An example plot of $\nu_{1,2,3,4}(f)$ is shown in Fig.~\ref{fig:fig1}(d), taking $B_0=10~{\rm mT}$, $\Omega=100~{\rm kHz}$, and $A_{zz}=A_{zx}=30~{\rm kHz}$ (corresponding to a $^{13}$C displaced $1~{\rm nm}$ from the NV center at a $30\degree$ polar angle, see~\ref{sec:SI_phop}). In the far-detuned limit, $|f-f_+|\gg\Gamma$, where $\Gamma$ is the power-broadened NV electron spin linewidth~\cite{DRE2011,JEN2013}, the dressed states are $\ket{m_s,m_i}$ eigenstates. However, near resonance, the levels undergo avoided crossings. On resonance, the frequency splitting between states $\ket{\psi_j}$ and $\ket{\psi_k}$, $\Delta\nu_{jk}$, depends sensitively on the hyperfine coupling parameters, $f_n$, and $\Omega$. If the microwave frequency is swept across resonance at a constant rate $\dot{f}$, an NV-$^{13}$C spin pair initially in $\ket{\psi_j}$ can diabatically hop to $\ket{\psi_k}$, with a probability that is roughly given by~\cite{ZAN2019}:
\begin{equation}
\label{eq:phop}
    P^{jk}_{\rm hop}(A_{zz},A_{zx},B_0,\Omega,\dot{f}) \approx \exp\!\left\{-\frac{\pi^2 (\Delta \nu_{jk})^2}{|\dot{f}|} \right\}.
\end{equation}
Numerical calculations of $P^{jk}_{\rm hop}(A_{zz},A_{zx},B_0,\Omega,\dot{f})$, based on density matrix modeling of Eq.~\eqref{eq:H} are presented in~\ref{sec:SI_phop}, showing qualitative agreement with Eq.~\eqref{eq:phop} in the regime where $\Delta \nu_{23}$ is largely determined by off-diagonal terms in Eq.~\eqref{eq:H}.

Equation~\eqref{eq:phop} indicates that a narrow avoided crossing, $\Delta\nu_{jk}^2\lesssim|\dot{f}|/(2\pi)$, leads to a higher probability of diabatic hopping. The mechanism for nuclear hyperpolarization can be understood based on the example plot of the dressed state levels in Fig.~\ref{fig:fig1}(d). Optical pumping polarizes NV centers into $\ket{0}$. The frequency of a constant-amplitude microwave field is ramped across $f_+$ from low to high frequency, Fig.~\ref{fig:fig1}(c). If the nuclear spin is initially in $\ket{\downarrow}$, the NV-$^{13}$C spin pair remains in $\ket{\psi_1}$, as the frequency gap to other states is too large for diabatic hopping, $\Delta\nu_{1k}\gtrsim\sqrt{|\dot{f}|/(2\pi)}$. In other words, the NV electron spin undergoes adiabatic fast passage (AFP), transitioning from $\ket{0}$ to $\ket{1}$, while the nuclear spin remains in $\ket{\downarrow}$. If the nuclear spin is instead initially in $\ket{\uparrow}$, the NV-$^{13}$C spin pair has a significant probability to undergo diabatic hopping from $\ket{\psi_2}$ to $\ket{\psi_3}$, owing to the small frequency splitting, $\Delta\nu_{23}\lesssim\sqrt{|\dot{f}|/(2\pi)}$. In this case, the nuclear spin may flip to $\ket{\downarrow}$ by the end of the sweep. 

The unidirectional microwave-frequency sweeps are repeated under continuous optical pumping. Only the direction of the sweep (in other words, the sign of $\dot{f}$) determines the final $^{13}$C state populations. This holds regardless of which NV center $f_\pm$ resonance is swept about~\cite{AJO2018}. By sweeping from low to high frequency ($\dot{f}>0$), each NV center is used to polarize a large number of nuclear spins into $\ket{\downarrow}$. The degree of the resulting nuclear hyperpolarization depends on several factors, including i) the fraction of neighboring $^{13}$C spins with hyperfine coupling parameters that provide large $P^{23}_{\rm hop}$, ii) the nuclear dipolar flip-flop rate, $\Gamma_{\rm ff}$, and iii) the longitudinal nuclear spin relaxation time, $T_{1,{\rm nuc}}$. In the diamonds studied here, with moderate nitrogen density and natural $^{13}$C abundance ($T_{1,{\rm nuc}}\gtrsim100~{\rm s}$, $\Gamma_{\rm ff}\approx100~{\rm s^{-1}}$), the bulk $^{13}$C polarization can exceed $5\%$~\cite{KAV2025} (\ref{app:johnson}). However, under our experimental conditions, the local $^{13}$C polarization within roughly $3~{\rm nm}$ of each NV center (bounded by $\Gamma_{\rm ff}$ and the polarization time $T_{\rm polar}\approx0.5~{\rm s}$) may be significantly higher than the bulk average~\cite{AJO2019,PAG2020}. A glossary of variables, along with descriptions and typical values, is provided in~\ref{sec:SI_glossary}.

\subsection{RF encoding of the $^{13}$C NMR frequency}
Following the polarization phase, RF pulses are applied that encode the $^{13}$C NMR frequency into changes in the nuclear-spin polarization. For example, a Ramsey sequence can be applied using a pair of near-resonant RF $\pi/2$ pulses with a detuning $\delta$ relative to $f_n$. When these pulses are separated by a free-precession interval $\tau$, the resulting nuclear-spin polarization is given by $\braket{I_z}\propto\cos{(2\pi\delta\tau)}$. Before and during the RF encoding phase, we typically leave the green laser light on for ${\sim}10~{\rm ms}$, such that NV centers largely remain in the $\ket{0}$ state where $^{13}$C spins are unaffected by hyperfine interaction.

\subsection{Optical detection of the $^{13}$C spin state}
Optical detection of $\braket{I_z}$ is carried out using the detection phase of the sequence depicted in Fig.~\ref{fig:fig1}(c). The introduction of an optical detection protocol for large ensembles of $^{13}$C spins in diamond is the primary novelty of this work. Here, microwave sweeping about $f_+$ is performed with alternating direction, and laser pulses are applied at the beginning of each sweep segment for NV electron spin readout and (at least partial) repolarization. Consider the case where the nuclear spin begins the detection phase in state $\ket{\downarrow}$, Fig.~\ref{fig:fig1}(d). A laser pulse initializes NV centers to $\ket{0}$. When microwaves are swept up in frequency ($\dot{f}>0$), the NV center undergoes AFP, and the final state $\ket{1,\downarrow}$ results in a lower fluorescence rate. However, when $\dot{f}<0$, there is a significant probability of diabatic hopping, resulting in a final state $\ket{0,\uparrow}$ that produces a higher fluorescence rate. In this way, the fluorescence rate alternates from low to high and repeats. If the nuclear polarization begins in the $\ket{\uparrow}$ state, the fluorescence instead alternates from high to low rates; sweeps with $\dot{f}>0$ result in a higher fluorescence rate while sweeps with $\dot{f}<0$ result in a lower fluorescence rate. 

While the diabatic hopping flips the nuclear spin state of one particular site, the cycle can repeat if another polarized nuclear spin replenishes this site's $\ket{\downarrow}$ polarization via spin diffusion or if the NV center interacts with a different hyperfine-coupled nuclear spin site. This provides a means for repetitive readout--each NV center can be used to optically detect the spin state of numerous $^{13}$C nuclei before spin relaxation processes dominate. By repeating cycles of microwave sweeps of alternating direction, the normalized amplitude of the resulting fluorescence oscillations, $\Delta F/F$, retaining sign, provides a direct mapping to $\braket{I_z}$.

\section{Results}
\label{sec:results}

\begin{figure*}[hbt]
   \centering
    \includegraphics[width=0.98\linewidth]{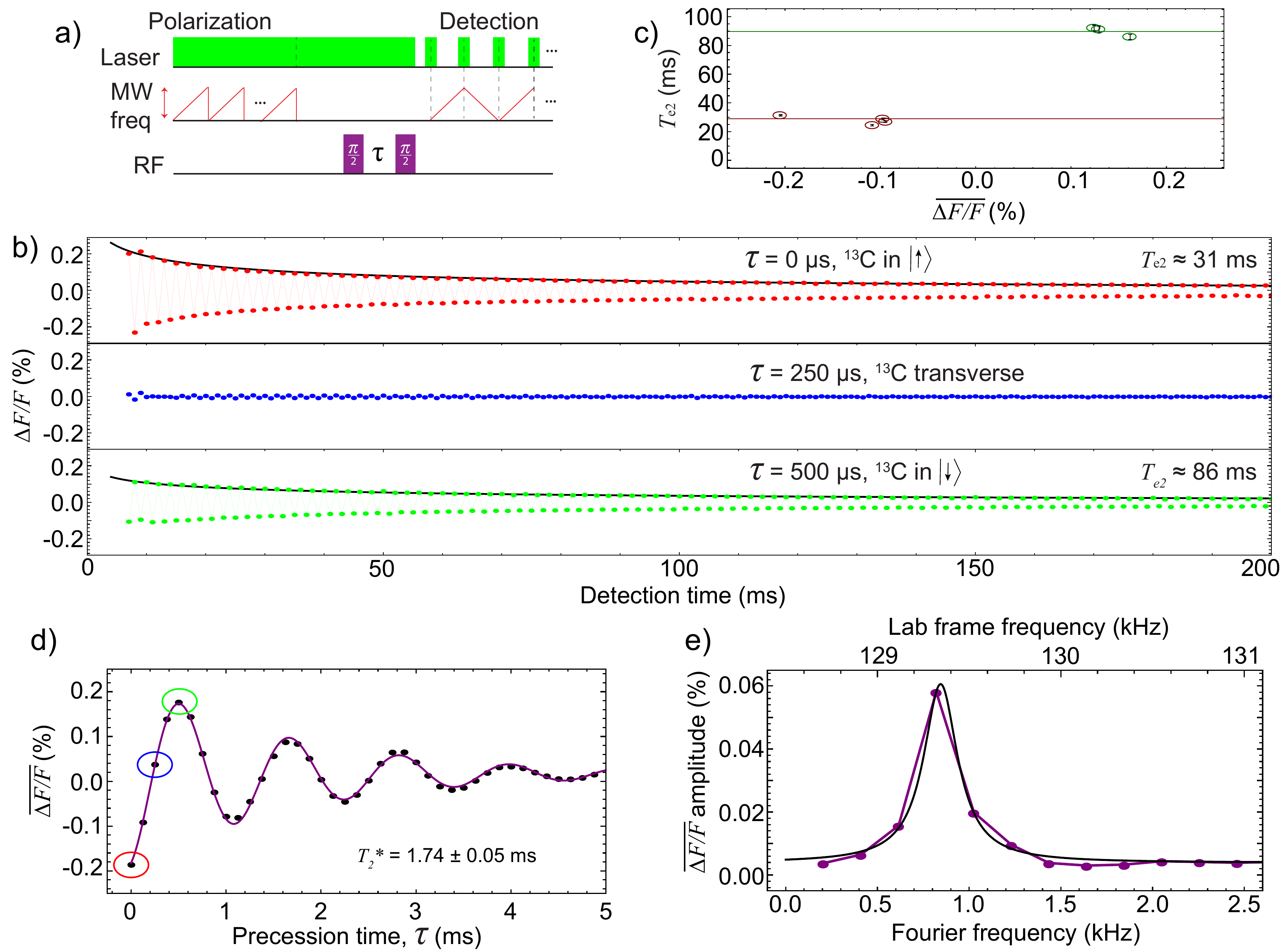}
    \caption{\textbf{Validation of $^{13}$C ODNMR readout.} (a) Pulse sequence for optically-detected $^{13}$C Ramsey spectroscopy (see also \ref{sec:SI_detailed_pulse_sequence}). (b) Fluorescence time traces during the detection phase for three values of $\tau$, corresponding to $\ket{\uparrow}$ (red, top), $\braket{I_z}\approx0$ (blue, middle), and $\ket{\downarrow}$ (green, bottom) after the RF encoding phase. Error bars (${\sim}0.001\%$) are smaller than the plot markers. Overlaid on the data (top and bottom only) is the fitted envelope decay function along with the $1/e^2$ decay time, $T_{e2}$. (c) $T_{e2}$ decay time as a function of readout amplitude ($\overline{\Delta F/F}$). The colored lines denote the typical $T_{e2}$ times for positive (green, $T_{e2}\approx90~{\rm ms}$) and negative (red, $T_{e2}\approx30~{\rm ms}$) readout amplitudes. Error bars represent fit uncertainty. (d) Ramsey interferogram, $\overline{\Delta F/F}(\tau)$ computed via Fourier processing of the $\Delta F/F$ time traces. Data points corresponding to the time traces in (b) are circled. A fit to Eq.~\eqref{eqn:Ramsey} reveals $T_2^{\ast}=1.74\pm0.05~{\rm ms}$ nuclear dephasing time. (e) $^{13}$C NMR spectrum obtained by taking a phased Fourier transform of the data in (d), along with Lorentzian fit. }
    \label{fig:fig2}
\end{figure*}

We performed optically-detected $^{13}$C Ramsey spectroscopy of diamond J5 (\ref{sec:SI_diamond_tables}) at $B_0=12~{\rm mT}$ to characterize the ODNMR readout. Figure~\ref{fig:fig2}(a) shows the Ramsey pulse sequence. The default duration of the pulse sequence is ${\sim}1~{\rm s}$ total: $0.5\mbox{--}1~{\rm s}$ for the polarization phase, ${\lesssim}10~{\rm ms}$ for the RF encoding phase, and ${\sim}0.4~{\rm s}$ for the detection phase. The microwave sweep segments are ${\sim}1~{\rm ms}$ long and laser pulses during the detection phase are applied for $100~{\rm \upmu s}$. During the RF encoding phase, the delay, $\tau$, between two slightly detuned ($\delta\approx1~{\rm kHz}$) nuclear-spin $\pi/2$ pulses is varied. For each value of $\tau$, the NV fluorescence time trace is typically averaged over $\geq8$ repetitions of the pulse sequence (\ref{sec:SI_baseline}). 

Figure~\ref{fig:fig2}(b) shows the averaged NV fluorescence signal time trace for three different values of $\tau$. Each point on these plots, $\Delta F/F$, corresponds to the average fluorescence signal collected during a given laser pulse, normalized to the mean fluorescence signal over the entire time trace. For $\tau=0$ (top, red), the nuclear spin ensemble has not precessed at all between the RF pulses and experiences an overall $\pi$ pulse, flipping the spins from $\ket{\downarrow}$ to $\ket{\uparrow}$. Oscillations in $\Delta F/F$ are observed, with an initial contrast of ${\sim}0.4\%$ peak-to-peak, that persist for detection times exceeding $0.2~{\rm s}$ after the RF encoding phase. For $\tau=250~{\rm \upmu s}$ (middle, blue), the nuclear spin ensemble precesses by ${\sim}90\degree$ after the first RF pulse, such that the second RF pulse (of the same phase) largely has no impact and $\braket{I_z}\approx0$. Consequently, no oscillations in  $\Delta F/F$ are observed. For $\tau=500~{\rm \upmu s}$ (bottom, green), the nuclear spin ensemble precesses by ${\sim}180\degree$, such that the final $\pi/2$ pulse leaves the nuclei in approximately the same state as the initial state (prior to the first $\pi/2$ pulse), $\ket{\downarrow}$. Here oscillations in $\Delta F/F$ are once again observed, except now the sign of the oscillations is opposite to that in the $\tau=0$ case.

The $\Delta F/F$ oscillations decay with detection time, but they do so markedly slower than a simple exponential decay would suggest and much slower than the NV longitudinal spin relaxation rate~\cite{JAR2012}. This is an important attribute, since each individual oscillation corresponds to the readout of at most one $^{13}$C spin per NV center, and thus the number of detectable oscillations dictates the degree of repetitive readout. Empirically, we find that the amplitude envelope of the oscillations is well fit to functions of the form: $\Delta F/F = a\,e^{-(t/T)^{\gamma}}+c$. Here, $a$ is the initial amplitude, $\gamma=0.3$ is a stretch factor that we set constant to avoid overfitting without sacrificing fit quality (\ref{sec:SI_spinstate}), $T$ is a characteristic decay time that we further parameterize by the $1/e^2$ decay time $T_{e2}=2^{1/\gamma} T$, and $c$ is a constant that accounts for a persistent fluorescence oscillation that is independent of $\tau$ (\ref{sec:SI_offset}). We attribute the small, persistent oscillation to asymmetries in NV AFP, as we have observed that fine-tuning the MW sweep central frequency can produce $c=0$. In all panels of Fig.~\ref{fig:fig2}(b), the constant oscillation ($c=0.02\%$) was subtracted from the data. Under this fitting procedure, we noticed that $T_{e2}$ is disparate for different signs of $\braket{I_z}$. Figure~\ref{fig:fig2}(c) plots $T_{e2}$ as a function of the average oscillation amplitude, $\overline{\Delta F/F}$, for the largest amplitude readouts. When $\overline{\Delta F/F}<0$, nuclear spins are primarily in $\ket{\uparrow}$ after the RF encoding phase, and they exhibit a shorter detection decay time, $T_{e2}\approx30~{\rm ms}$. When $\overline{\Delta F/F}>0$, nuclear spins are primarily in $\ket{\downarrow}$ after the RF pulses, and they exhibit a longer detection decay time $T_{e2}\approx90~{\rm ms}$. While the exact decay times depend sensitively on choice of empirical fit function, this behavior is qualitatively evident in the data without fitting (\ref{sec:SI_spinstate}). The decay profiles of the $\Delta F/F(t)$ readout oscillations are likely influenced by $^{13}$C depolarization due to optically-induced fluctuations in the $^{13}$C-NV hyperfine interactions~\cite{DUT2007,JIA2008,JIA2009}~(\ref{app:spectator}). Future work may apply detailed numerical modeling to investigate the optical detection signal decay.

To generate a Ramsey interferogram, we use a Fourier processing method that does not rely on curve fitting. For a given value of $\tau$, we take the real part of the Fourier transform of the first $60~{\rm ms}$ of the $\Delta F/F$ time trace and extract the peak amplitude, $\overline{\Delta F/F}$, retaining sign. Figure~\ref{fig:fig2}(d) shows the $\overline{\Delta F/F}(\tau)$ Ramsey interferogram obtained in this manner. These data are fit to the function:
\begin{equation}
\label{eqn:Ramsey}
    \overline{\Delta F/F}(\tau) = A\cos{(2\pi \delta\tau)}\,e^{-\tau/T_2^{\ast}},
\end{equation}
where $|A|$ is the ODNMR Ramsey visibility, $\delta$ is the RF detuning, and $T_2^{\ast}$ is the nuclear spin dephasing time. For the fit to data in Fig.~\ref{fig:fig2}(d), we find $|A|=0.21\pm 0.01\%$, $\delta=863\pm 3~{\rm Hz}$ and $T_2^{\ast}=1.74\pm0.05~{\rm ms}$, which are fairly typical values for this diamond. 

Figure~\ref{fig:fig2}(e) shows the $^{13}$C ODNMR spectrum, obtained by taking a phased Fourier transform of the Ramsey interferogram in Fig.~\ref{fig:fig2}(d). The data are well fit by a Lorentzian function, revealing an NMR central frequency $f_n=129.34\pm 0.04~{\rm kHz}$ and full-width-at-half-maximum (FWHM) linewidth $224\pm 14~{\rm Hz}$ (uncertainty is fit standard error). 

We find that the diamond $^{13}$C NMR spectra obtained by our ODNMR method are similar to those obtained from the bulk using inductive coil detection. In~\ref{sec:SI_coilandNV}, we show a side-by-side comparison of ODNMR and coil-detected NMR spectra obtained at $B_0=20~{\rm mT}$. The ODNMR spectrum using polarization buildup time $T_{\rm polar}\approx0.5~{\rm s}$ has a similar central frequency and lineshape as that obtained by inductive coil detection using $T_{\rm polar,c}=200~{\rm s}$ polarization buildup, aside from the coil-based spectral peak being ${\sim}2$ times narrower. This highlights two key advantages of our method: i) the $^{13}$C diamond ODNMR spectra are somewhat immune to hyperfine coupling distortions, reflecting the bulk NMR spectra, since NV centers are polarized in $\ket{0}$ during the RF encoding phase, and ii) the measurement duty cycle of our ODNMR method, $\phi\approx T_{e2}/(T_{\rm polar}+T_{e2})\approx10^{-1}$, is approximately four orders of magnitude larger than that of conventional coil detection, $\phi\approx T_2^{\ast}/T_{\rm polar,c}\approx10^{-5}$.
 
\subsection{ODNMR Ramsey visibility}
\label{sec:readout_contrast}

\begin{figure*}[t]
    \centering
    \includegraphics[width=0.98\linewidth]{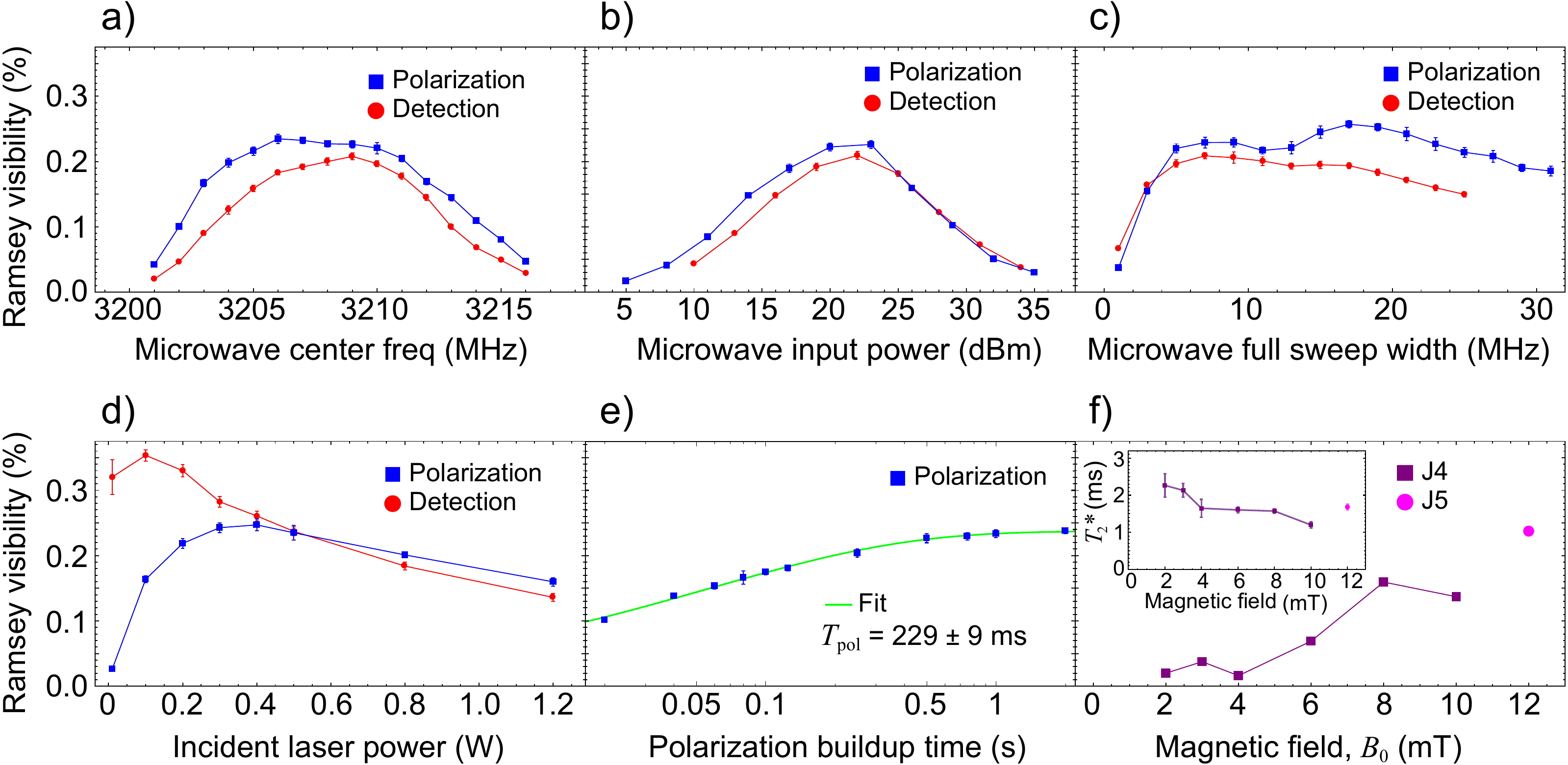}
    \caption{\textbf{Parameter Sweeps.}  (a) ODNMR Ramsey visibility, $|A|$, as a function of the central frequency of the microwave sweep.  (b) Visibility as a function of the microwave power input to the wire loop (effectively varying $\Omega^2$). (c) Visibility as a function of microwave sweep full-span frequency (effectively varying $\dot{f}$). (d) Dependence of visibility on the incident peak laser power. (e) Visibility as a function of the polarization time. (f) Visibility as a function of magnetic field, $B_0$. Measurements for $B_0=2\mbox{--}10~{\rm mT}$ were taken with diamond J4, and the measurement at $B_0=12~{\rm mT}$ was taken with diamond J5. Inset: $T_2^{\ast}$ versus $B_0$, extracted from mono-exponential decay fits. Additional details are provided in the main text and \ref{sec:SI_baseline}. Error bars represent fit uncertainty and do not incorporate the effects of slow drifts in experimental parameters. }
    \label{fig:fig3}
\end{figure*}

A primary figure of merit for the $^{13}$C spin readout fidelity is the initial visibility of the Ramsey fringes, $|A|$. To explore how $|A|$ depends on various experimental parameters, we vary one of the following parameters during the polarization or detection phases, while holding all other parameters at baseline values: microwave sweep central frequency, microwave input power, microwave sweep width, and incident laser power. These experiments are conducted by analyzing $^{13}$C Ramsey interferograms of diamond J5, as in Fig.~\ref{fig:fig2}(d), with baseline parameter values listed in~\ref{sec:SI_baseline}. 

When varying the microwave sweep central frequency, Fig.~\ref{fig:fig3}(a), for both the polarization and detection phases, we observe a broad resonance shape in $|A|$ that mimics the NV $f_+$ ODMR spectrum. The hyperfine structure associated with the intrinsic $^{14}$N is largely obscured due to the relatively large sweep width. The peak ODNMR Ramsey visibility, $|A|\approx0.25\%$, is achievable for a microwave sweep central frequency in a broad range of several MHz about $f_+$.

When varying the microwave power ($\propto\Omega^2$) for both detection and polarization phases, Fig.~\ref{fig:fig3}(b), we again observe a resonance shape in $|A|$. The behavior when $\Omega$ is large can be understood by analyzing Eqs.~\eqref{eq:H} and \eqref{eq:phop}. When $\Omega$ is large compared to the hyperfine and Larmor terms in Eq.~\eqref{eq:H}, this produces a relatively large avoided crossing gap, $\Delta\nu_{23}\approx\Omega$ (\ref{app:delta_nu_omega}). If $\Omega^2\gg|\dot{f}|/(2\pi)$, then the hopping probability is small and polarization transfer is rare; see Eq.~\eqref{eq:phop}. This behavior is clearly seen in numerical density-matrix simulations (\ref{sec:SI_phop}). However, if $\Omega^2\ll|\dot{f}|/(2\pi)$, Eq.~\eqref{eq:phop} loses validity, as the time required for a coherent spin flip greatly exceeds the effective interaction time during the sweep. In this case, the NV center remains in its initial state, and polarization transfer is once again rare.

 In order to vary the microwave frequency sweep rate, $\dot{f}$, we vary the span of the microwave frequency sweep while holding the sweep time constant, Fig.~\ref{fig:fig3}(c). On both polarization and detection phases, when the full sweep width is $\lesssim2~{\rm MHz}$, the ODNMR Ramsey visibility is poor, indicating inefficient polarization transfer. We attribute this to microwaves interacting with only a fraction of the entire inhomogeneously-broadened, $^{14}$N-hyperfine-split NV spin resonance manifold. For larger $|\dot{f}|$, the transfer is relatively efficient, and we find $|A|\approx0.25\%$ for a wide range of values, corresponding to $\sqrt{|\dot{f}|/(2\pi)}\approx20\mbox{--}80~{\rm kHz}$. We interpret this flat response as being due to the presence of a wide range of NV-$^{13}$C hyperfine coupling interactions. In this regime, many NV-$^{13}$C pairs satisfy the condition $\Delta\nu_{23}^2\lesssim|\dot{f}|/(2\pi)$, leading to substantial diabatic $P^{23}_{\rm hop}$ hopping probability. This is supported by numerical density-matrix simulations (\ref{sec:SI_phop}). Combined with nuclear spin diffusion, this behavior enables polarization transfer with a large number of $^{13}$C nuclei. 

Next, we recorded the ODNMR Ramsey visibility as a function of laser power, Fig.~\ref{fig:fig3}(d). Here, the continuous-wave laser power in the polarization phase and the peak power of laser pulses in the detection phase were varied via analog control of the acousto-optic modulator. In the polarization phase, for laser powers below ${\sim}0.1~{\rm W}$ (corresponding to intensity of ${\sim}10~{\rm W/cm^2}$), the ODNMR Ramsey visibility is poor. In this range, the optical excitation rate is comparable to or smaller than the microwave sweep repetition rate (and also not much larger than the NV longitudinal spin relaxation rate), and thus NV centers are not adequately polarized in $\ket{0}$. For laser powers above ${\sim}0.1~{\rm W}$, the Ramsey visibility is relatively constant (|$A|\approx0.25\%$) until the laser power is increased beyond ${\sim}0.5~{\rm W}$. In the high light intensity regime, the ODNMR Ramsey visibility once again drops, an observation that was previously attributed to the disruption of spin coherence during microwave sweeps~\cite{AJO2018,ZAN2019}. 

For the detection phase, the dependence of the ODNMR Ramsey visibility on light intensity is somewhat different from that in the polarization phase. In the detection phase, the laser beam is pulsed with a $10\%$ duty cycle. For peak laser powers below ${\sim}0.1~{\rm W}$ (peak intensity: ${\sim}10~{\rm W/cm^2}$), $|A|$ is at a maximum, exceeding $0.3\%$, corresponding to $\Delta F/F$ oscillations exceeding $0.6\%$ pk-pk. This indicates that complete initialization of NV centers into $\ket{0}$ is not a strict requirement for maximizing Ramsey visibility.  However, as the peak laser power is increased beyond ${\sim}0.2~{\rm W}$, the Ramsey visibility steadily drops. We observed a similar behavior when holding the peak laser power constant and varying the laser excitation duty cycle instead (\ref{sec:SI_pulselength}). We also observed a drop in Ramsey visibility when holding the laser power and duty cycle constant but shrinking the beam size. These results indicate a maximum ``photon budget'' for the number of optical cycles the NV center can undergo during the detection phase. Since laser light is mostly off during microwave sweeps, the average-light-intensity-dependent loss of signal cannot be attributed to instantaneous disruption of spin coherence but rather to slower nuclear depolarization. In~\ref{app:optical_depol}, we propose a mechanism for these depolarization dynamics based on optical fluctuation of the electron spin state of transverse-hyperfine-coupled NV centers, which leads to stochastic rotations of the $^{13}$C quantization axes~\cite{DUT2007,JIA2008,JIA2009}. Future work may apply detailed numerical modeling to analyze the readout photon budget.

\begin{figure*}[hbt]
    \centering
    \includegraphics[width=0.98\linewidth]{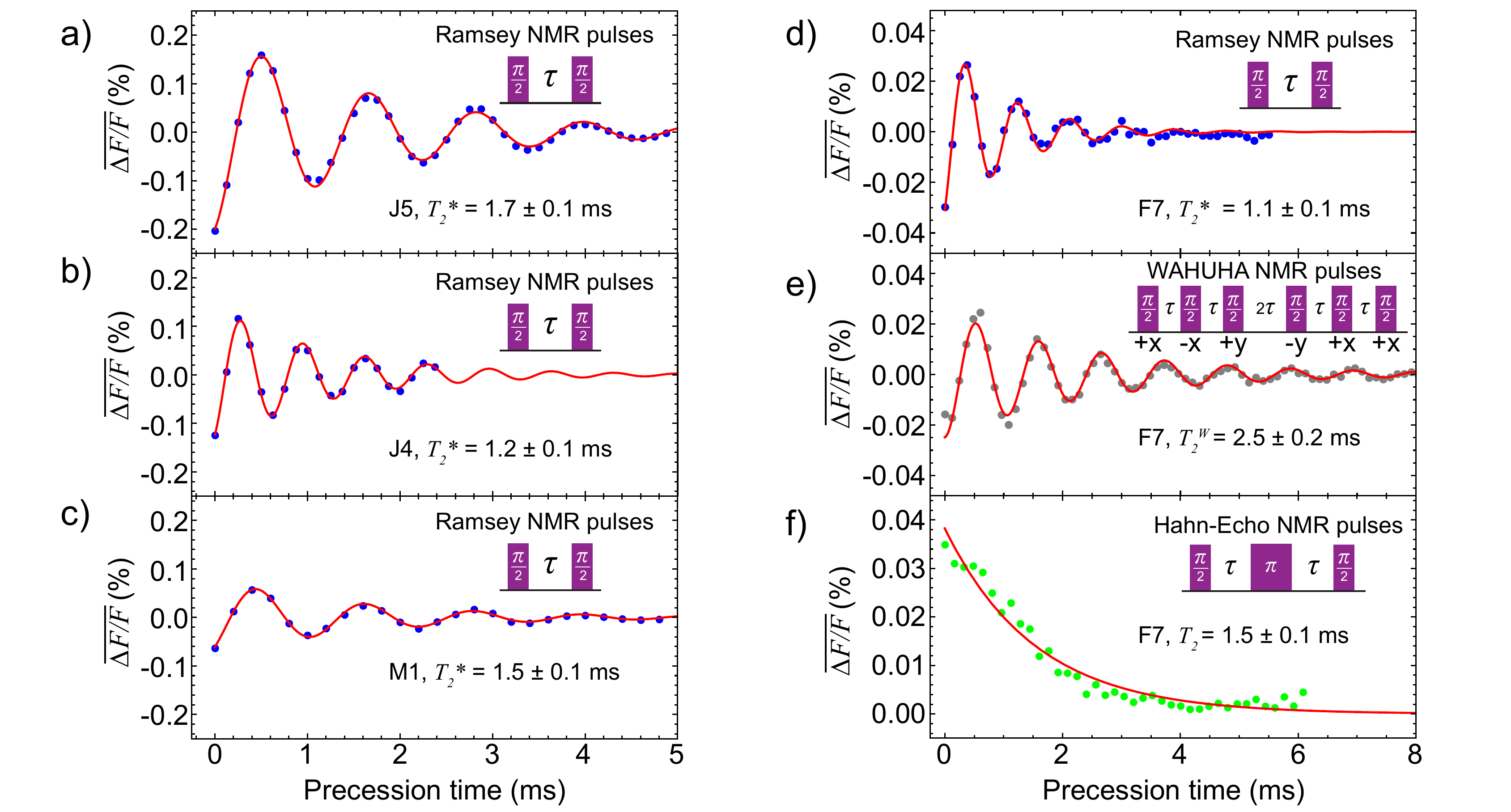}
    \caption{\textbf{Nuclear spin dephasing.} (a)-(d) Ramsey interferograms carried out with various different diamonds. Insets include the Ramsey pulse sequence, the diamond used, and the fitted $^{13}$C nuclear $T_2^{\ast}$ decay time obtained from an exponentially damped sine curve. (e) WAHUHA decay curve for diamond F7. Aside from pulse sequence, other experimental parameters are the same as for (d), including the fit function. The fitted decay time under this pulse sequence, $T_2^W$, is also shown. (f) Hahn echo decay curve using resonant pulses, along with mono-exponential decay fit. \ref{sec:SI_baseline} lists parameters for (a)-(f). }
    \label{fig:fig4}
\end{figure*}

Next, we varied the time of the polarization phase of the experiment to explore its impact on the ODNMR Ramsey visibility. Figure~\ref{fig:fig3}(e) shows that $|A|$ grows with the polarization time, but largely saturates for times $\gtrsim{0.5}~{\rm s}$. We find that the Ramsey visibility after polarization time $t$ can be described by an empirical function (\ref{sec:SI_pol_buildup_fitting}): $|A| = A_{\rm sat} \left (1-e^{-(t/T_p)^{0.5}}\right )$. A fit to the data in Fig.~\ref{fig:fig3}(e) yields a saturation visibility $A_{\rm sat}=0.237 \pm 0.002\%$ and a characteristic $1-1/e^2$ polarization buildup time of $T_{\rm pol} = 4T_p=229\pm9~{\rm ms}$. The sub-exponential polarization buildup curve reflects the ensemble average over several different processes, including the diabatic hopping rates of different NV-$^{13}$C pairs and nontrivial spin diffusion dynamics~\cite{AJO2019,PAG2020}.

Finally, we explored the dependence of $|A|$ on the magnetic field strength, $B_0$. Specifically, we explored the technologically-relevant low-field regime where requirements on field stabilization are less stringent. Figure~\ref{fig:fig3}(f) shows the experimental dependence of $|A|(B_0)$. This panel incorporates data from two versions of the apparatus: an electromagnet with diamond J5 for $B_0\geq12~{\rm mT}$ and a Helmholtz coil with diamond J4 for $B_0\leq10~{\rm mT}$~(\ref{sec:SI_diamond_tables}, \ref{sec:SI_exp_layout}). While the trends are not obvious, it can be inferred from Fig.~\ref{fig:fig3}(f) (see also~\ref{sec:SI_higher_field_odnmr}) that the ODNMR Ramsey visibility reaches a plateau for $B_0\gtrsim8~{\rm mT}$, but it rapidly decays for lower fields. This behavior is partly attributed to the microwave Rabi frequency $\Omega\approx100~{\rm kHz}$ used here. Through density matrix simulations (\ref{sec:SI_phop}), we observe an interplay between $\Omega$ and $\gamma_{\rm nuc}B_0$ terms in Eq.~\eqref{eq:H}, such that significant $P^{23}_{\rm hop}$ diabatic hopping is only possible when $B_0\gtrsim\Omega/\gamma_{\rm nuc}$. Reducing the microwave power might allow for lower field operation, but as noted in the discussion of Figs.~\ref{fig:fig3}(b,c), there are limits associated with the microwave sweep rate needed for diabatic hopping and, ultimately, the NV decoherence rates. We also explored the nuclear spin dephasing time as a function of $B_0$ and found that it is largely constant throughout the $B_0=2{\mbox{--}}12~{\rm mT}$ range, see inset to Fig.~\ref{fig:fig3}(f).

\subsection{Nuclear spin dephasing times}
\label{sec:nuclear_dephasing_times}
We used our ODNMR method to characterize the coherence properties of the weakly coupled $^{13}$C nuclei contributing to the signal. Figure~\ref{fig:fig4}(a-d) shows ODNMR Ramsey interferograms of four natural isotopic abundance diamonds with nitrogen concentrations approximately in the $0.5\mbox{--}10~{\rm ppm}$ range and estimated NV density in the $0.1\mbox{--}1~{\rm ppm}$ range (\ref{sec:SI_diamond_tables}). We find that the ODNMR Ramsey visibility is correlated with impurity density, with the higher impurity density diamonds providing larger visibilities. This could be related to the correlation of NV$^-$:NV$^0$ charge-state ratio with nitrogen density, which tends to provide higher spin-dependent fluorescence contrast in diamonds with higher nitrogen density~\cite{ACO2009}. However, in all diamonds, we observe a nuclear spin dephasing time in the range $T_2^{\ast}=1.1\mbox{--}1.7~{\rm ms}$. 

We further studied the dephasing times of diamond F7 (${\rm [N]}\approx0.5~{\rm ppm}$, ${\rm [NV]}\approx0.1~{\rm ppm}$) under different pulse sequences. The Ramsey dephasing time of this diamond is similar to the others, $T_2^{\ast}=1.1\pm0.1~{\rm ms}$, see Fig.~\ref{fig:fig4}(d). Under a WAHUHA homonuclear dipolar decoupling sequence~\cite{WAU1968}, the dephasing time increases modestly to $T_2^W=2.5\pm0.2~{\rm ms}$, Fig.~\ref{fig:fig4}(e). This suggests that $^{13}$C-$^{13}$C dipolar coupling has a contribution to the Ramsey dephasing, but another mechanism dominates. Since this diamond has very low nitrogen concentration, the P1 paramagnetic impurities are unlikely to be the source. The Hahn echo relaxation time, $T_2=1.5\pm0.1~{\rm ms}$, Fig.~\ref{fig:fig4}(f), exhibits little effect from a refocusing pulse, suggesting a decoherence source that fluctuates on relatively short timescales. 

We expect that longitudinal spin relaxation of neighboring NV centers $T_{1,{\rm nv}}\approx3~{\rm ms}$ is the dominant source of dephasing in this study. NV centers are initialized into $\ket{0}$ during the RF encoding phase of the experiment, but they stochastically hop to $\ket{\pm1}$ on the $T_{1,{\rm nv}}$ timescale, which induces uncontrolled variation in the $^{13}$C NMR frequencies. Since each $^{13}$C spin undergoes a different NMR frequency shift, owing to variation in its hyperfine tensor, the ensemble-averaged effect is to induce dephasing. This is a fundamental liability of our ODNMR approach, as it relies on weak hyperfine coupling to ground-state NV centers, but it may be mitigated in the future through NV microwave decoupling protocols during the RF encoding phase~\cite{CHE2018,KUA2025}. Alternatively, future studies may explore the use of color centers with ground-state spin singlet and metastable triplet states~\cite{PEZ2024,SAK2023,TAT2026}.

\section{\label{sec:Discussion}Discussion and outlook}
We have introduced a method for ODNMR that allows for probing a large number of diamond $^{13}$C spins. Compared to prior diamond ODNMR experiments using the intrinsic $^{14}$N nuclear spin associated with NV centers, our method allows for the measurement of a larger number of nuclear spins and operation in a broad range of magnetic fields. To characterize the relevance of our approach to future precision measurements, recall that the minimum detectable change in spin precession frequency is given by:
\begin{equation}
\label{eqn:sens}
    \delta f \approx \frac{1}{2\pi \mathcal{F} \sqrt{N_p T_2^{\ast}\xi t}}.
\end{equation}
Here, $T_2^{\ast}\approx2~{\rm ms}$ is the nuclear spin dephasing time and we now include a sensing duty cycle factor, $\xi\approx T_2^{\ast}/(T_{\rm polar}+T_{\rm read})\approx5\times10^{-3}$, where $T_{\rm polar}\approx 0.2~{\rm s}$ is the polarization time and $T_{\rm read}\approx0.2~{\rm s}$ is the readout time. The number of polarized nuclear spins in our experiment can be estimated as $N_p\approx N_{\rm rep}N_{nv}\approx10^{16}$, where $N_{nv}\approx10^{14}$ is the number of aligned NV centers interrogated and $N_{\rm rep}\approx10^2$ is the number of repetitive readouts during $T_{\rm read}$. Assuming photon-shot-noise-limited detection, the readout fidelity is estimated as $\mathcal{F}\approx|A|\sqrt{\eta}\approx2.5\times10^{-4}$, where $|A|\approx3\times10^{-3}$ is the Ramsey visibility and $\eta\approx7\times10^{-3}$ is the number of photons detected per $^{13}$C nuclear spin per readout. 

Taken together, we estimate that our current experiments could obtain a frequency shift precision $\delta f\approx2\times10^{-3}~{\rm Hz}$ for $t=1~{\rm s}$ total acquisition time, if they were limited by photon shot noise. This corresponds to an angle random walk of ${\sim}0.7^{\circ}/\sqrt{s}$. If realized, this sensitivity would be nearly an order of magnitude improvement over prior $^{14}$N ODNMR gyroscope experiments~\cite{JAR2021}, but there is much room for improvement. The number of polarized spins could be increased by ${\sim}10$x by using a larger diamond and higher laser power (holding intensity constant). The readout fidelity could be improved by ${\sim}30$x by improving the photon detection efficiency and using a diamond with higher NV spin-dependent fluorescence contrast~\cite{ARA2026}. These relatively straightforward improvements would provide $\delta f\approx2\times10^{-5}~{\rm Hz}$ for $t=1~{\rm s}$. More speculative improvements may come from optimizing the photon budget, through a better understanding of the readout-phase optical power dependence, and potentially using the intrinsic $^{14}$N nuclear spin as a local memory~\cite{SOS2021}. Finally, large gains in sensitivity would come from increasing $T_2^{\ast}$, potentially through microwave decoupling~\cite{CHE2018,KUA2025} or the use of ground-state spin singlet color centers~\cite{PEZ2024,SAK2023,TAT2026}.

In summary, we introduced a method for optical detection of large ensembles of coherent $^{13}$C nuclear spins in diamond. This method is based on microwave-swept dynamics that bidirectionally transfer polarization between NV electron and $^{13}$C nuclear spins. We used it to demonstrate ODNMR spectroscopy at low magnetic fields ($\lesssim10~{\rm mT}$) and ambient temperature. With further optimization, this approach may find application in fundamental physics tests or in compact rotation sensors.

\begin{acknowledgments}
We gratefully acknowledge advice and support from D.~Ferschweiler, C.~Meriles, J.~Damron, A.~McDowell, T.~Ivanov, G.~Birdwell, and D.~Thrasher. \\
\textbf{Competing interests.} A.J. and D.B. have financial interests in the company ODMR Technologies. The authors declare that they have no other competing interests.\\
\textbf{Author contributions.} A.J., M.D.A, J.S., and V.M.A. conceived the idea and designed the experiments. M.D.A. built the main experimental apparatus and Y.S., A.B., D.L., B.A.R., and A.J.P. assisted with apparatus construction and data collection. D.L., C.R., J.C., M.G., S.C., V.M., D.B., and S.L. helped with theoretical analysis and data interpretation. J.S. and M.D.A. wrote the control and automation software. M.D.A. acquired and analyzed the primary data and wrote the initial manuscript draft. V.M.A. supervised the project. All authors helped edit the manuscript. \\
\textbf{Funding.} This work was supported by the National Science Foundation (CHE-1945148, OIA-1921199), Army Research Lab (W911NF-23-2-0092), National Institutes of Health (R42GM145129), and the Moore Foundation (grant DOI 10.37807/GBMF12968). The work of D.B. was supported in part by the Cluster of Excellence ``Precision Physics, Fundamental Interactions, and Structure of Matter'' (PRISMA++ EXC 2118/2) funded by the German Research Foundation (DFG) within the German Excellence Strategy (Project ID 390831469).
\end{acknowledgments}

\clearpage
\appendix
\setcounter{equation}{0}
\setcounter{section}{0}
\makeatletter
\renewcommand{\thetable}{A\arabic{table}}
\renewcommand{\theequation}{A\Roman{section}-\arabic{equation}}
\renewcommand{\thefigure}{A\arabic{figure}}
\renewcommand{\thesection}{Appendix~\Roman{section}}
\makeatother

\section{Coil based readout fidelity}
\label{app:johnson}
Here, we consider the readout fidelity of inductive detection of polarized spin-1/2 nuclei of gyromagnetic ratio $\gamma_{\rm nuc}$ and polarized spin density $\rho_{\rm nuc}$. In the common scenario where spin projection noise is not the dominant source of noise, the signal-to-noise ratio for a single readout (${\rm SNR_{ss}}$) is fundamentally limited by Johnson noise in the coil resonator, given by~\cite{HOU1976}:
\begin{equation}
   {\rm SNR_{ss}} \approx K \epsilon M_0 \sqrt{\frac{\pi\mu_0 Q \gamma_{\rm nuc}B_0 V_c}{2 k_B T_c \Delta f}},
   \label{eq:SNR}
\end{equation}
where $K\approx1$ is a numerical factor based on coil geometry, $\epsilon\approx1$ is the fill factor, $M_0 = \rho_{\rm nuc}\gamma_{\rm nuc}h/2$ is the sample magnetization with Planck constant $h$, $\mu_0$ is the vacuum permeability, $Q$ is the resonator quality factor, $B_0$ is the magnetic field, $V_c$ is the coil volume, $k_B$ is the Boltzmann constant, $T_c$ is the coil temperature, and $\Delta f\approx1/T_2^{\ast}$ is the acquisition bandwidth for a single readout.

Consider an NMR signal, $S(f)$, obtained by demodulating a free induction decay signal down to nearly DC and taking the out-of-phase quadrature. The signal $S(f)$ is then a dispersive Lorentzian of peak-to-trough amplitude $S_{\rm max}$ and peak-to-trough width $\Gamma=1/(\pi T_2^{\ast})$. For a single readout, the uncertainty in the absolute NMR frequency is approximately given by:
\begin{equation}
\label{eq:taylor}
    \delta f_{\rm ss}\approx \frac{\Delta S}{\frac{\partial S}{\partial f}},
\end{equation}
where $\Delta S\approx S_{\rm max}/{\rm SNR_{ss}}$ is the noise in the signal. Taking the approximation ${\partial S}/{\partial f}\approx 2S_{\rm max}/\Gamma$, we rewrite Eq.~\eqref{eq:taylor} as:
\begin{equation}
\label{eq:df}
    \delta f_{\rm ss}\approx\frac{\Gamma}{2\cdot{\rm SNR_{ss}}}\approx\frac{1}{2\pi T_2^*\cdot{\rm SNR_{ss}}}.
\end{equation}
When averaging multiple measurements, each of duration $T_2^{\ast}/\xi$ (where $\xi$ is the sensing duty cycle), for a total time $t$, the overall NMR frequency uncertainty is given by:
\begin{equation}
\label{eq:dftot}
    \delta f=\delta f_{\rm ss}\sqrt{\frac{T_2^{\ast}}{\xi t}}\approx\frac{1}{2\pi\,{\rm SNR_{ss}}\sqrt{T_2^{\ast}\xi t}}.
\end{equation}

In comparing Eq.~\eqref{eq:dftot} with Eq.~\eqref{eqn:sens} of the main text, we see that they are of identical form when the readout fidelity is expressed as:
\begin{equation}
    \mathcal{F}=\frac{\rm SNR_{ss}}{\sqrt{N_{\rm p}}} \approx \sqrt{\frac{\pi \mu_0 h^2 T_2^{\ast} Q \rho_{\rm nuc} \gamma_{\rm nuc}^3 B_0}{8 k_B T_c}}.
    \label{eq:fidelity}
\end{equation}
Taking the following representative values: $Q=100$, $T_c = 300~{\rm K}$, $\gamma_{\rm nuc}=10.7~{\rm MHz/T}$, $B_0=10~{\rm mT}$, and assuming an idealized value for the dephasing time of $T_2^{\ast} = 1~{\rm s}$, we find an upper bound on readout fidelity for inductive detection of $\mathcal{F}\approx2.5\times10^{-3}$ for natural-abundance diamond $^{13}$C nuclear spins with $5\%$ polarization. For thermal polarization of natural abundance diamond $^{13}$C spins, Eq.~\eqref{eq:fidelity} implies a Johnson-noise-limited fidelity $\mathcal{F}\approx10^{-6}$. For $100\%$ polarization in $100\%$ $^{13}$C diamond, we find $\mathcal{F}\approx10^{-1}$.

Quite a few idealizations have been made in deriving these values (noiseless amplifiers, perfect fill factors, ideal coil dimensions, ideal coil resonator behavior, etc.). Real experiments will always have a lower SNR, and thus these readout fidelity values should be taken as virtually unreachable upper bounds. In particular, for Johnson-noise-limited coil detection, the readout fidelity is a function of the bulk polarized spin density, $\rho_{\rm nuc}$, see Eq.~\eqref{eq:fidelity}. This is in contrast with the photon-shot-noise-limited readout fidelity of our ODNMR method (Sec.~\ref{sec:Discussion}), which has only an indirect dependence on the local polarized $^{13}$C spin density. In order to make a realistic comparison of the coil readout fidelity with our ODNMR method, we must first consider how the respective signals depend on polarization time. 

\subsection{Polarization buildup: ODNMR versus coil}
To quantify the bulk polarization buildup, we conducted traditional inductive NMR spectroscopy at a bias field of $B_0 = 20~{\rm mT}$. The same resonant coil used to apply RF pulses is used as an inductive detector. The polarization phase of the experiment is the same as used in our ODNMR protocol, Fig.~\ref{fig:fig1}(c). After the polarization phase, we apply a resonant RF $\pi/2$ pulse to initialize the free induction decay of hyperpolarized $^{13}$C nuclear spins. The coil receive signal is amplified, filtered, and recorded using an oscilloscope. The NMR signal amplitude is extracted by fitting the Fourier transform of the free induction decay to a Lorentzian function. 

\begin{figure}[bh]
    \centering
    \includegraphics[width=0.99\linewidth]{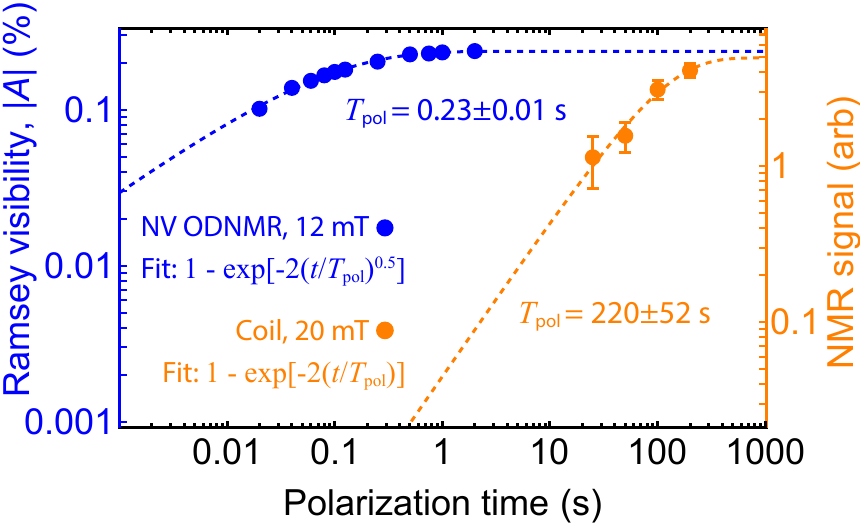}
    \caption{\textbf{Polarization buildup.} ODNMR Ramsey visibility and coil NMR signal amplitude versus polarization time. The ODNMR data are replotted from Fig.~\ref{fig:fig3}(e). Exponential fit functions (see text) are shown as dashed lines and the fitted $1-1/e^2$ polarization buildup times are annotated.}
    \label{fig:coil_vs_diamond}
\end{figure}

Figure~\ref{fig:coil_vs_diamond} shows the inductive NMR signal amplitude as a function of polarization time. At the longest polarization time $T_{\rm polar,c}=200~{\rm s}$, we observe a coil NMR signal strength that corresponds to roughly $5\%$ bulk $^{13}$C nuclear spin polarization. This estimate is based on calibration using the NMR signal from a $^{13}$C enriched glycerol sample with the same setup. The coil NMR data are fit to a simple exponential function, revealing the characteristic $1-1/e^2$ polarization buildup time, $T_{\rm pol}=220\pm52~{\rm s}$. For comparison, we also show in Fig.~\ref{fig:coil_vs_diamond} the ODNMR Ramsey visibility as a function of polarization time (obtained at $B_0=12~{\rm mT}$), reproduced from Fig.~\ref{fig:fig3}(e) of the main text. The bulk polarization buildup for the coil NMR signal is much slower (by a factor of ${\sim}10^3$) than that of the ODNMR signal. 

\subsection{Comparison of readout fidelities under constant sensing duty cycle}
\label{app:johnsonexpt}
We now return to our analysis of the Johnson-noise-limited readout fidelity under inductive detection, Eq.~\eqref{eq:fidelity}. Since $\rho_{\rm nuc}$ depends on the polarization time, $T_{\rm polar,c}$, it also depends on the sensing duty cycle $\xi\approx T_2^{\ast}/T_{\rm polar,c}$. In order to make a fair comparison to the readout fidelity inferred from ODNMR experiments, we require that the same sensing duty cycle $\xi=5\times10^{-3}$ is applied. Adopting the realistic value of $^{13}$C dephasing time observed in our experiments, $T_2^{\ast}=2~{\rm ms}$, we thus use $T_{\rm polar,c}\approx0.4~{\rm s}$, which is the measurement time used in calculations for ODNMR readout fidelity. The polarized nuclear spin density is then given by $\rho_{\rm nuc}=\rho_{\rm nuc,max}(1-e^{-2 T_{\rm polar,c}/T_{\rm pol}})=3.5\times10^{-4}~{\rm nm^{-3}}$, where $\rho_{\rm nuc,max}=0.1~{\rm nm^{-3}}$ is the natural abundance $^{13}$C polarized spin density at $5\%$ polarization and $T_{\rm pol}=220~{\rm s}$ is the fitted value in Fig.~\ref{fig:coil_vs_diamond}. Taking the same representative values as used previously: $Q=100$, $T_c = 300~{\rm K}$, $\gamma_{\rm nuc}=10.7~{\rm MHz/T}$, $B_0=10~{\rm mT}$, and now using the realistic values for $\rho_{\rm nuc}=3.5\times10^{-4}~{\rm nm^{-3}}$ and $T_2^{\ast}=2~{\rm ms}$, we find that the Johnson-noise-limited readout fidelity under inductive detection is $\mathcal{F}\approx7\times10^{-6}$.
\\

\section{Diamond table}
\label{sec:SI_diamond_tables}
Table ~\ref{table:SI_diamondtable} outlines the diamonds used in this experiment and properties associated with each diamond including: dimensions, nitrogen density, NV density, electron irradiation fluence, irradiation energy, and the figures in the paper where the diamond was used to acquire data. 

\begin{table}[h]
  \centering
  \renewcommand{\arraystretch}{1.3} 
  \resizebox{\columnwidth}{!}{
  \begin{tabular}{l|c|c|c|c|c|c|l}
    \toprule
    \textbf{Diamond} & \textbf{Dim.} & \textbf{[N]} & \textbf{[NV]} & \textbf{Fluence} & \textbf{Surface} & \textbf{Energy} & \textbf{Figures} \\
    \textbf{Name} & \textbf{(mm)} & \textbf{(ppm)} & \textbf{(ppm)} & \textbf{($10^{17}$/cm$^2$)} & \textbf{cut} &\textbf{(MeV)} & \\
    \midrule
    J5 & \makecell{2.5$\times$2.3\\$\times$0.8} & ${\sim}10$ & 1 & 6 & (100) & 1 & \makecell[l]{\ref{fig:fig2}(b-e), \ref{fig:fig3}(a-f), \ref{fig:fig4}(a), \\\ref{fig:SI_coilandNV},\ref{fig:SI_pulselength},\ref{fig:SI_pol_buildup_fitting},\ref{fig:SI_higher_field_odnmr}}
    \\
    \hline
    J4 & \makecell{3.5$\times$3.4\\$\times$1.1} & 3 & 0.3 & 2 & (100) & 1 & \makecell[l]{\ref{fig:fig3}(f), \\ \ref{fig:fig4}(b)} \\
    \hline
    M1 & \makecell{3$\times$3\\$\times$0.5} & ${\sim}1$ & 0.3 & --- & (100) &--- & \ref{fig:fig4}(c) \\
    \hline
    F7 & \makecell{0.6$\times$0.3\\$\times$0.1} & 0.5 & 0.1 & --- & (110) & --- & \ref{fig:fig4}(d-f) \\
    \bottomrule
  \end{tabular}
  }
  \caption{\textbf{Diamond properties.} For J4, J5, and M1, [N] is listed as reported by the vendors. For M1, [NV] is listed as reported by the vendor. For F7, [NV] is inferred from NV $T_2$ measurements~\cite{BAU2020}. For J5 and J4, [NV] is estimated based on relative fluorescence brightness. M1 and F7 were irradiated and annealed by vendors with unspecified parameters. J5 and J4 were annealed in vacuum at $800^{\circ}~{\rm C}$ for 4 hours and $1100^{\circ}~{\rm C}$ for 2 hours.}
  \label{table:SI_diamondtable}
\end{table}

Diamond J5 was used to take most of data shown throughout the manuscript, including Fig.~\ref{fig:fig2}(b-e), Fig.~\ref{fig:fig3}(a-e), one data point on Fig.~\ref{fig:fig3}(f), and Fig.~\ref{fig:fig4}(a). Additionally, J5 was used during the initial stages of the experiment operating at higher fields, $B_0=12\mbox{--}50~{\rm mT}$. Although we did not study the ODNMR effects systematically at fields above 12 mT, we did not find much difference in performance in this field range (\ref{sec:SI_higher_field_odnmr}).

The experiment was rebuilt with a Helmholtz coil to produce a uniform, temporally-stable low field, and during this time J5 was replaced with diamond J4. The J4 diamond was used for the low field measurements in most of Fig.~\ref{fig:fig3}(f). J4 was also used for Fig.~\ref{fig:fig4}(b). We then tested our experimental technique on the M1 and F7 diamonds. M1 was used for Fig.~\ref{fig:fig4}(c) and F7 was used for Fig.~\ref{fig:fig4}(d-f).

\section{Experimental layout}
\label{sec:SI_exp_layout}
A diagram of the experimental setup is shown in Fig.~\ref{fig:fig1}(a). Here, we provide additional details.

An epifluorescence optical setup is used to illuminate the diamond and collect and detect the resulting NV fluorescence. For excitation, a $532~{\rm nm}$ Lighthouse Photonics Sprout-D laser is used that provides a variable power up to $5~{\rm W}$. The beam is shrunk with telescoping lenses and passes through an acousto-optic modulator (Brimrose, TeO$_2$) used for analog control over the laser power. Two RF signals ($80~{\rm MHz}$) of variable amplitude are generated by two independent channels of a function generator and pass through a single-pole double-throw switch (Mini-Circuits, ZASWA-2-50DR+). A TTL signal applied to the switch toggles between the two RF signals to allow for different amplitudes for the polarization and detection phase. The output of the switch passes through a second identical switch which is used to turn on and off the RF signal (via a second TTL pulse signal) to form the laser pulses used in the detection phase of the experiment. The output of the second switch passes through an amplifier and is connected to the transducer of the AOM. In this way, we have the ability to rapidly toggle between two laser powers and also to pulse the laser beam on and off.

After the AOM, the beam passes through an anamorphic prism pair to elongate one dimension of the beam cross section by a factor of 4. The beam also passes through a waveplate which offers polarization control of the laser light. A final pair of lenses is used to project the beam to a ${\sim}2\times0.5~{\rm mm^2}$ cross section that is incident on one of the side faces of the diamond. The second lens in this lens pair is a 2-inch-diameter 0.6 numerical-aperture lens that is also used to collect fluorescence from the diamond. In between this lens pair there is a dichroic mirror that is used to separate red-shifted NV fluorescence from the laser light. The fluorescence passes through the dichroic mirror and is focused by another lens onto an amplified photodetector (Thorlabs PDA36A2). This optics configuration is effective at collecting NV fluorescence from a ${\sim}2\times2~{\rm mm^2}$ region in the focal plane, which is comparable to the largest dimension of the laser beam illumination area.

To control the microwave and RF delivery timing, shown in Fig.~\ref{fig:fig2} and Fig.~\ref{fig:fig4}, we use a SpinCore Pulseblaster ESR-PRO 500 MHz timing card to send TTL signals to control switches and triggers to equipment used in the experiment. The Pulseblaster card is controlled either by the native PulseBlaster Interpretor software or by an additional LabView wrapper used to automate a full Ramsey experiment.

Two microwave signal generators (SRS SG384) are used to drive NV spin transitions with continuous frequency sweeping--one for the polarization phase and another for the detection phase. For polarization, the internal asymmetric ramp of the signal generator is used for frequency sweeping. For detection, we use a separate function generator to generate triangle waveforms which are sent to the external frequency modulation port of the microwave signal generator. The signals from both microwave signal generators are sent to a single-pole double-throw switch (Mini-Circuits, ZASWA-2-50DR+) that allows us to toggle between them using TTL pulses. The output of the switch passes through a second identical switch which is used to turn on and off the microwave signal (via a second TTL pulse signal). The output is amplified (Mini-Circuits ZHL-16W-43-S+) and connected to the microwave delivery wire that is wrapped around the diamond.

For RF generation, we used an arbitrary waveform generator (AWG) as a source for pulses. For Ramsey spectroscopy, the signal generator constantly outputs the RF signal and we use switches to form $\pi/2$ pulses. This ensures that the pulses are of the same phase. The pulsed RF signal is passed through an RF amplifier (Tomco BT00250-AlphaS) whose output is blanked when pulses are off. The amplifier output is fed to the RF coil wrapped around the diamond. For WAHUHA and Hahn-Echo pulse sequences, we generate the pulse sequences using the AWG function of the signal generator to provide precise phase and amplitude control. For RF signals below 100 kHz (corresponding to magnetic fields $<10~\rm{mT}$), an audio amplifier is used instead of the Tomco RF amplifier. Also, for the data in Fig.~\ref{fig:SI_coilandNV}, there is an additional resonant circuit connected to the RF coil. This circuit is responsible for amplifying the RF transmit pulses, amplifying the coil-based NMR receive signal, and isolating the transmit and receive via a $\pi$ network~\cite{SIL2023}.

Signals from the coil receiver and photodetector are digitized and saved via an oscilloscope. 

For fields at or above $12~{\rm mT}$, we use an electromagnet to generate a bias field. The magnet is passively shimmed such that field gradients do not noticeably broaden the NMR lines below $50~{\rm mT}$. We observe similar NMR linewidths when operating the electromagnet at both $50~{\rm mT}$ and $12~{\rm mT}$. For fields at or below $12~{\rm mT}$, we generated our bias field with a Helmholtz coil pair. The nearly identical $T_2$ and $T_2^{\ast}$ values of Figs.~\ref{fig:fig4}(d,f) provide evidence that magnetic field gradients are negligible.

The diamond is water cooled through thermal contact to an AlN substrate which itself is connected to a hollow piece of brass with water running through it. This assembly along with the coils wrapped around the diamond are contained within an aluminum box for RF shielding purposes. The aluminum box has holes drilled for optical and electrical access.

\subsection{Detailed pulse sequence}
\label{sec:SI_detailed_pulse_sequence}

\begin{figure}[hbt]
   \centering
    \includegraphics[width=0.98\linewidth]{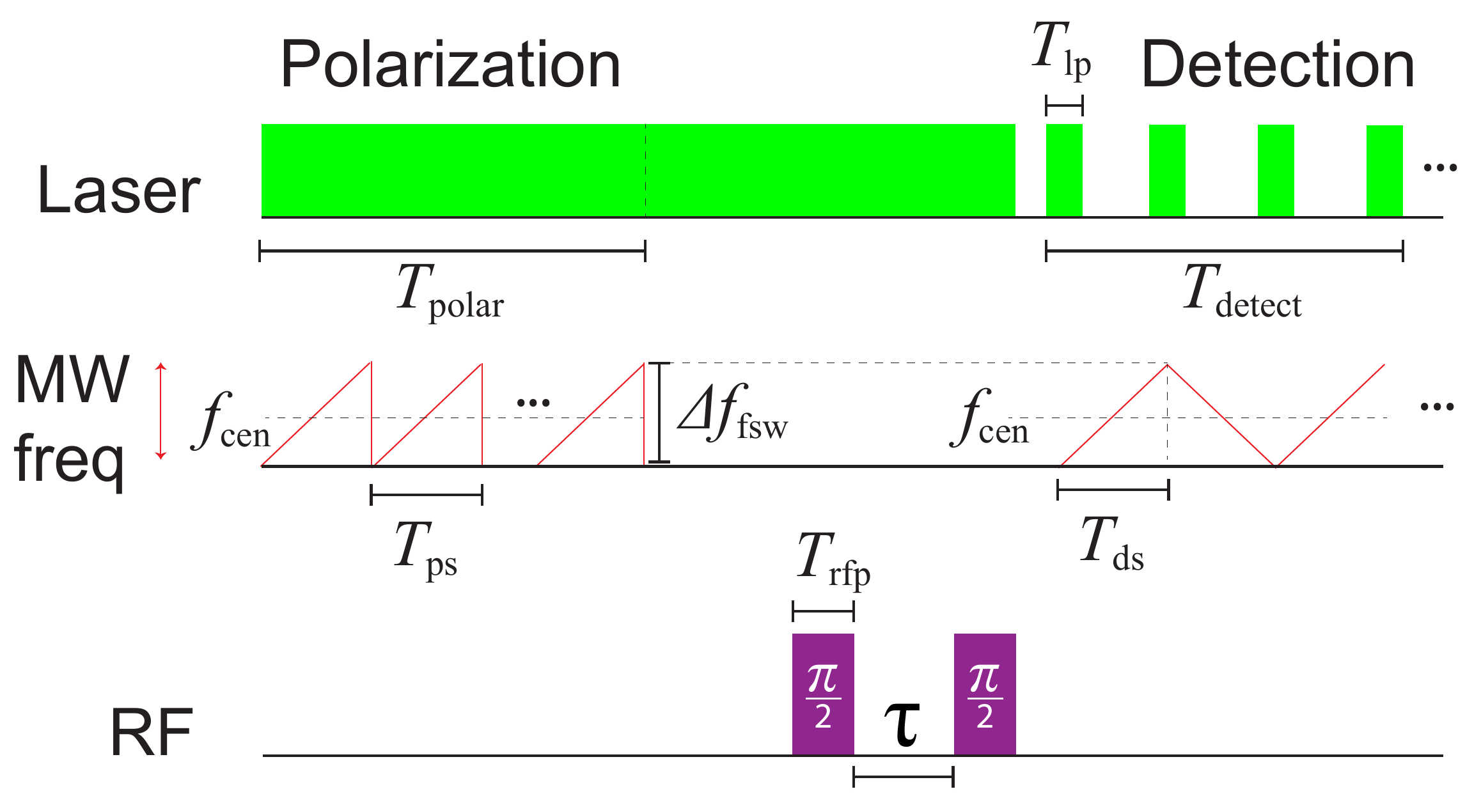}
    \caption{
    \textbf{Detailed pulse sequence.} Annotated version of the ODNMR pulse sequence. See text for typical values.}
    \label{fig:SI_detailed_pulse}
\end{figure}

Figure~\ref{fig:SI_detailed_pulse} provides a detailed view of the standard ODNMR detection sequence. Typical experimental parameters are as follows. $T_{\rm lp}=100~{\rm \upmu s}$ is the duration of a laser pulse. $T_{\rm polar} \approx 0.5~{\rm s}$ is the polarization-phase time. $T_{\rm detect}=400~{\rm ms}$ is the detection-phase time. $T_{\rm ps}=0.83~{\rm ms}$ is the time of a microwave sweep during the polarization phase. $T_{\rm ds}=1~{\rm ms}$ is the time of a microwave sweep during the detection phase. $T_{\rm rfp} \approx 50~{\rm \upmu s}$ is the duration of a radio frequency pulse. $\Delta f_{\rm fsw}=9~{\rm MHz}$ is the full sweep width of a microwave frequency sweep. $f_{\rm cen} \approx 3209~{\rm MHz}$ is the microwave center frequency. $\tau \leq 10~{\rm ms}$ is the maximum time of the RF encoding phase.

\section{Diabatic hopping simulations}
\label{sec:SI_phop}

\subsection{Distribution of hyperfine parameters}
\label{sec:SI_dist}
Here, we estimate the distribution of hyperfine coupling parameters between a typical NV center and its surrounding $^{13}$C atoms. We make the simplifying assumption that the hyperfine coupling arises purely from spin dipole-dipole interaction, neglecting the small fraction of $^{13}$C with larger coupling due to Fermi contact interaction. We place $^{13}$C spins randomly within a sphere of radius $R=3~{\rm nm}$ and compute the distance $r$ from the sphere's origin. The number of $^{13}$C spins placed in a given simulation run is $N_{tot}=4\pi R^3\rho/3$, where $\rho=1.9~{\rm nm^{-3}}$ is the $^{13}$C density in natural isotopic abundance diamond. 

\begin{figure}[htb]   
\includegraphics[width=0.98\columnwidth]{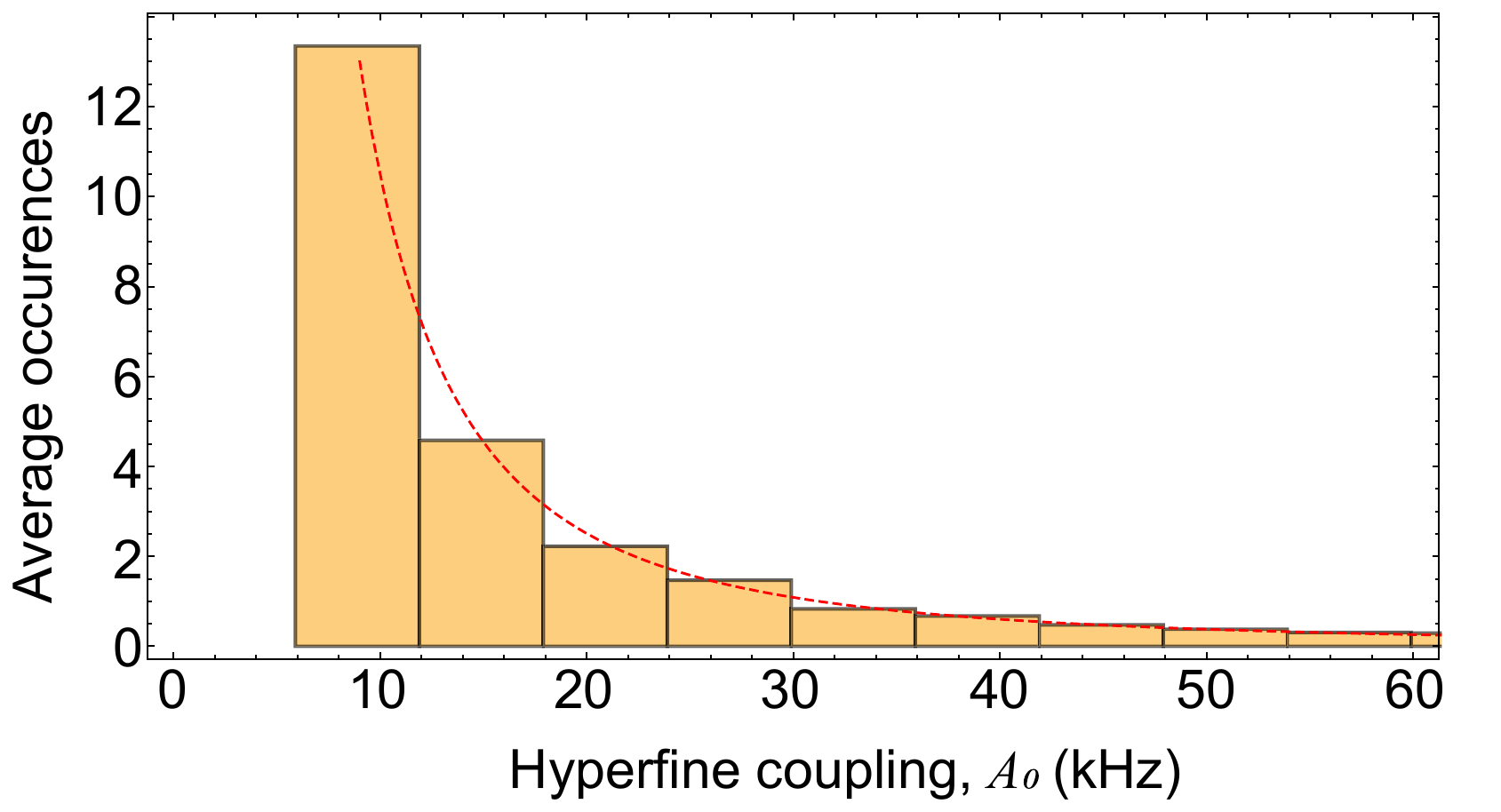}
    \caption{\textbf{NV-$^{13}$C hyperfine distribution.} Histogram of the number of $^{13}$C of a given hyperfine magnitude, $A_0$, for a typical NV center. The calculation uses randomly placed spins according to the $^{13}$C density and assumes purely dipolar hyperfine coupling, see Eq.~\eqref{eq:ao}. The dashed red line indicates the $1/A_0^2$ dependence, see Eq.~\eqref{eqn:hfshell}.} 
    \label{fig:SI_histogram}
\end{figure}

The resulting distribution of $r$ is then converted to a distribution in hyperfine coupling magnitude, $A_0$, according to: 
\begin{equation}
\label{eq:ao}
    A_0 = \frac{\mu_0}{4\pi} \frac{h\,\gamma_{\rm nv} \, \gamma_{nuc}}{r^3}\equiv k_0/r^3,
\end{equation}
where $\mu_0=4\pi \times 10^{-7}~{\rm T^2\, m^3/J}$ is the vacuum permeability, $h=6.626\times10^{-34}~{\rm J/Hz}$ is Planck's constant, $\gamma_{\rm nv}=28.03~{\rm GHz/T}$ is the NV gyromagnetic ratio, $\gamma_{\rm nuc}=10.71~{\rm MHz/T}$ is the $^{13}$C gyromagnetic ratio, and $k_0=\mu_0 h \gamma_{\rm nv}\gamma_{nuc}/(4\pi)=19.9~{\rm kHz\,nm^3}$ is a coupling coefficient used to simplify expressions. The simulations are repeated $10^4$ times and the resulting histograms of $A_0$ are averaged together to represent the ensemble average. 

Figure~\ref{fig:SI_histogram} shows the resulting histogram, using bin widths of $\delta\!A=6~{\rm kHz}$. The occurrences exhibit a $1/A_0^2$ dependence. This can also be shown analytically by integration using Eq.~\eqref{eq:ao} and taking the limit as $\delta\!A\rightarrow0$. In this case, the number of $^{13}$C spins with hyperfine coupling magnitude between $A_0$ and $A_0+\delta\!A$ is:
\begin{equation}
\label{eqn:hfshell}
    N_{\rm shell} = \frac{4}{3}\, \pi\, \frac{\rho\, k_0\, \delta\!A}{A_0^2}.
\end{equation}
The $1/A_0^2$ distribution is critical for interpreting the range of hyperfine couplings of $^{13}$C-NV spin pairs that contribute to the observed ODNMR signals. Specifically, lower hyperfine coupling magnitudes are more likely to contribute, owing to the much larger number of available $^{13}$C spins, so long as the other conditions for diabatic hopping (described in the main text) are met. 

\subsection{ODNMR readout mechanism}
\label{app:delta_nu_omega}
\begin{figure}[hbt]
    \centering
    \includegraphics[width=0.95\linewidth]{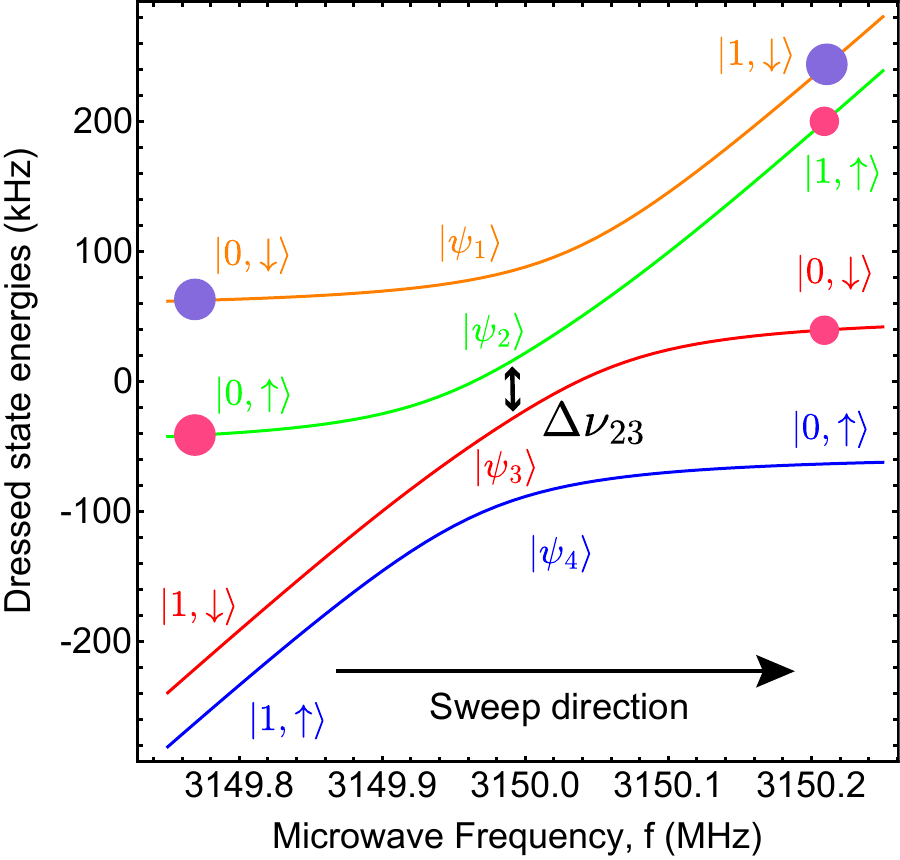}
    \caption{\textbf{Detailed dressed-state diagram.} Example dressed-state level diagram depicting the frequency splitting $\Delta \nu_{23}$ and the population distribution before and after an upwards microwave ($\dot{f}>0$) sweep of initial states $\ket{0,\downarrow}$ (purple) or $\ket{0, \uparrow}$ (pink). This example assumes $P^{23}_{\rm hop}\approx0.5$.}
    \label{fig:Larger_gap_annotation}
\end{figure}

As discussed in Sec.~\ref{sec:expt} of the main text, the ODNMR readout mechanism relies on the presence of a nuclear-spin-selective probability of observing a diabatic hop as the microwave frequency is swept. Figure~\ref{fig:Larger_gap_annotation} provides a detailed annotated version of Fig.~\ref{fig:fig1}(d) of the main text to further clarify the concept. We assume the NV electron spin state is initialized into $\ket{0}$ at the beginning of every microwave frequency sweep. Generally, the probability of performing a diabatic hop is high if the sweep rate, $\dot{f}$, and the frequency splitting between two coupled dressed eigenstates, $\Delta\nu_{ij}$, satisfy the condition $\Delta\nu_{ij}\lesssim\sqrt{|\dot{f}|/2\pi}$, see Eq.~\eqref{eq:phop}. The frequency splittings $\Delta \nu_{ij}$ are determined by the parameters in the Hamiltonian, Eq.~\eqref{eq:H}, including the magnetic field $B_0$, the microwave Rabi frequency $\Omega$, and the hyperfine coupling parameters $A_{zz}$ and $A_{zx}$. In order to engineer a diabatic hopping probability that depends on the nuclear spin state, one must select parameters that result in an ordering of frequency splittings: $\Delta \nu_{14}>\sqrt{|\dot{f}|/2\pi}\gtrsim\Delta\nu_{23}$. In the limit that the microwave Rabi frequency is much larger than the hyperfine coupling and nuclear Larmor terms of the Hamiltonian, the energy levels are dominated by the electron spin Hamiltonian. In this regime, the frequency splittings are $\Delta \nu_{14}\approx\Delta \nu_{23}\approx\Omega$, as in the textbook example of dressed states for a driven spin-1/2 system~\cite{SUT1997}.

\subsection{Diabatic hopping probability}
\label{app:diabatic_hop}

\begin{figure*}[htb]   
\includegraphics[width=0.8\linewidth]{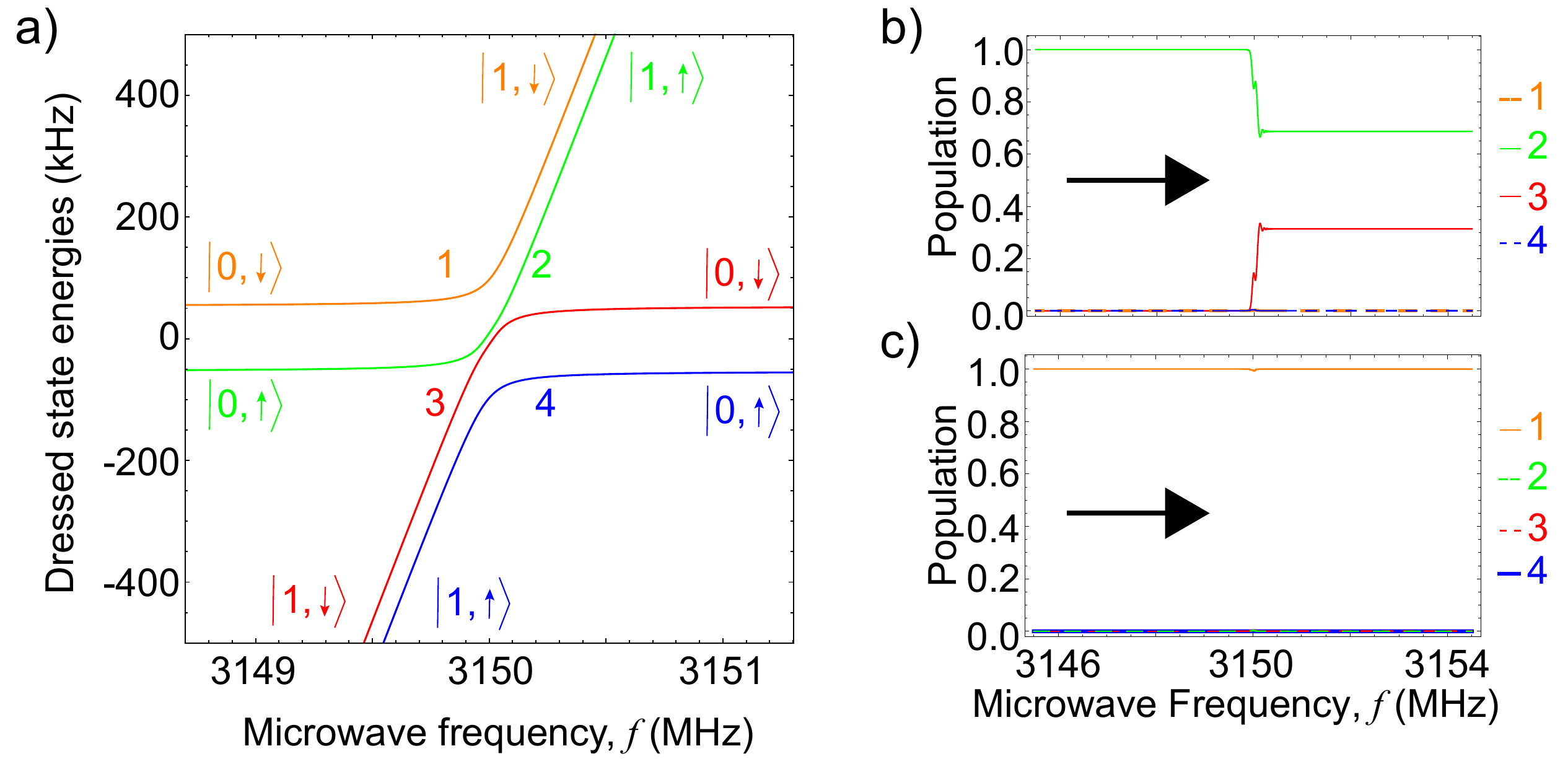}
    \caption{\textbf{Simulating diabatic hopping dynamics.} (a) Dressed-state energy diagram from Fig.~\ref{fig:fig1}(d). (b) Simulated dressed-state populations during a microwave sweep, with initial population in $\ket{\psi_2}$. (c) Same as (b) but with initial population in $\ket{\psi_1}$.}
    \label{fig:SI_sim_explain}
\end{figure*}

Following Ref.~\cite{ZAN2019}, we numerically computed the probability to transfer population between hyperfine dressed states using density matrix modeling. We only consider one NV-$^{13}$C pair, so we can reduce the spatial dimensions by forcing the spins to lie in the xz-plane, with polar angle $\theta$ between them. Therefore, under a suitable secular approximation~\cite{ZAN2019}, we only consider $A_{zz}$ and $A_{zx}$ components of the hyperfine tensor. Considering only spin dipole-dipole hyperfine interaction, these components are given by $A_{zz}=A_0[1+3\cos(2\theta)]/2$ and $A_{zx}=3A_0\sin(2\theta)/2$.

Under a microwave drive of frequency $f$ and Rabi frequency $\Omega$, the Hamiltonian of Eq.~\eqref{eq:H} in the main text can be written in matrix form as:
\begin{equation}
\label{eq:hamiltonian_matrix}
\renewcommand{\arraystretch}{2.2}
H {=} \begin{array}{c@{\;}c}
 & \begin{array}{cccc}
 \hspace{-3mm}
    \makebox[3.2em][c]{\small$\ket{0,\uparrow}$} &
   \makebox[2.2em][c]{\small$\ket{0,\downarrow}$} &
   \makebox[4.5em][c]{\small$\ket{1,\uparrow}$} &
   \makebox[6.5em][c]{\small$\ket{1,\downarrow}$}
   \end{array} \\
\begin{array}{c}
  \mbox{\small$\bra{0,\uparrow}$}  \\
  \mbox{\small$\bra{0,\downarrow}$} \\
  \mbox{\small$\bra{1,\uparrow}$}  \\
  \mbox{\small$\bra{1,\downarrow}$}
\end{array}
&
\begin{pmatrix}
  -\tfrac{f_n}{2}   & 0                 & \tfrac{\Omega}{2} & 0 \\
  0                 & \tfrac{f_n}{2}    & 0                 & \tfrac{\Omega}{2} \\
  \tfrac{\Omega}{2} & 0                 & \scriptstyle f{-}f_{+}{-}\tfrac{f_n}{2}{+}\tfrac{A_{zz}}{2} & \tfrac{A_{zx}}{2} \\
  0                 & \tfrac{\Omega}{2} & \tfrac{A_{zx}}{2} & \scriptstyle f{-}f_{+}{+}\tfrac{f_n}{2}{-}\tfrac{A_{zz}}{2}
  
\end{pmatrix} \vspace{3mm}
\end{array}
\end{equation}

To track dynamics, we let the system start with population in either $\ket{0,\uparrow}$ or $\ket{0,\downarrow}$, due to optical pumping of the NV center. We express the drive frequency as $f(t)=f_{+}-\Delta f/2 + \dot{f}\,t$, where $\Delta f$ is the full sweep span and $\dot{f}$ is a constant sweep rate. We then discretely increment $f$ in $1~{\rm kHz}$ increments from $f_{+}-\Delta f/2$ to $f_{+}+\Delta f/2$, corresponding to steps of $\Delta t=111~{\rm ns}$ from $t=0$ to $t=1~{\rm ms}$ for the baseline sweep rate $|\dot{f}|=9~{\rm MHz/ms}$. At each step, we apply the unitary time evolution operator $U=e^{-2 \pi i H (\Delta t)}$, with $H$ evaluated at the corresponding value of $f(t)$, and calculate the new state populations. The new populations are then input into the next step for evolution.

We run two sets of simulations to gain an understanding of the full dynamics of polarization transfer. One set initializes the system to $\ket{0,\downarrow}$ (dressed state $\ket{\psi_1}$) prior to sweeping, and the other set initializes the system into $\ket{0,\uparrow}$ (dressed state $\ket{\psi_2}$). By evaluating the population at the end of each sweep, we can evaluate all diabatic hopping probabilities: $P^{12}_{\rm hop}$, $P^{13}_{\rm hop}$, $P^{14}_{\rm hop}$, $P^{21}_{\rm hop}$, $P^{23}_{\rm hop}$, and $P^{24}_{\rm hop}$.

Figure~\ref{fig:SI_sim_explain}(a) is a plot of the dressed state energies as a function of applied microwave frequency, reproduced from Fig.~\ref{fig:fig1}(d) of the main text. Figure~\ref{fig:SI_sim_explain}(b) shows the population of states for each step of a simulation initialized in $\ket{0,\uparrow}$ (dressed state $\ket{\psi_2}$). By the end of the sweep, the population in $\ket{\psi_2}$ has partially transferred to $\ket{\psi_3}$; specifically, $P^{23}_{\rm hop}= 0.31$ and $P^{21}_{\rm hop}=P^{24}_{\rm hop}=0$. Fig.~\ref{fig:SI_sim_explain}(c) shows the population of states at each step of a simulation initialized in $\ket{0,\downarrow}$ (dressed state $\ket{\psi_1}$). By the end of the sweep, the population in $\ket{\psi_1}$ has not transferred to any other state, as $P^{12}_{\rm hop}=P^{13}_{\rm hop}=P^{14}_{\rm hop}=0$. For these example plots in Fig.~\ref{fig:SI_sim_explain}, the input parameters include: $B_0=10~{\rm mT}$, $A_{zz}=A_{zx}=30~{\rm kHz}$, $\Omega=100~{\rm kHz}$, and $|\dot{f}|=9~{\rm MHz/ms}$.

We run this simulation while varying relevant parameters of the Hamiltonian, Eq.~\eqref{eq:hamiltonian_matrix}. Throughout nearly the entire range studied here, we find that the only significantly non-zero diabatic hopping probability is $P^{23}_{\rm hop}$. We thus focus our discussion on this nuclear-spin-flipping diabatic hopping probability.

\begin{figure}[htb]   
\includegraphics[width=\columnwidth]{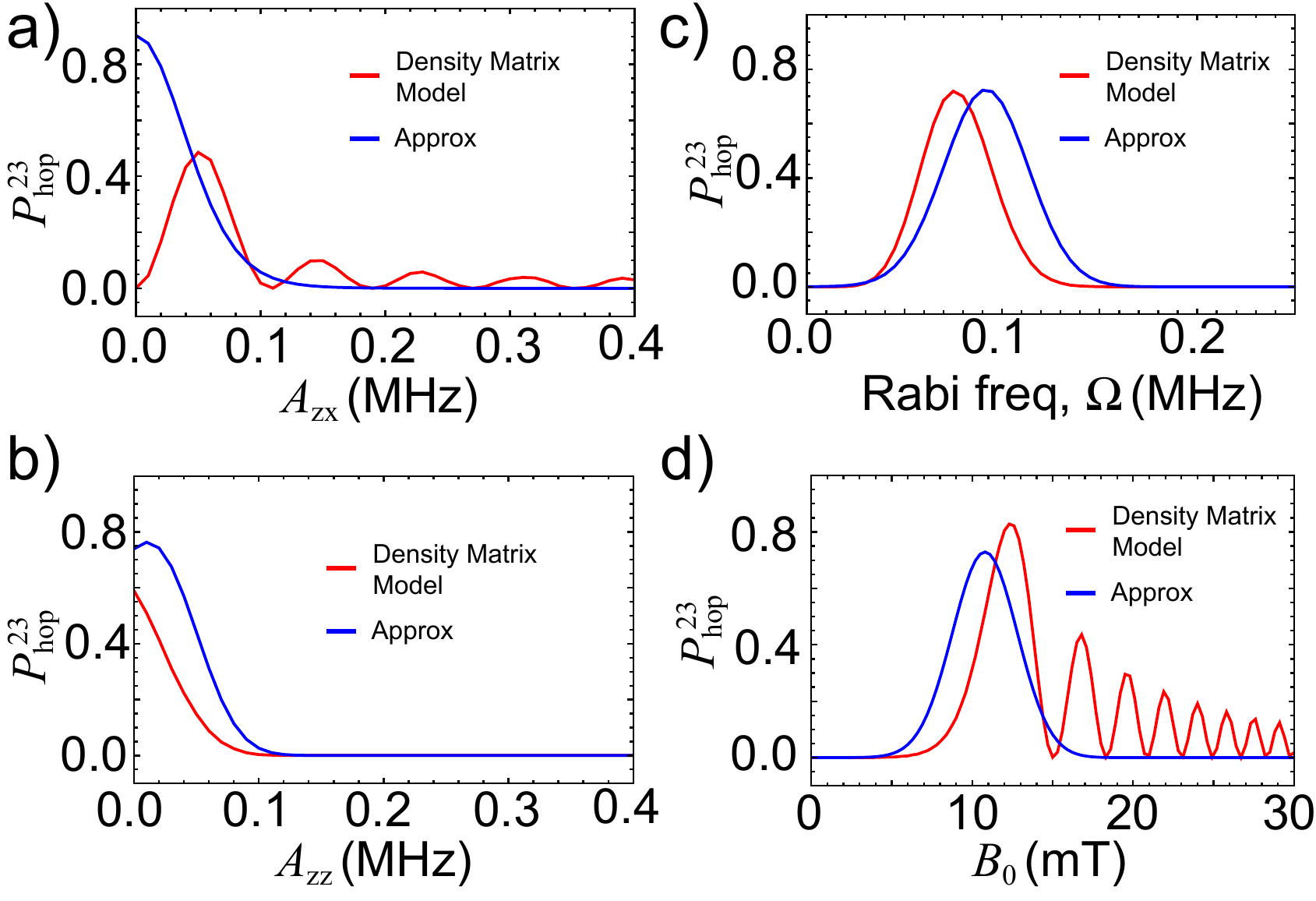}
    \caption{\textbf{One dimensional parameter sweeps.} Diabatic hopping probability, $P^{23}_{\rm hop}$, as a function of (a) $A_{zx}$, (b) $A_{zz}$, (c) $\Omega$, and (d) $B_0$ based on the density matrix model described in \ref{sec:SI_phop}. Red curves correspond to simulation and blue curves correspond to Eq.~\eqref{eq:phop}. The baseline values for all parameters when not being swept are: $B_0=10~{\rm mT}$, $A_{zz}=A_{zx}=30~{\rm kHz}$, $\Omega=100~{\rm kHz}$, and $|\dot{f}|=9~{\rm MHz/ms}$.}
    \label{fig:SI_1d_sims}
\end{figure}

\begin{figure*}[htb]   
\includegraphics[width=0.98\linewidth]{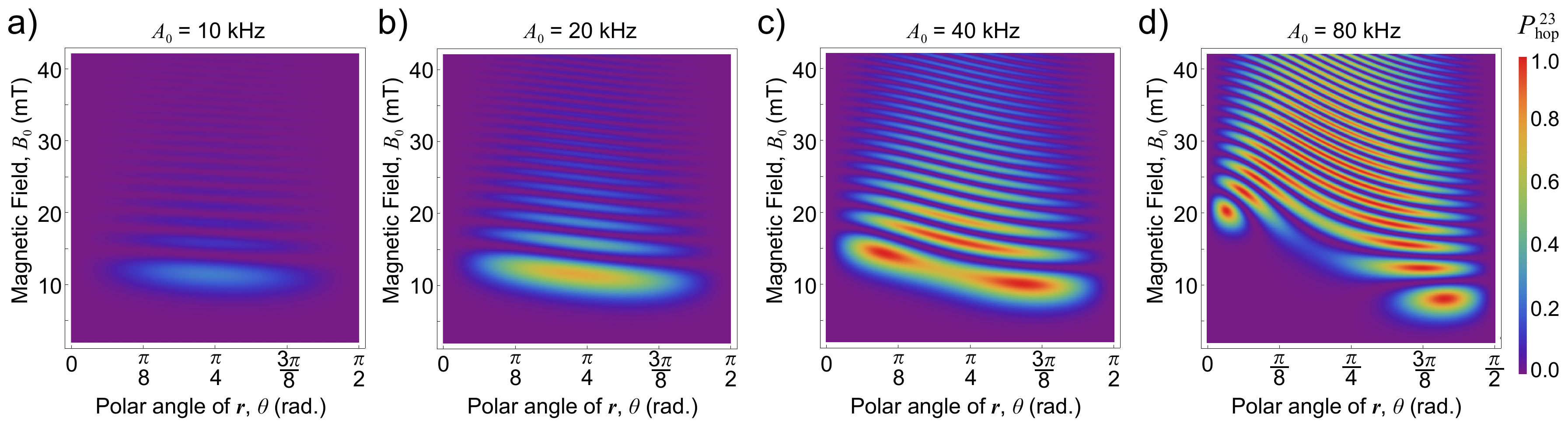}
    \caption{\textbf{Two dimensional parameter sweeps - $B_0$ and $\theta$.} Diabatic hopping probability, $P^{23}_{\rm hop}$, as a function of $B_0$ and $^{13}$C-NV displacement vector polar angle $\theta$. The $A_{zx}$ and $A_{zz}$ hyperfine parameter magnitudes are cyclic for polar-angle intervals spanning $\pi/2$, resulting in redundancies in $P^{23}_{\rm hop}$, so only $\theta=[0,\pi/2]$ is shown. Array plots are shown for four hyperfine coupling magnitudes: (a) $A_0{=}10~\rm{kHz}$, (b) $A_0{=}20~\rm{kHz}$, (c) $A_0{=}40~\rm{kHz}$, (d) $A_0{=}80~\rm{kHz}$. For all plots, $\Omega=100~{\rm kHz}$ and $|\dot{f}|=9~{\rm MHz/ms}$.}
    \label{fig:SI_2d_sims_field}
\end{figure*}

Figure~\ref{fig:SI_1d_sims} shows the results of ``one-dimensional sweeps'', where we vary only one parameter variable at a time and extract $P^{23}_{\rm hop}$. Specifically, Fig.~\ref{fig:SI_1d_sims}(a) shows results for varying $A_{zx}$, Fig.~\ref{fig:SI_1d_sims}(b) varies $A_{zz}$, Fig.~\ref{fig:SI_1d_sims}(c) varies $\Omega$, and Fig.~\ref{fig:SI_1d_sims}(d) varies $B_0$. The base values for the parameters when they aren't being swept are:  $B_0=10~{\rm mT}$, $A_{zz}=A_{zx}=30~{\rm kHz}$, $\Omega=100~{\rm kHz}$, and $|\dot{f}|=9~{\rm MHz/ms}$. We compare the numerical density-matrix simulation results with the approximate formula provided in Eq.~\eqref{eq:phop} of the main text and find some qualitative similarities. However, the formula in Eq.~\eqref{eq:phop} does not capture the rich oscillatory behavior observed in simulations. The simulations are qualitatively consistent with the ODNMR visibility behavior observed in experimental parameter sweeps, Fig.~\ref{fig:fig3} of the main text. These include the resonance-like dependence on microwave power, Fig.~\ref{fig:fig3}(b), and the threshold behavior with respect to external field strength, Fig.~\ref{fig:fig3}(f).

In order to gain insight into the subset of $^{13}$C spins contributing to the ODNMR signal for different magnetic fields, we perform two dimensional parameter sweeps and record the diabatic hopping probability, $P^{23}_{\rm hop}$. We fix the hyperfine coupling magnitude at a particular value, corresponding to a spherical shell of a particular radius, see Eq.~\eqref{eq:ao}. We then vary the $^{13}$C-NV displacement vector polar angle $\theta$ (corresponding to traversing a great circle of the spherical shell) and also vary the magnetic field strength $B_0$. We repeat this process for several values of $A_0$. Figure~\ref{fig:SI_2d_sims_field} shows the resulting maps of $P^{23}_{\rm hop}(\theta,B_0)$ for $A_0=\{10,20,40,80\}~{\rm kHz}$, holding constant $\Omega=100~{\rm kHz}$ and $|\dot{f}|=9~{\rm MHz/ms}$.

The array plots in Fig.~\ref{fig:SI_2d_sims_field} show some similarities, regardless of $A_0$. First, polar angles of $\theta=N\pi/2$, where $N$ is an integer, always result in $P^{23}_{\rm hop}=0$. This is due to the $A_{zx}\propto\sin(2\theta)$ dependence of the transverse hyperfine coupling. For these values of $\theta$, $A_{zx}=0$ and there are no off-diagonal matrix elements in the Hamiltonian Eq.~\eqref{eq:hamiltonian_matrix} that can alter the nuclear spin state. Second, for magnetic fields $B_0\lesssim5~{\rm mT}\approx\Omega/\gamma_{\rm nuc}$, $P^{23}_{\rm hop}$ approaches zero for all values of $A_0$ in the $10{\mbox{--}}80~{\rm kHz}$ range studied in Fig.~\ref{fig:SI_2d_sims_field}. In this relatively low hyperfine coupling regime, the nuclear-spin-conserving AFP process dominates over diabatic hopping when the field is small $B_0\ll\Omega/\gamma_{\rm nuc}$. 

The largest difference among the array plots in Fig.~\ref{fig:SI_2d_sims_field} is the behavior at higher fields, $B_0\gtrsim10~{\rm mT}$. When the hyperfine coupling magnitude is large, there is significant $P^{23}_{\rm hop}$ diabatic hopping probability even at large $B_0$. In this regime, the precise probability oscillates with both $B_0$ and $\theta$, indicating a rich interplay among terms in the Hamiltonian, Eq.~\eqref{eq:hamiltonian_matrix}, but the probability approaches unity for a large portion of parameter space.

Another notable feature in Fig.~\ref{fig:SI_2d_sims_field} is that the maximum diabatic hopping probability begins to diminish for hyperfine coupling magnitudes $A_0\lesssim10~{\rm kHz}$. While the probability density of $^{13}$C spins in this hyperfine coupling range is much larger, owing to the $1/A_0^2$ probability density (\ref{sec:SI_dist}), the diminished $P^{23}_{\rm hop}$ probabilities indicate that these spins are less likely to contribute \textit{directly} to the ODNMR readout signal. However these spins can still contribute \textit{indirectly} through spin diffusion. In order to account for the large number ($\gtrsim10^2$) of oscillations in the ODNMR readout signals (Fig.~\ref{fig:fig2} of the main text), we expect spin diffusion must play a role, governed by the nuclear spin flip-flop rate $\Gamma_{\rm ff}$ (\ref{sec:SI_glossary}). For example, there are typically only ${\sim}25$ $^{13}$C spins within $r=1.5~{\rm nm}$ of an NV center, corresponding to hyperfine coupling magnitudes $A_0\gtrsim5~{\rm kHz}$. In order to account for the larger number of nuclear spin angular momentum quanta transferred during ODNMR readouts ($\gtrsim10^2$), $^{13}$C spins with $A_0\lesssim5~{\rm kHz}$ must be involved in the process somehow. Given the low diabatic hopping probability of these very weakly coupled $^{13}$C spins, we expect their contribution is indirect through nuclear spin diffusion.

\begin{figure}[htb]   
\includegraphics[width=0.78\columnwidth]{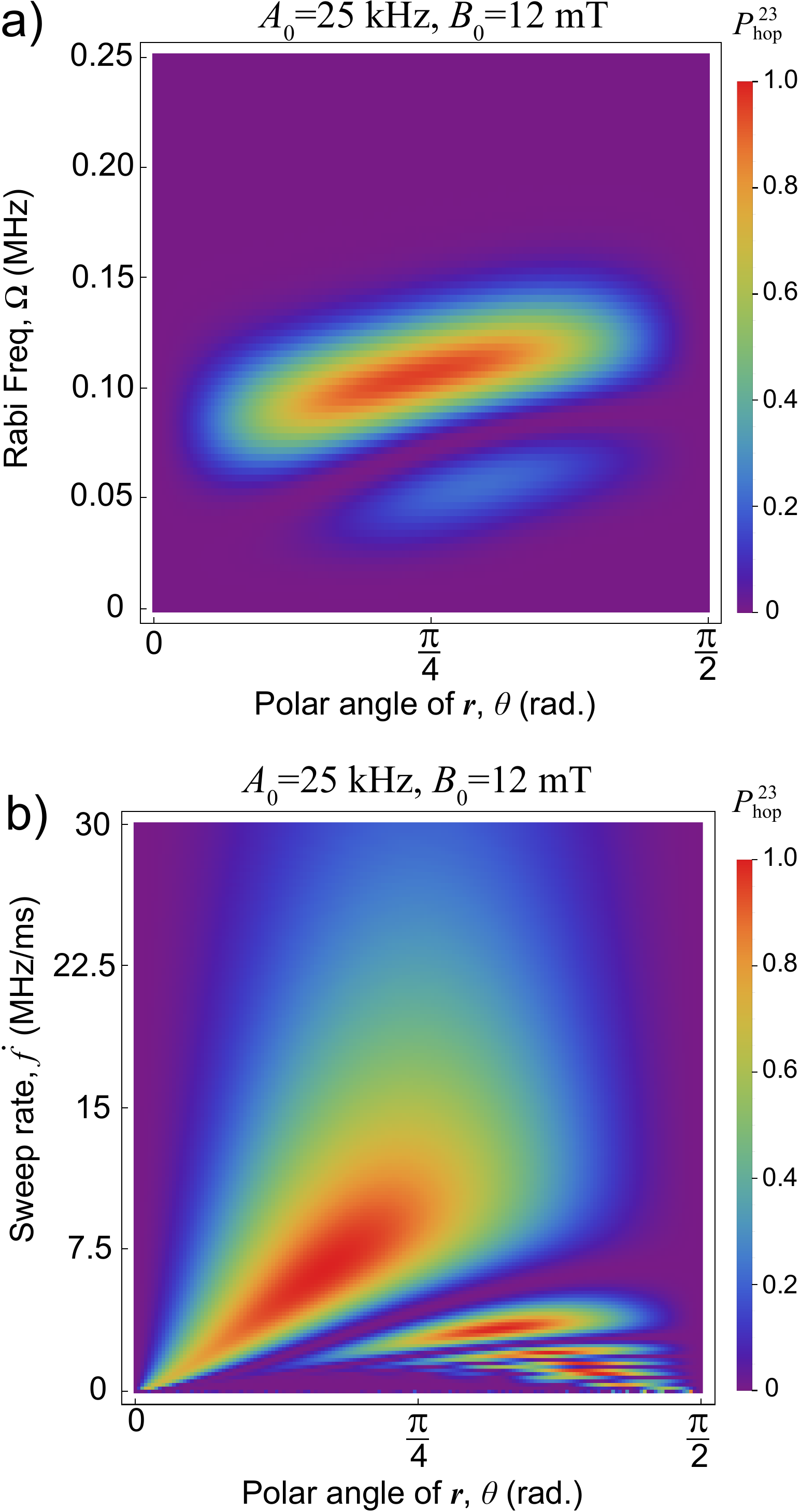}
    \caption{\textbf{Sweeps of $\Omega,\dot{f},\theta$.} (a) Diabatic hopping probability $P^{23}_{\rm hop}$ vs. Rabi frequency $\Omega$ and polar angle $\theta$.  (b) $P^{23}_{\rm hop}(\dot{f},\theta)$. For (a) and (b), $A_0=25~\rm{kHz}$ and $B_0=12~\rm{mT}$.}
    \label{fig:SI_2d_sims_notfield}
\end{figure}

We also varied the microwave Rabi frequency $\Omega$ and sweep rate $\dot{f}$ to examine its impact on the diabatic hopping probabilities. Figure~\ref{fig:SI_2d_sims_notfield}(a) is a two dimensional array plot of $P^{23}_{\rm hop}$ as a function of $\Omega$ and $\theta$. The ``resonance-like'' behavior observed in Fig.~\ref{fig:fig3}(b) of the main text and Fig.~\ref{fig:SI_1d_sims}(c) is also seen in this array plot. However additional oscillatory structure is evident for some angles of $\theta$. Figure~\ref{fig:SI_2d_sims_notfield}(b) shows an array plot of $P^{23}_{\rm hop}(\theta,\dot{f})$. Here there is additional support for the ``double-bump'' structure observed in Fig.~\ref{fig:fig3}(c) and in a prior work on single-crystal diamond~\cite{KAV2025}. As inferred from Fig.~\ref{fig:SI_2d_sims_notfield}(b), the sharp peak at lower sweep rate can be attributed to $^{13}$C spin with relatively large transverse hyperfine components ($|A_{zx}|/|A_{zz}|\gtrsim1$), while the broader peak at larger $|\dot{f}|$ can be attributed to spins with a larger axial hyperfine component ($|A_{zx}|/|A_{zz}|\lesssim1$). 

\subsection{Number of addressed nuclear spins}
In Sec.~\ref{sec:results} of the main text, we claim that the number of detectable oscillations in a $\Delta F/F(t)$ ODNMR readout dictates the number of nuclear spins per NV center that are addressed. To see why this is the case, and why the oscillations could not be explained by the shuttling of polarization back and forth from a single $^{13}$C spin, we consider the following idealized scenario. We assume a single perfect NV-$^{13}$C pair, where the diabatic hop probability under microwave sweeping is $P^{23}_{\rm hop}=P^{32}_{\rm hop}=1$ and all other microwave-induced transitions are assumed to be completely adiabatic, flipping only the NV electron spin. We assume that the NV center is fully polarized to $\ket{0}$ between each microwave sweep segment. For a microwave sweep segment of increasing frequency ($\dot{f}>0$), the transition selection rules are:
\begin{align*}
&\ket{0,\downarrow}\rightarrow\ket{1,\downarrow} \\
&\ket{0,\uparrow}\rightarrow\ket{0,\downarrow}.
\end{align*}
For a microwave sweep segment of decreasing frequency ($\dot{f}<0$), the transition selection rules are:
\begin{align*}
&\vert 0, \downarrow\rangle \rightarrow \vert 0, \uparrow\rangle\\
&\vert 0, \uparrow\rangle \rightarrow\vert 1, \uparrow \rangle.
\end{align*}
Suppose the nuclear spin is initially in state $\ket{\uparrow}$, and we now perform the detection phase of the ODNMR readout sequence, as depicted in Fig.~\ref{fig:fig1}(c) of the main text. The NV-$^{13}$C pair state will thus evolve as:
\begin{equation*}
    \ket{0,\uparrow} \xrightarrow{\dot{f}>0} \ket{0,\downarrow} \xrightarrow{\dot{f}<0} \ket{0,\uparrow} \xrightarrow{\dot{f}>0} \ket{0,\downarrow} ...
\end{equation*}
In this case, fluorescence level will always remain high and will not oscillate, as the NV center is always in $\ket{0}$.

If the nuclear spin is initially in state $\ket{\downarrow}$, the NV-$^{13}$C pair state will evolve as:
\begin{equation*}
    \ket{0,\downarrow} \xrightarrow{\dot{f}>0} \ket{1,\downarrow} \xrightarrow{\rm init} \ket{0,\downarrow} \xrightarrow{\dot{f}<0} \ket{0,\uparrow} \xrightarrow{\dot{f}>0} \ket{0,\downarrow}...
\end{equation*}
In this case, the fluorescence will oscillate once and then remain high. 

Thus, the transfer of a single nuclear spin quantum can only lead to at most one oscillation in NV fluorescence. The only way for the fluorescence to oscillate back and forth ${\sim}10^2$ times, as observed in our experiments (Fig.~\ref{fig:fig2}(b) of the main text), is for there to be a transfer of ${\sim}10^2$ nuclear spin quanta. In principle, the \textit{direct} transfer could occur due to one ``special'' NV-$^{13}$C pair combined with nuclear spin diffusion from the polarized bath to the one ``special'' $^{13}$C spin. But this still corresponds to the readout of ${\sim}10^2$ nuclei per NV because the bath must transfer ${\sim}10^2$ spin quanta from different nuclear spins to the ``special'' $^{13}$C spin in order for ${\sim}10^2$ oscillations to occur. However, as discussed in \ref{app:diabatic_hop}, since many NV-$^{13}$C pairs exhibit substantial $P^{23}_{\rm hop}$ diabatic hopping probability, we expect that there is typically more than one ``special'' $^{13}$C spin that participates in \textit{direct} transfer.

In the actual experiment, the $\Delta F/F(t)$ signal is an ensemble average over many NV centers, each with its own unique configuration of $^{13}$C spins with sub-unity $P^{23}_{\rm hop}$ diabatic hopping probabilities. Moreover, the $\Delta F/F(t)$ readout oscillations do not exhibit the maximum possible fluorescence contrast, and they undergo nontrivial decay profiles. Thus, the number of detectable oscillations in a $\Delta F/F(t)$ ODNMR readout is only a rough approximation to the number of nuclear spins per NV center that are detected. Nevertheless, the value of ${\sim}10^2$ is a reasonable order-of-magnitude estimate. 

\section{Parameter Tables}

\subsection{Glossary}
\label{sec:SI_glossary}
Tables~\ref{tab:g1} and \ref{tab:g2} provide a glossary of rates and parameters referred to throughout the manuscript, along with their typical values.

\begin{table}[htbp]
\centering
\scriptsize
\renewcommand{\arraystretch}{1.2}
\caption{$^{13}$C nuclear spin dynamics parameters.}
\label{tab:nuclear_spin}
\begin{tabular}{lll}
\toprule
\textbf{Symbol} & \textbf{Description} & \textbf{Value} \\
\midrule
$T_2^{\ast}$ & $^{13}$C Ramsey dephasing time & ${\sim}1\mbox{--}2~{\rm ms}$ \\
$T_2^W$      & $^{13}$C dephasing under WAHUHA & ${\sim}3~{\rm ms}$ \\
$T_2$        & $^{13}$C Hahn echo decay & ${\sim}2~{\rm ms}$ \\
$T_{1,{\rm nuc}}$ & $^{13}$C longitudinal spin relaxation & ${\gtrsim}100~{\rm s}$ \\
\midrule
$T_{e2}$     & Detection envelope $1/e^2$ decay time & ${\sim}30\mbox{--}90~{\rm ms}$ \\
\midrule
$T_{\rm pol}$ & $^{13}$C $1-1/e^2$ polarization buildup time & ${\sim}100\mbox{--}300~{\rm ms}$ \\
$\Gamma_{\rm ff}$ & $^{13}$C nuclear dipolar flip-flop rate & ${\sim}100~{\rm s^{-1}}$ \\
\midrule
$\gamma_{\rm nuc}$ & $^{13}$C gyromagnetic ratio & $10.7~{\rm MHz/T}$ \\
\bottomrule
\end{tabular}
\label{tab:g1}
\end{table}

\begin{table}[htbp]
\centering
\scriptsize
\renewcommand{\arraystretch}{1.02}
\caption{NV center properties and parameters.}
\label{tab:NV}
\begin{tabular}{lll}
\toprule
\textbf{Symbol} & \textbf{Description} & \textbf{Value} \\
\midrule
$D$           & NV axial zero-field splitting           & $2.87~{\rm GHz}$ \\
$\gamma_{\rm nv}$ & NV electron gyromagnetic ratio      & $28.03~{\rm GHz/T}$ \\
$f_+$         & NV $\ket{0}\leftrightarrow\ket{1}$ frequency at $B_0=10~{\rm mT}$ & ${\sim}3150~{\rm MHz}$ \\
$T_{1,{\rm nv}}$ & NV longitudinal spin relaxation     & ${\sim}3~{\rm ms}$ \\
\midrule
$A_{zz}$      & Axial hyperfine coupling, sim. baseline & ${\sim}30~{\rm kHz}$ \\
$A_{zx}$      & Transverse hyperfine coupling, sim. baseline & ${\sim}30~{\rm kHz}$ \\
\midrule
$\Omega$      & Microwave Rabi frequency (baseline)    & ${\sim}100~{\rm kHz}$ \\
$\dot{f}$     & Microwave frequency sweep rate         & ${\sim}9~{\rm GHz/s}$ \\
\midrule
$T_{\rm seq}$ & Total pulse sequence duration          & ${\sim}1~{\rm s}$ \\
$T_{\rm polar}$ & Polarization phase duration  & ${\sim}0.5\mbox{--}1~{\rm s}$ \\
$T_{\rm detect}$ & Detection phase duration               & ${\sim}0.4~{\rm s}$ \\
$T_{\rm read}$ & Readout time used in sensitivity estimate & ${\sim}0.2~{\rm s}$ \\
\bottomrule
\end{tabular}
\label{tab:g2}
\end{table}

\subsection{Baseline Values}
\label{sec:SI_baseline}
Below, we present tables outlining the baseline values for the measurement procedure for all major plots in the main text. For all plots (unless otherwise mentioned), the microwave sweep width is $9~{\rm MHz}$ for both polarization and detection phases. The microwave sweep time is $1~{\rm ms}$ for each sweep direction in the detection phase and $0.833~{\rm ms}$ for the polarization phase. The laser excitation power is $400~{\rm mW}$ for both polarization and detection phases. The absolute microwave power values stated are estimated, but the relative values are correct. We measure before an amplifier and estimate losses.

Table~\ref{table:SI_baseline_fig3} shows the baseline experimental parameters used for the experiments presented in Fig.~\ref{fig:fig3} and Table~\ref{table:SI_baseline_fig24} shows the baseline values for Figs.~\ref{fig:fig2}, \ref{fig:fig4}, and \ref{fig:SI_coilandNV}.

\begin{table}[!htbp]
  \centering
  \scriptsize
  \renewcommand{\arraystretch}{1.1}
  \setlength{\tabcolsep}{1pt}
  \begin{tabular}{l|c|c|c|c|c|c}
    \makecell{Plot} & \makecell{Microwave \\ central \\ frequency \\ (MHz)} & \makecell{Polariz. \\ time \\ (s)} & \makecell{MW \\ power \\ (dBm)} & \makecell{Avg. \\ per $\tau$ \\ value} & \makecell{Number \\ of Ramsey \\ points} & \makecell{Total expt \\ acquisition \\ time (s)} \\
    \midrule
    Fig.~\ref{fig:fig3}(a) & 3208 & 0.5-1 & 23 & 8 & 20 & 160 \\
    \hline
    Fig.~\ref{fig:fig3}(b) & 3208 & 0.5-1 & 23 & 8 & 20 & 160 \\
    \hline
    Fig.~\ref{fig:fig3}(c) & 3208 & 0.5-1 & 23 & 8 & 20 & 160 \\
    \hline
    Fig.~\ref{fig:fig3}(d) & 3208 & 0.5-1 & 23 & 8 & 20 & 160 \\
    \hline
    Fig.~\ref{fig:fig3}(e) & 3208 & 0.02-2 & 23 & 8 & 20 & 160 \\
    \hline
    Fig.~\ref{fig:fig3}(f) & varies & 0.5-1 & varies & 8-64 & 10-20 & 80 - 1280 \\
  \end{tabular}
  \caption{\textbf{Baseline Values for Fig.~\ref{fig:fig3}.} The range 0.5-1 s in the polarization time column means the polarization time is in this range but the exact value was not recorded.}
  \label{table:SI_baseline_fig3}
\end{table}

\begin{table}[!htbp]
  \centering
  \scriptsize
  \renewcommand{\arraystretch}{1.1}
  \setlength{\tabcolsep}{1pt}
  \begin{tabular}{l|c|c|c|c|c|c}
    \makecell{Plot} & \makecell{Magnetic \\ field \\ $B_0$ (mT)} & \makecell{Polariz. \\ time \\ (s)} & \makecell{MW \\ power \\ (dBm)} & \makecell{Avg. \\ per $\tau$ \\ value} & \makecell{Number \\ of Ramsey \\ points} & \makecell{Total expt \\ acquisition \\ time (s)} \\
    \midrule
    Fig.~\ref{fig:fig2}(d) & 12 & 1 & 23 & 256 & 40 & 10240 \\
    \hline
    Fig.~\ref{fig:fig4}(a) & 12 & 1 & 23 & 256 & 40 & 10240 \\
    \hline
    Fig.~\ref{fig:fig4}(b) & 10 & 4 & unknown & 16 & 20 & 1280 \\
    \hline
    Fig.~\ref{fig:fig4}(c) & 8 & 0.25-1 & 25 & 16 & 25 & 400 \\
    \hline
    Fig.~\ref{fig:fig4}(d) & 8 & 0.25-1 & 33 & 8 & 45 & 360 \\
    \hline
    Fig.~\ref{fig:fig4}(e) & 8 & 0.25-1 & 33 & 16 & 80 & 1280 \\
    \hline
    Fig.~\ref{fig:fig4}(f) & 8 & 0.25-1 & 33 & 16 & 39 & 624 \\
    \hline
    Fig.~\ref{fig:SI_coilandNV}(a) & 20 & 200 & 20 & 640 & N/A & 128000 \\
    \hline
    Fig.~\ref{fig:SI_coilandNV}(c) & 20 & 0.5 & 20 & 64 & 33 & 2112 \\
  \end{tabular}
  \caption{\textbf{Baseline Values for Figs.~\ref{fig:fig2},~\ref{fig:fig4}, and~\ref{fig:SI_coilandNV}.} The range 0.25-1 s in the polarization time column means the value is in this range but the exact value was not recorded.}
  \label{table:SI_baseline_fig24}
\end{table}

\section{ODNMR readout oscillations}
\label{sec:SI_spinstate}

\subsection{Fitting process and robustness checks}
\label{sec:SI_fit_readout}
Here, we describe the fitting procedure for the envelope decay of the direct ODNMR readout $\Delta F/F(t)$, and we elaborate on robustness checks used to validate claims in Sec.~\ref{sec:results} of the main text. We focus our description on the data shown in Fig.~\ref{fig:fig2}(b), as it had the longest averaging times. We select the $\Delta F/F(t)$ curves presented in Fig.~\ref{fig:fig2}(b), which correspond to $^{13}$C in $\ket{\uparrow}$ or $\ket{\downarrow}$ states, as they have the largest amplitude. 

We first subtract the persistent ``offset oscillation'' from the data. As discussed in Sec.~\ref{sec:results} of the main text and elaborated further in \ref{sec:SI_offset}, this offset oscillation can be attributed to AFP of the NV electron spin and it does not characterize $^{13}$C nuclear spin dynamics. This is evidenced by the observations that the offset is present regardless of the Ramsey free-precession time $\tau$, it persists indefinitely as a function of readout time $t$, and it can be nullified by precise tuning of the microwave central frequency (\ref{sec:SI_offset}).

\begin{figure}[htb]
   \centering
    \includegraphics[width=0.85\linewidth]{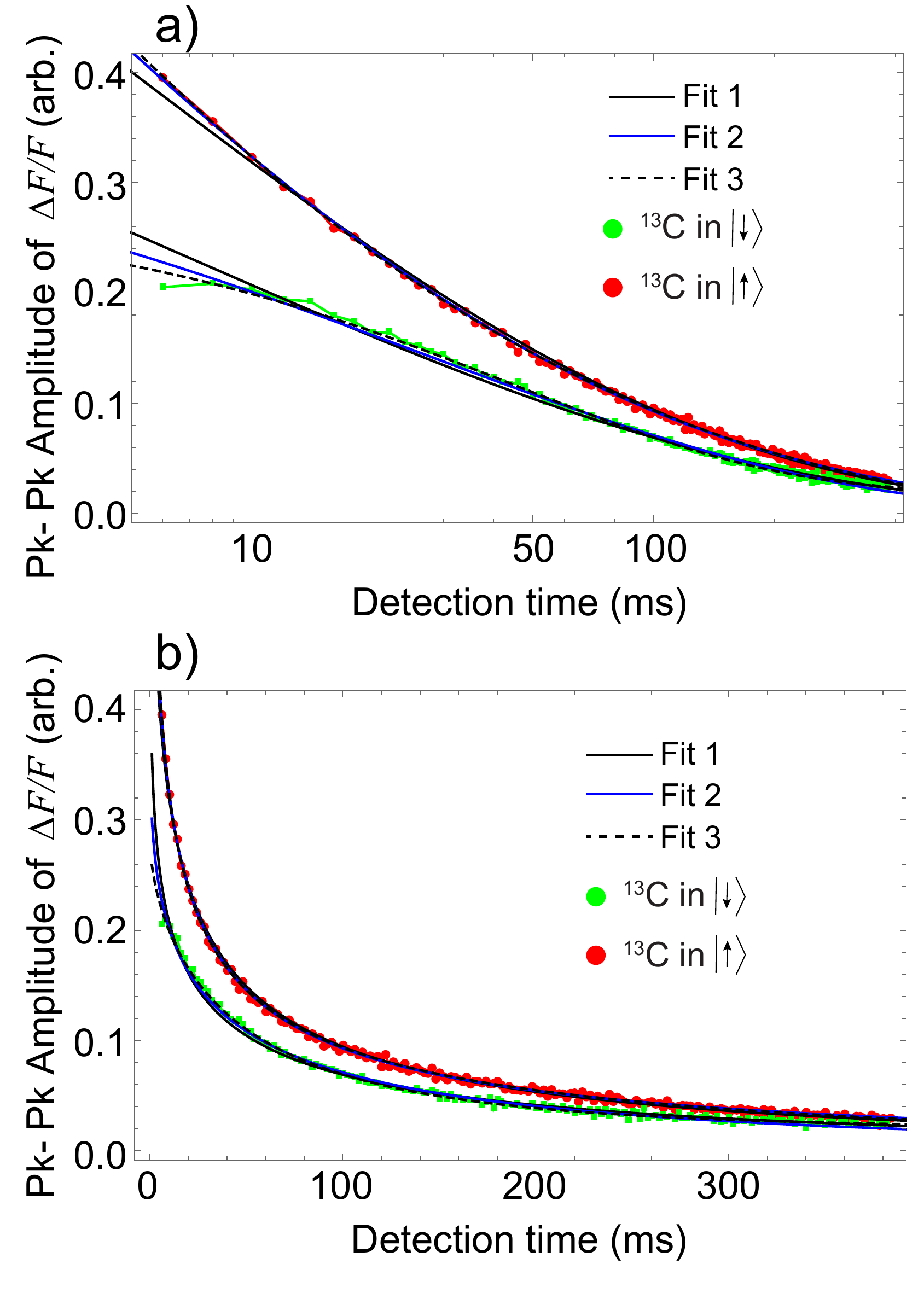}
    \caption{\textbf{Fitting $\Delta F/F(t)$ data.} $\Delta F/F(t)$ for $^{13}$C in $\ket{\uparrow}$ and $\ket{\downarrow}$ along with fits to Eqs.~\eqref{eq:readoutfits}. (a) Log horizontal scale. (b) Linear horizontal scale.} 
    \label{fig:SI_3_fits_decay}
\end{figure}

Next, we extract the ``envelope curves'' of the decay, and eliminate the oscillations, by taking every other point in the $\Delta F/F(t)$ time traces. We then take the absolute value of each curve so the envelope curves have the same sign for better comparison. The resulting envelope curves are shown in Fig.~\ref{fig:SI_3_fits_decay} with logarithmic (a) or linear (b) horizontal axis. 

We fit the envelope curves to three plausible empirical functions:
\begin{subequations}
\label{eq:readoutfits}
\begin{align}
f_1(t) &= a \, e^{-(t/T)^{0.3}} \\
f_2(t) &= a \, e^{-(t/T)^{\gamma}} \\
f_3(t) &= a \, e^{-(t/T)^{\gamma}} + c.
\end{align}
\end{subequations}
Here, $a$ represents the initial amplitude, $T$ is a characteristic decay time, $\gamma$ is the exponential stretch factor, and $c$ is a vertical offset. In functions where $\gamma$ is a free parameter, there is a strong correlation between $T$ and $\gamma$ indicating overfitting. We thus further parameterize the decay time based on the $1-1/e^2$ time, $T_{e2} = 2^{1/\gamma}\,T$. Table~\ref{tab:si_data_stretch_fitting} shows the fit results and uncertainties. Note that none of these fit conditions is the same as the one we use for Fig.~\ref{fig:fig2}(c) of the main text. There we first subtract off a constant oscillation and then fix $\gamma=0.3$. However in all cases, we find that $T_{e2}$ is significantly longer for the $\ket{\downarrow}$ polarization.

\begin{table}[h]
\centering
\renewcommand{\arraystretch}{1.25}
\begin{tabular}{llcc}
\hline\hline
Fit & \multicolumn{1}{c}{Parameter} & $^{13}$C $|\!\uparrow\rangle$ & $^{13}$C $|\!\downarrow\rangle$ \\
\hline
\multirow{2}{*}{Fit 1: $f_1(t)$}
 & \multicolumn{1}{c}{$a$}                    & $1.08 \pm 0.01$ & $0.63 \pm 0.01$ \\
 & \multicolumn{1}{c}{$T_{e2}$ (ms)}           & $51.4 \pm 0.6$ & $72 \pm 2$ \\
\hline
\multirow{3}{*}{Fit 2: $f_2(t)$}
 & \multicolumn{1}{c}{$a$}                    & $1.76 \pm 0.08$ & $0.39 \pm 0.01$ \\
 & \multicolumn{1}{c}{$T_{e2}$ (ms)}           & $20 \pm 3$ & $147 \pm 15$ \\
 & \multicolumn{1}{c}{$\gamma$}               & $0.24 \pm 0.01$ & $0.41 \pm 0.01$ \\
\hline
\multirow{4}{*}{Fit 3: $f_3(t)$}
 & \multicolumn{1}{c}{$a$}                    & $3.1 \pm 0.6$ & $0.26 \pm 0.01$ \\
 & \multicolumn{1}{c}{$T_{e2}$ (ms)}           & $5.7 \pm 5.3$ & $133 \pm 6$ \\
 & \multicolumn{1}{c}{$\gamma$}               & $0.18 \pm 0.02$ & $0.65 \pm 0.020$ \\
 & \multicolumn{1}{c}{$c$}                    & $-0.012 \pm 0.004$ & $0.019 \pm 0.001$ \\
\hline\hline
\end{tabular}
\caption{Fit parameters, based on Eqs.~\eqref{eq:readoutfits}, of curves in Fig.~\ref{fig:SI_3_fits_decay}.}
\label{tab:si_data_stretch_fitting}
\end{table}

All three fit functions in Eqs.~\eqref{eq:readoutfits} fit well to the data, as seen in Fig.~\ref{fig:SI_3_fits_decay}. Thus, considering these functions are empirical and not based on the underlying physics, we selected a function with few fit parameters to explore the behavior of $T_{e2}$ as a function of amplitude. The results were presented in Fig.~\ref{fig:fig2}(c) of the main text, where a clear trend of $T_{e2}$ being larger for $\overline{\Delta F/F}>0$ ($^{13}$C in $\ket{\downarrow}$) was observed. However, as can be seen from Table~\ref{tab:si_data_stretch_fitting}, this trend holds for all three additional fit functions. It can also be observed directly by rescaling the data, as will be discussed next in \ref{sec:SI_directvis}.

\subsection{Visualizing ODNMR readout decay}
\label{sec:SI_directvis}

\begin{figure*}[hbt]
   \centering
    \includegraphics[width=0.85\linewidth]{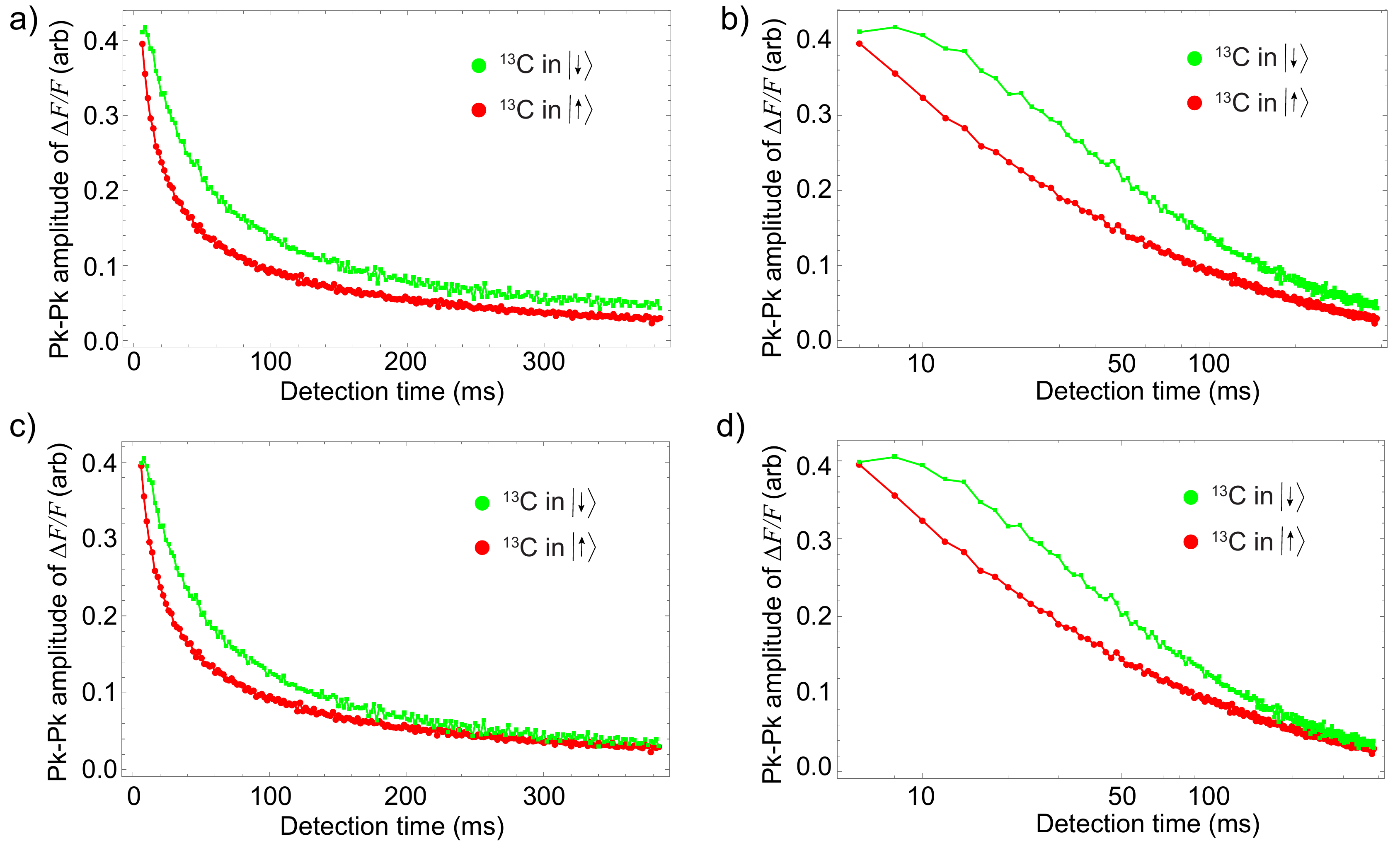}
    \caption{\textbf{$^{13}$C spin-state-dependent decay.} All figures show envelope curves extracted from $\Delta F/F(t)$ ODNMR readout oscillations in Fig.~\ref{fig:fig2}(b) of the main text. (a) The $\ket{\downarrow}$ is multiplied by 2. (b) Same as (a) but on a log horizontal scale. (c) Same as (a) but a constant offset of 0.006 is subtracted from the $\ket{\downarrow}$ curve. (d) Same as (c) but on a log horizontal scale.}
    \label{fig:SI_spinstate}
\end{figure*}

As seen in the prior section (\ref{sec:SI_fit_readout}), the fit parameters associated with ODNMR decay oscillations are strongly correlated with the choices made in the empirical model (offsets, stretch factor, etc.). As a reassurance that the unexpected variation in $T_{e2}$ times observed in Fig.~\ref{fig:fig2}(c) of the main text is not the result of a fitting artifact, we plot the envelope curve data directly for visual inspection. 

Figure~\ref{fig:SI_spinstate} shows the envelope curve data, extracted from the readout oscillation data as described in \ref{sec:SI_fit_readout}, under different modifications. In Fig.~\ref{fig:SI_spinstate}(a), the ODNMR readout envelope curves for $^{13}$C in $\ket{\downarrow}$ and $\ket{\uparrow}$ are shown on a linear scale. The $\ket{\downarrow}$ curve is multiplied by a factor of two to account for the lower signal amplitude, presumably due to dephasing owing to the $0.5~{\rm ms}$ Ramsey free precession time used to obtain this readout envelope curve. Figure~\ref{fig:SI_spinstate}(b) shows the same curves with a logarithmic horizontal axis. Figure~\ref{fig:SI_spinstate}(c) shows the curves on linear axes, but now a small offset is subtracted from the $\ket{\downarrow}$ curve such that the two curves nearly meet for detection times $t\gtrsim400~{\rm ms}$. Figure~\ref{fig:SI_spinstate}(d) shows the same curves as (c) on a logarithmic horizontal axis. In all cases a clear difference in the decay profiles is observed and the $\ket{\downarrow}$ curve appears to decay more slowly. Thus, even in the absence of a precise model to fit these envelope curves, the qualitative decay behavior is confirmed directly from the data themselves. 

\subsection{Spectator spins and future work}
\label{app:spectator}
The polarization asymmetry in the ODNMR $\Delta F/F(t)$ readout oscillations may be, in part, due to the presence of $^{13}$C nuclei coupled to NV centers that were (unintentionally) in $\ket{\pm1}$ during the RF encoding phase. If their hyperfine coupling is strong enough, then these $^{13}$C spins could retain their initial polarization state, unaffected by the RF pulses. These ``spectator spins'' may also depolarize faster due to optically-induced fluctuations of their strongly coupled NV centers during readout~\cite{DUT2007,JIA2008,JIA2009}~(see~\ref{app:optical_depol}). Future work may study these dynamics as a function of the initial polarization state (prior to the RF encoding phase). For example, polarization into $\ket{\uparrow}$ could be realized by sweeping microwaves in the opposite direction during the polarization phase of the experiment. Additionally, the depolarization dynamics may be studied as a function of polarization time to infer the role of spin diffusion.

\section{Ramsey offset}
\label{sec:SI_offset}
\begin{figure}[hbt]
    \centering
    \includegraphics[width=\linewidth]{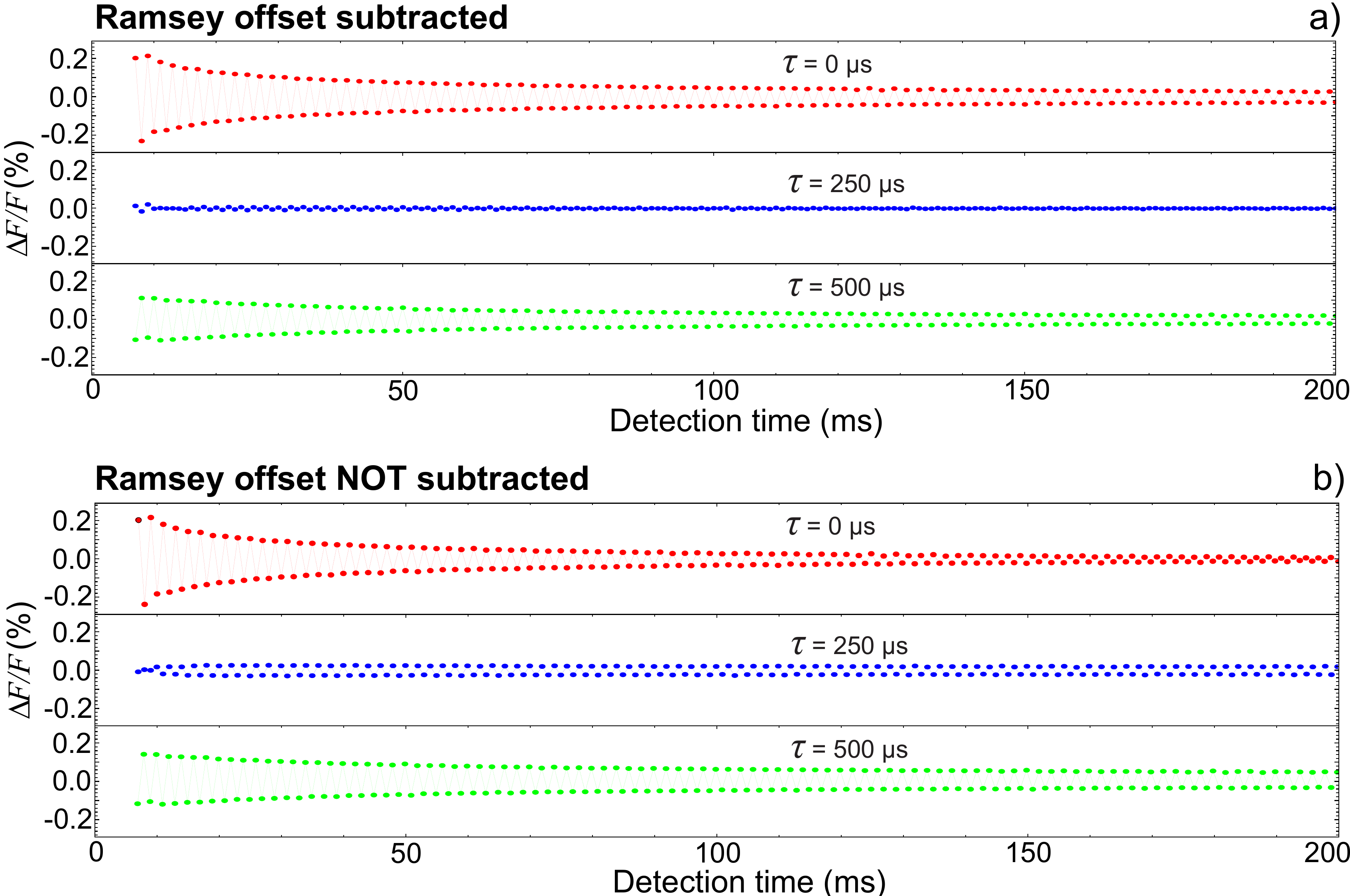}
    \caption{\textbf{Offset subtraction.} Fluorescence time traces as in Fig.~\ref{fig:fig2}(b) with the persistent offset oscillation $c$ (a) subtracted and (b) not subtracted.}
    \label{fig:sub_and_no_sub_ramsey}
\end{figure}

As discussed in Sec.~\ref{sec:results} of the main text, the ODNMR readout oscillations exhibit a constant-amplitude oscillation in addition to the $\braket{I_z}$-dependent decaying oscillation. This persistent offset oscillation, $c$, does not directly affect the extracted Ramsey visibility, and it does not impact the interpretation of the ODNMR readout mechanism. However, depending on the precise experimental parameters, it can be a prominent feature in the $\Delta F/F(t)$ time traces. We believe this offset
oscillation is due to AFP of the NV electron spin. This is evidenced by the observations that the
constant-amplitude oscillation is present regardless of the Ramsey free-precession
time $\tau$, it persists indefinitely as a function of readout
time $t$, and it is typically much smaller in amplitude than the nuclear-spin-dependent oscillations (see Fig.~\ref{fig:sub_and_no_sub_ramsey}). The final reason is that this offset can be nullified by precise tuning of the microwave central frequency.

As mentioned in the main text, we have two methods of analyzing the ODNMR readout oscillation time traces. The first involves fitting the envelope curves of the $\Delta F/F(t)$ oscillations to a stretched exponential function, as described in \ref{sec:SI_spinstate}. Prior to implementing this method, we subtract the constant-amplitude oscillation from the $\Delta F/F(t)$ data. This is justified because the oscillation is indeed nearly constant in amplitude regardless of $t$ or $\tau$. For instance, for an initial value of $\braket{I_z}$ that results in a $\Delta F/F(0)$ oscillation of opposite sign of that of the persistent oscillation, the overall oscillation of the $\Delta F/F(t)$ data flips sign once the constant oscillation is larger than that of the $\braket{I_z}$-dependent decaying $\Delta F/F$ portion. For an initial value of $\braket{I_z}$ that results in a $\Delta F/F(0)$ oscillation of the same sign as that of the persistent oscillation, the signal never dies off completely. Figure~\ref{fig:sub_and_no_sub_ramsey} shows the same $\Delta F/F(t)$ fluorescence oscillations as in Fig.~\ref{fig:fig2}(b) of the main text before and after subtracting the constant offset oscillation $c=0.02\%$.

\begin{figure}[htb]   
\includegraphics[width=0.9\columnwidth]{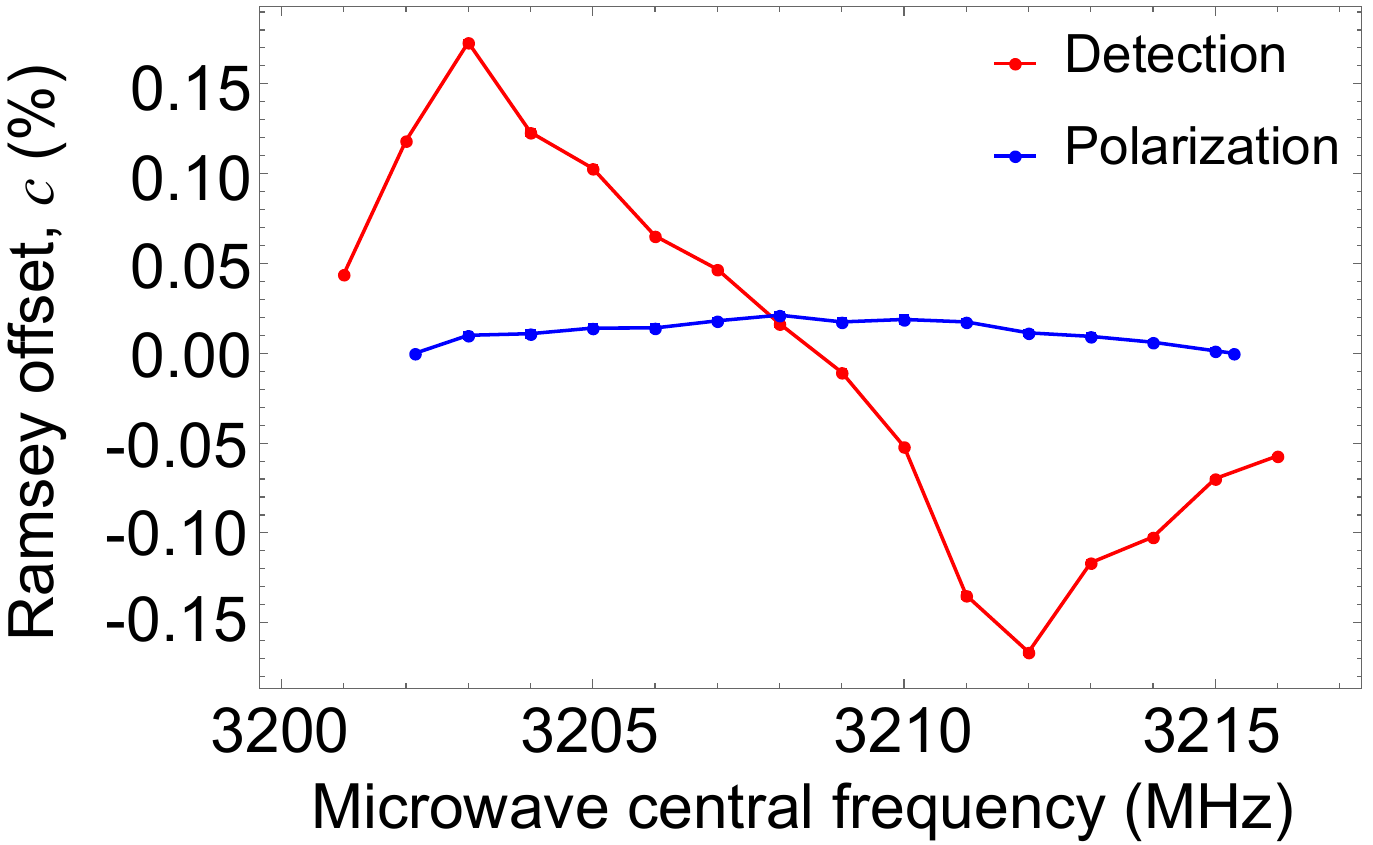}
    \caption{\textbf{Ramsey Offset.} Fitted offset of Ramsey interferograms, $c$, as a function of microwave central frequency for both the polarization and detection phases. During the detection phase, the offset is nullified when the microwave sweep is centered about $f_+$, as expected for an AFP process.} 
    \label{fig:SI_offset}
\end{figure}

The other way we analyze the ODNMR readout oscillation time traces does not involve fitting. This is our primary method used throughout the main text, including for Figs.~\ref{fig:fig2}(d-e), Fig.~\ref{fig:fig3}, and Fig.~\ref{fig:fig4}. Here, the constant oscillation is not necessarily subtracted directly from the data. For a given Ramsey free precession interval $\tau$, we take the amplitude of the real part of the Fourier transform of the first $60~{\rm ms}$ of $\Delta F/F(t)$, retaining sign. This value is referred to as $\overline{\Delta F/F}(\tau)$. When processing the ODNMR readout curves in this manner, the persistent constant-amplitude oscillation shows up as a vertical shift of all the data in a $\overline{\Delta F/F}(\tau)$ Ramsey interferogram. We can then add an offset term, $c$, to Eq.~\eqref{eqn:Ramsey} to extract the constant oscillation amplitude from fits to Ramsey interferograms. 

Figure~\ref{fig:SI_offset} shows the fitted offset as a function of microwave central frequency for either the polarization phase or the detection phase. The offset has no dependence on the microwave sweep properties during the polarization phase, as expected for an $\braket{I_z}$-independent AFP process during the detection phase. However, the offset has a strong dependence on the microwave central frequency for the detection phase, appearing approximately as the derivative of the NV ODMR lineshape. As seen in Fig.~\ref{fig:SI_offset}, the offset can be zeroed out by precisely tuning the microwave frequency to the center of the $f_+$ NV ODMR manifold during the detection phase.

\section{Coil data and NV data}
\label{sec:SI_coilandNV}
In early versions of the experiments, we used an RF coil in a single-sided transceiver mode to acquire bulk diamond $^{13}$C NMR spectra. This allowed us to quantify the degree of nuclear hyperpolarization and compare the NMR spectra directly to the NV ODNMR detection method. This was primarily done at higher magnetic field $B_0\approx20\mbox{--}50~{\rm mT}$, where the inductive NMR signal-to-noise ratio is better.

\begin{figure*}[hbt]
   \centering
    \includegraphics[width=0.8\linewidth]{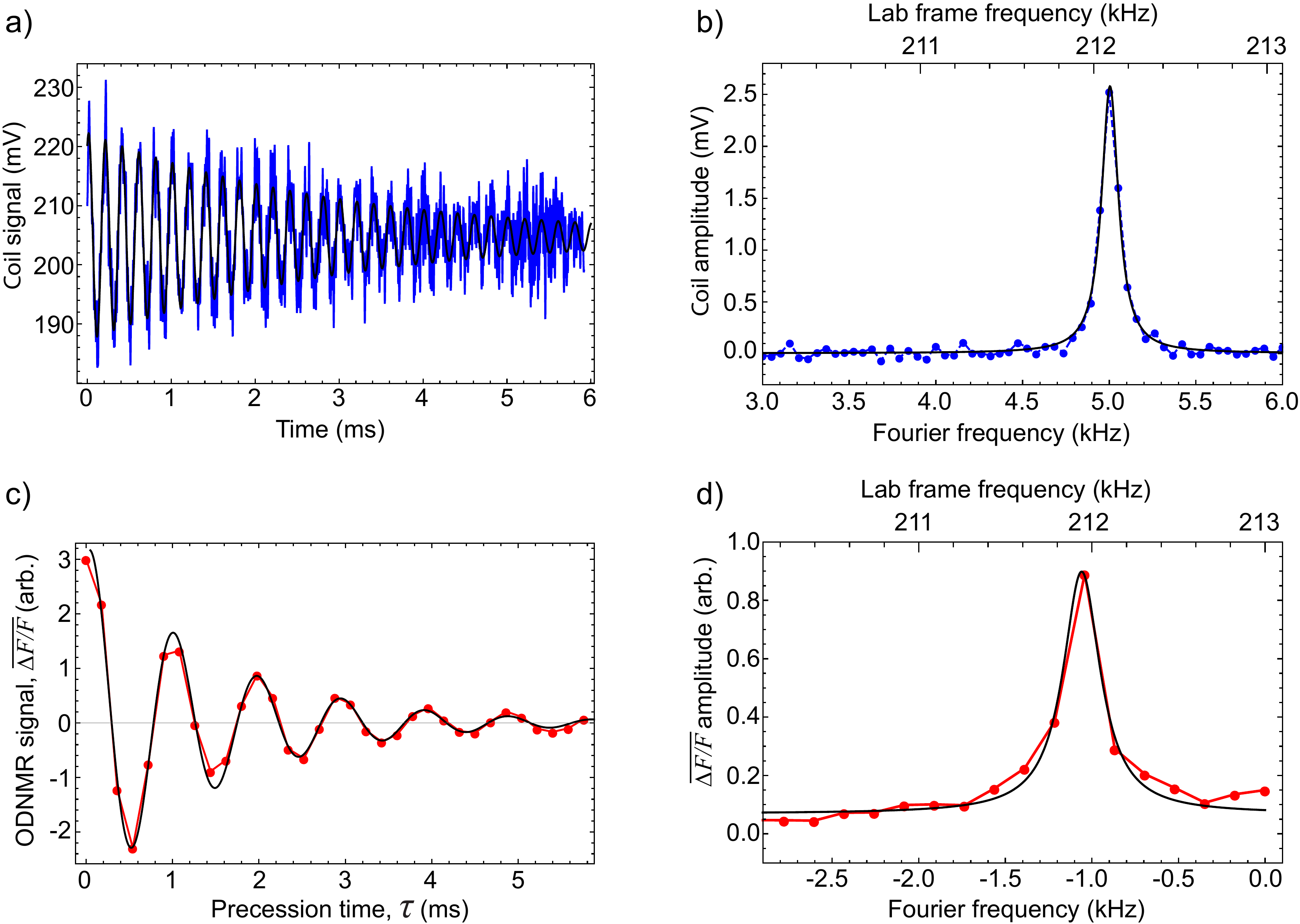}
    \caption{\textbf{Comparison of ODNMR and coil-based detection.} (a) Coil-detected free induction decay of $^{13}$C in diamond (down-converted using a mixer). (b) Fourier transform of the data in (a). (c) ODNMR Ramsey interferogram of $^{13}$C in diamond obtained via the alternative lock-in amplifier method during the detection phase (see text). (d) Fourier transform of the data in (c). The experimental parameters used to obtain these plots are included in \ref{sec:SI_baseline}.}
    \label{fig:SI_coilandNV}
\end{figure*}

Figure~\ref{fig:SI_coilandNV}(a) shows the time-domain free induction decay obtained by coil RF detection at $B_0=19.8~{\rm mT}$. This curve is obtained by continuous laser excitation and asymmetric microwave sweeping for a time $T_{\rm polar,c}=200~{\rm s}$, which was found to be a characteristic saturation time for coil-based detection. The received coil signal is mixed with a synchronized $217~{\rm kHz}$ signal to produce oscillations at ${\sim}5~{\rm kHz}$. The data are fit to an exponentially-decaying sinusoidal function, revealing $T_2^{\ast}=3.0\pm0.1~{\rm ms}$. The phased Fourier transform of these data, representing the NMR spectrum, is plotted in Fig.~\ref{fig:SI_coilandNV}(b). This curve is fit to a Lorentzian function, revealing a fitted FWHM of $0.12\pm0.01~{\rm kHz}$.

We also acquired an ODNMR signal at the same bias field, $B_0=19.8~{\rm mT}$. For these data, we used an earlier version of the detection protocol based on lock-in detection (see also~\ref{sec:SI_higher_field_odnmr}). In this case, the polarization phase is identical to that in the main text, with duration $T_{\rm polar}=0.5~{\rm s}$. However, unlike in the main-text protocol, for the detection phase we follow a protocol similar to the ``Ramsey-$M_z$'' method of Ref.~\cite{SMI2025}. The laser is on continuously during the detection phase, the microwave frequency is modulated sinusoidally about $f_+$, and the NV fluorescence signal is sent to a lock-in amplifier and demodulated at the sweep rate (after a phase adjustment). This allows us to extract a relative amplitude of the readout oscillations, retaining sign, similar to the pulsed protocol presented throughout the main text. We also periodically apply RF $\pi$ pulses to the $^{13}$C nuclei to rapidly (${\sim}100~{\rm Hz}$) modulate the sign of $\braket{I_z}$, suppressing the effect of uncontrolled drifts in the signal. By incrementing the free precession time during the RF encoding phase and recording the detection-phase lock-in signal amplitude (retaining sign) at each value of $\tau$, we record the Ramsey interferogram shown in Fig.~\ref{fig:SI_coilandNV}(c). The data are fit to an exponentially-decaying sinusoidal function, revealing $T_2^{\ast}=1.5\pm0.1~{\rm ms}$. The phased Fourier transform of these data, representing the NMR spectrum, is plotted in Fig.~\ref{fig:SI_coilandNV}(d). This curve is fit to a Lorentzian function, revealing a fitted FWHM of $0.26\pm0.03~{\rm kHz}$. 

The ODNMR and coil-based spectra are similar, indicating that our method is largely addressing the same bulk NMR signal. However there are some differences. The NMR central frequencies differ by ${\sim}100~{\rm Hz}$. We tentatively attribute this to drift of the leading $B_0$ field, as the data were taken on different days. More significantly, the $T_2^{\ast}$ dephasing time in the coil based acquisition is ${\sim}2$ times longer than in the ODNMR case. A possible explanation is as follows. To acquire the coil-based spectra the polarization phase is much longer, $T_{\rm polar,c}=200~{\rm s}$, so there has been time for substantial spin diffusion to remote $^{13}$C sites. Thus, a decent fraction of nuclei probed in the coil-detection experiment have very small hyperfine coupling. In this case, the coupling may be small enough ($\lesssim100~{\rm Hz}$) that any NV longitudinal relaxation does not alter the precession frequency enough to cause dephasing. 

The observed $^{13}$C dephasing time in the coil-based experiment is consistent with relaxation due to $^{13}$C-$^{13}$C dipolar coupling, $T_2^{\ast}|_{\rm dip}\approx3~{\rm ms}$. The $^{13}$C dephasing time observed during NV ODNMR is expected to be influenced by a combination of $^{13}$C-$^{13}$C dipolar dephasing and the effect of NV longitudinal spin relaxation. Based on the results of the WAHUHA dipolar decoupling sequence, Fig.~\ref{fig:fig4}(e) of the main text, and the typical room-temperature NV $T_1$ times~\cite{JAR2012}, we expect the contribution of NV longitudinal spin relaxation to the $^{13}$C dephasing time is $T_2^{\ast}|_{\rm NV T1}\approx2.5~{\rm ms}$. The expected ODNMR dephasing time is thus $T_2^{\ast}|_{\rm tot}\approx(1/T_2^{\ast}|_{\rm dip} + 1/T_2^{\ast}|_{\rm NV T1})^{-1}\approx1.4~{\rm ms}$. This is consistent with the $^{13}$C dephasing times observed in ODNMR experiments of $T_2^{\ast}\approx1.5~{\rm ms}$.

\section{Detection laser pulse length}
\label{sec:SI_pulselength}
In Fig.~\ref{fig:fig3}(d) of the main text, we observe a notable decay in the ODNMR Ramsey visibility with laser excitation power during the detection phase. To study this further, we also kept the laser power constant and varied the pulse length during the detection phase. 

\begin{figure}[hbt]
   \centering
    \includegraphics[width=0.8\linewidth]{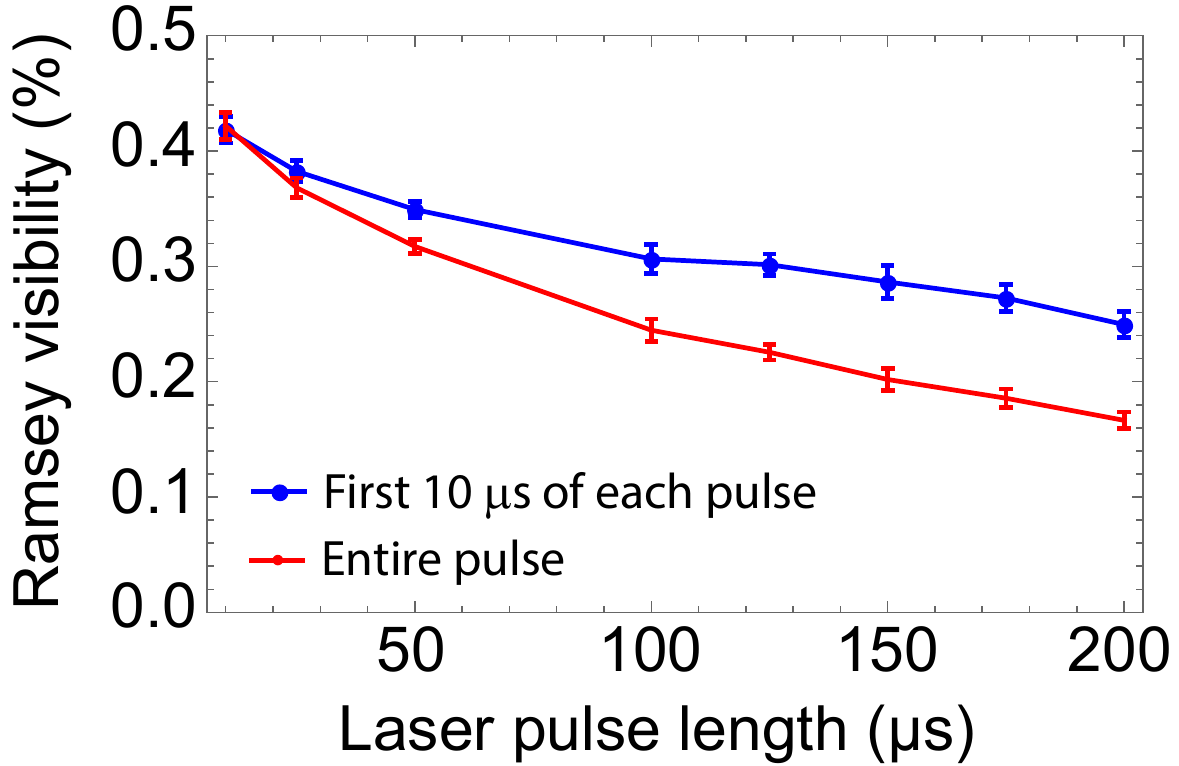}
    \caption{\textbf{Laser pulse length.} ODNMR Ramsey visibility vs. detection-phase laser pulse length. (blue) The first $10~{\rm \upmu s}$ of the fluorescence signal of each pulse is used for data processing. (red) The fluorescence signal of the entire pulse is used for data processing. Error bars represent fit uncertainty.} 
    \label{fig:SI_pulselength}
\end{figure}

Figure~\ref{fig:SI_pulselength} shows the results of this experiment. We process the data in two ways. In the first case, we keep only the first $10~{\rm \upmu s}$ of the laser pulse when determining the fluorescence signal (blue) and in the second case we average together all of the fluorescence signal within the pulse (red). In either case, we observe a decrease in ODNMR Ramsey visibility with pulse area. 

In additional experiments (not shown), we varied the repetition period of the laser pulses (with corresponding variation in microwave sweep rate). We also observed a drop in the ODNMR visibility with decreasing repetition period. 

\begin{figure}[hbt]
    \centering
    \includegraphics[width=0.8\linewidth]{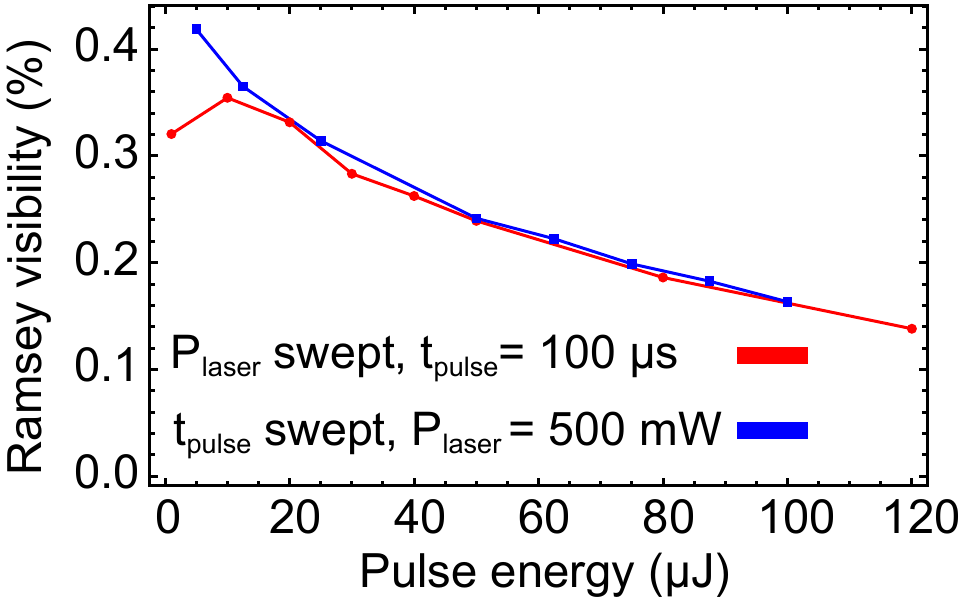}
    \caption{\textbf{Pulse energy.} Ramsey visibility as a function of laser pulse energy. In red, the pulse duration was fixed at $t_{\rm pulse}=100~{\rm \mu s}$. In blue, the laser power was kept fixed at $P_{\rm laser} = 500~{\rm mW}$ and the pulse duration was swept.}
    \label{fig:photon_budget_dose}
\end{figure}

Finally, we replot the laser power dependence data of Fig.~\ref{fig:fig3}(d) and Fig.~\ref{fig:SI_pulselength}(red) on the same plot to show the dependence of the ODNMR Ramsey visibility as a function of laser pulse energy. Figure~\ref{fig:photon_budget_dose} shows how the Ramsey visibility changes with the laser pulse energy delivered. The pulse-energy dependence is quite similar regardless of whether the pulse duration or intensity is varied. This suggests that two-photon processes, such as NV$^0$/NV$^-$ charge state conversion, are unlikely to be the primary contributors to the observed optically-induced loss of signal.

These data, taken together, support the observation that there is a fixed dose of light fluence that each NV center can absorb before the onset of $^{13}$C depolarization. This corresponds to a fixed photon budget that reduces the observed readout fidelity. 

\subsection{Mechanism of optical nuclear depolarization}
\label{app:optical_depol}
A detailed quantitative model for optically-induced $^{13}$C depolarization may be a focus of future work, but here we describe a plausible mechanism. The concept is that the optical $^{13}$C depolarization is due to stochastic hopping of the hyperfine-coupled NV electron-spin state during laser pulse readouts. This is an effect that has been studied since early experiments involving single NV centers coupled to $^{13}$C nuclear spins; see, for example, \cite{DUT2007,JIA2008,JIA2009}. During the optical pulse, there is a non-zero probability for NV centers to repopulate in the $\ket{\pm1}$ ground state sublevels. When NV centers are in $\ket{\pm1}$, the transverse hyperfine term leads to a rotation of the $^{13}$C quantization axis with respect to the B-field axis ($\hat{z}$). This sudden rotation of the quantization axis induces precession of the $^{13}$C spin vector about the new quantization axis and away from $\hat{z}$. Since these spin dynamics happen stochastically at different times for $^{13}$C spins, each with their own unique quantization-axis tilt arising from the wide range of hyperfine tensors, the net effect is to depolarize the ensemble nuclear spin state. For $^{13}$C spins with larger transverse hyperfine coupling, the effect is more pronounced and can lead to depolarization after just a few optical cycles. More weakly coupled nuclear spins can survive for many optical cycles before depolarization.

\begin{figure}[hbt]
    \centering
    \includegraphics[width=\linewidth]{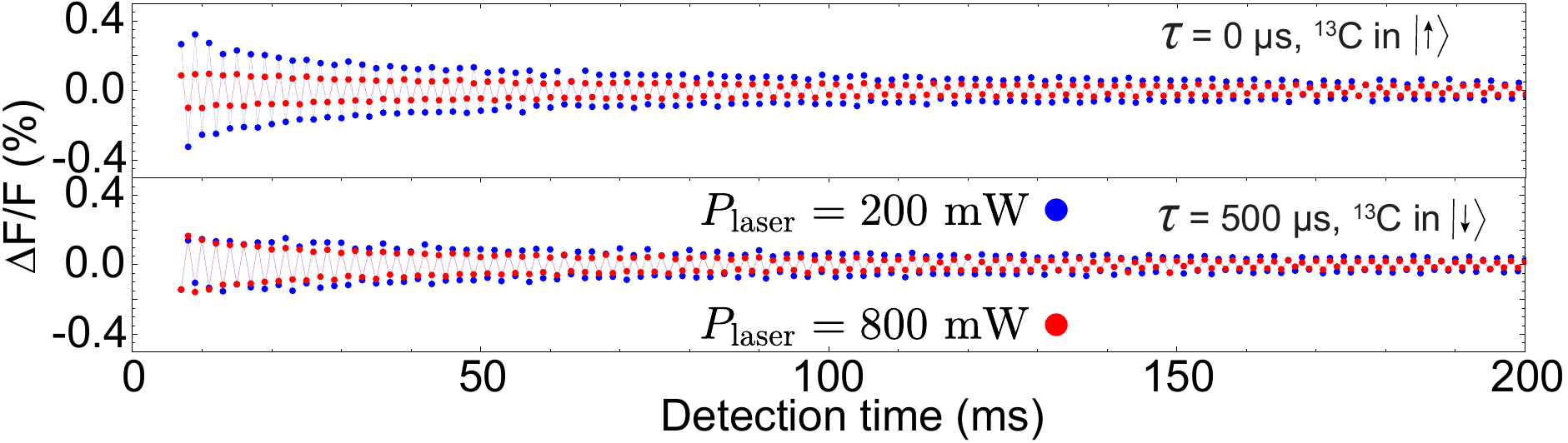}
    \caption{\textbf{Optical depolarization.} Fluorescence time traces for 2 different laser powers ($200~{\rm mW}$ and $800~{\rm mW}$) at two different Ramsey free precession time values ($\tau = 0$ and $\tau = 500~{\rm \mu s}$), corresponding to nuclear spins prepared in $\ket{\uparrow}$ and $\ket{\downarrow}$. The amplitude of $\Delta F/F$ fluorescence oscillations is smaller for the higher laser power, particularly for the $\tau=0$ ($\ket{\uparrow}$) case.}
    \label{fig:200_800mW}
\end{figure}

One prediction that comes from the above hypothesized mechanism is that nuclei with larger transverse hyperfine coupling parameters are most susceptible to light-induced depolarization. Figure~\ref{fig:200_800mW} shows the $\Delta F/F(t)$ readout fluorescence oscillations for nuclear spin polarization (following the RF encoding stage) along $\ket{\uparrow}$ and $\ket{\downarrow}$ at high ($0.8~{\rm W}$) and low ($0.2~{\rm W}$) laser powers. In all cases, the nuclei are initialized into $\ket{\downarrow}$ prior to the RF encoding phase. For subsequent $\ket{\uparrow}$ polarization ($\tau=0$), the largest difference in $\Delta F/F(t)$ for two laser powers is in the first few readout oscillations. This is consistent with the idea that high light power causes a more immediate depolarization of strongly coupled nuclei. For $\ket{\downarrow}$ polarization ($\tau=500~{\rm \upmu s}$), the $\Delta F/F(t)$ curves are similar for both laser powers. This may indicate that strongly-coupled nuclei are not contributing much to the signal regardless of power. This is not inconsistent with the ``spectator bath'' hypothesis (\ref{app:spectator}) for the polarization asymmetry of $\Delta F/F(t)$ curves, but future experiments are required to fully test this hypothesis.

\section{Polarization buildup fitting}
\label{sec:SI_pol_buildup_fitting}
In Fig.~\ref{fig:fig3}(e) of the main text, we present a polarization buildup curve acquired at bias field $B_0=12~{\rm mT}$. We fit this curve to a stretched exponential buildup function with a stretch factor of $\beta=0.5$. This choice is motivated by additional data fitting, conducted using $\beta$ as a free parameter, for polarization buildup curves taken at different fields. 

The general polarization buildup curve is described by the fit function:
\begin{equation}
\label{eq:buildupgen}
   |A|(t) =  A_{\rm sat}(1-e^{-(\frac{t}{T_p})^\beta}),
\end{equation}
where $A(t)$ is the Ramsey visibility, $A_{\rm sat}$ is the maximum amplitude, $t$ is the duration of the polarization phase (also called polarization buildup time), $T_p$ is a characteristic rise time, and $\beta$ is the exponential stretch factor. Due to the correlation between fit parameters $\beta$ and $T_p$, we further parameterize the rise time by the buildup curve's $1-1/e^2$ time, $T_{\rm pol} = 2^{1/\beta}\,T_p$. 

\begin{figure}[hbt]
   \centering
    \includegraphics[width=0.98\linewidth]{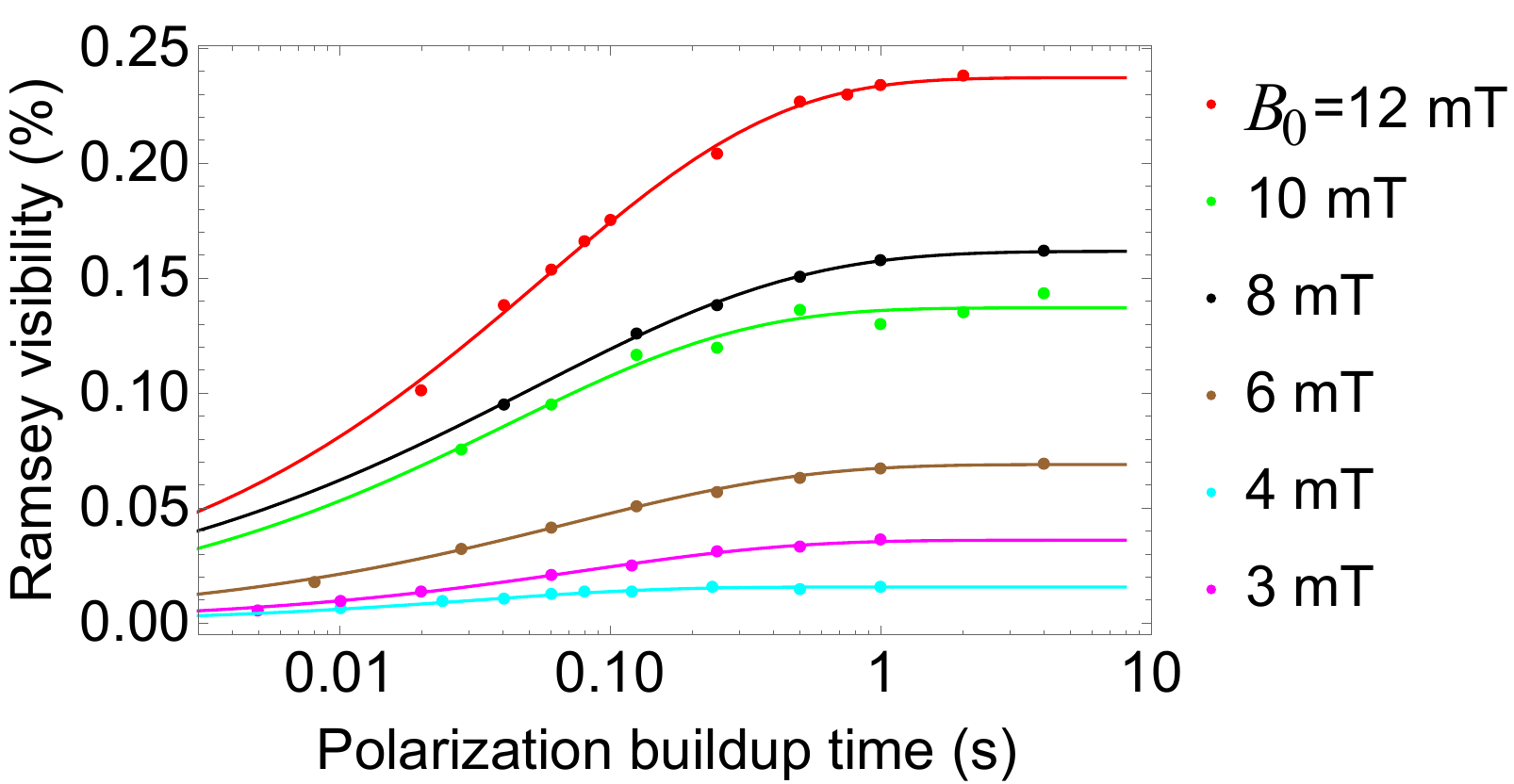}
    \caption{\textbf{Polarization buildup curves for different $B_0$.} Polarization buildup data for various magnetic fields, $B_0$, along with fits to Eq.~\eqref{eq:buildupgen}.} 
    \label{fig:SI_pol_buildup_fitting}
\end{figure}

Figure~\ref{fig:SI_pol_buildup_fitting} shows the polarization buildup curves for magnetic fields in the range $B_0=3\mbox{--}12~{\rm mT}$ along with fits to Eq.~\eqref{eq:buildupgen}, showing good agreement. The results of the fit for the list of bias fields are shown in Table~\ref{tab:si_pol-buildup_fitting}. The fit results indicate that $\beta=0.5$ is close to the ideal choice for this fit function for all fields. Moreover, we find that $T_{\rm pol}$ does not vary much with magnetic field.

\begin{table}[htbp]
\begin{ruledtabular}
\begin{tabular}{cccc}
  $B_0$ (mT) & $A_{\rm sat} (\%)$ & $T_{\rm pol}$ (ms) & $\beta$ \\
\hline
12 & $0.237 \pm 0.002$ & $226 \pm 19$ & $0.50 \pm 0.02$ \\
10 & $0.137 \pm 0.003$ & $171 \pm 50$ & $0.49 \pm 0.09$ \\
 8 & $0.162 \pm 0.001$ & $250 \pm 22$ & $0.44 \pm 0.02$ \\
 6 & $0.069 \pm 0.001$ & $286 \pm 23$ & $0.50 \pm 0.02$ \\
 4 & $0.016 \pm 0.001$ & $95 \pm 17$ & $0.63 \pm 0.07$ \\
 3 & $0.036 \pm 0.001$ & $274 \pm 42$ & $0.56 \pm 0.04$ \\
\end{tabular}
\end{ruledtabular}
\caption{Fit parameters $A_{\rm sat}$, $\beta$, and $1-1/e^2$ time
$T_{\rm pol} = 2^{1/\beta}\,T_p$ for different applied fields $B_0$.
Uncertainties represent the fit standard error.}
\label{tab:si_pol-buildup_fitting}
\end{table}

\section{Higher field ODNMR}
\label{sec:SI_higher_field_odnmr}

The data in Fig.~\ref{fig:fig3}(f) of the main text provide some evidence for a threshold effect, whereby the ODNMR Ramsey visibility plateaus above $B_0\gtrsim8~{\rm mT}$ and goes to zero for lower fields. Those data also support the claim that $T_2^{\ast}$ is largely independent of field in this regime. However the data are limited in this figure. 

\begin{figure}[hbt]
   \centering
    \includegraphics[width=0.73\linewidth]{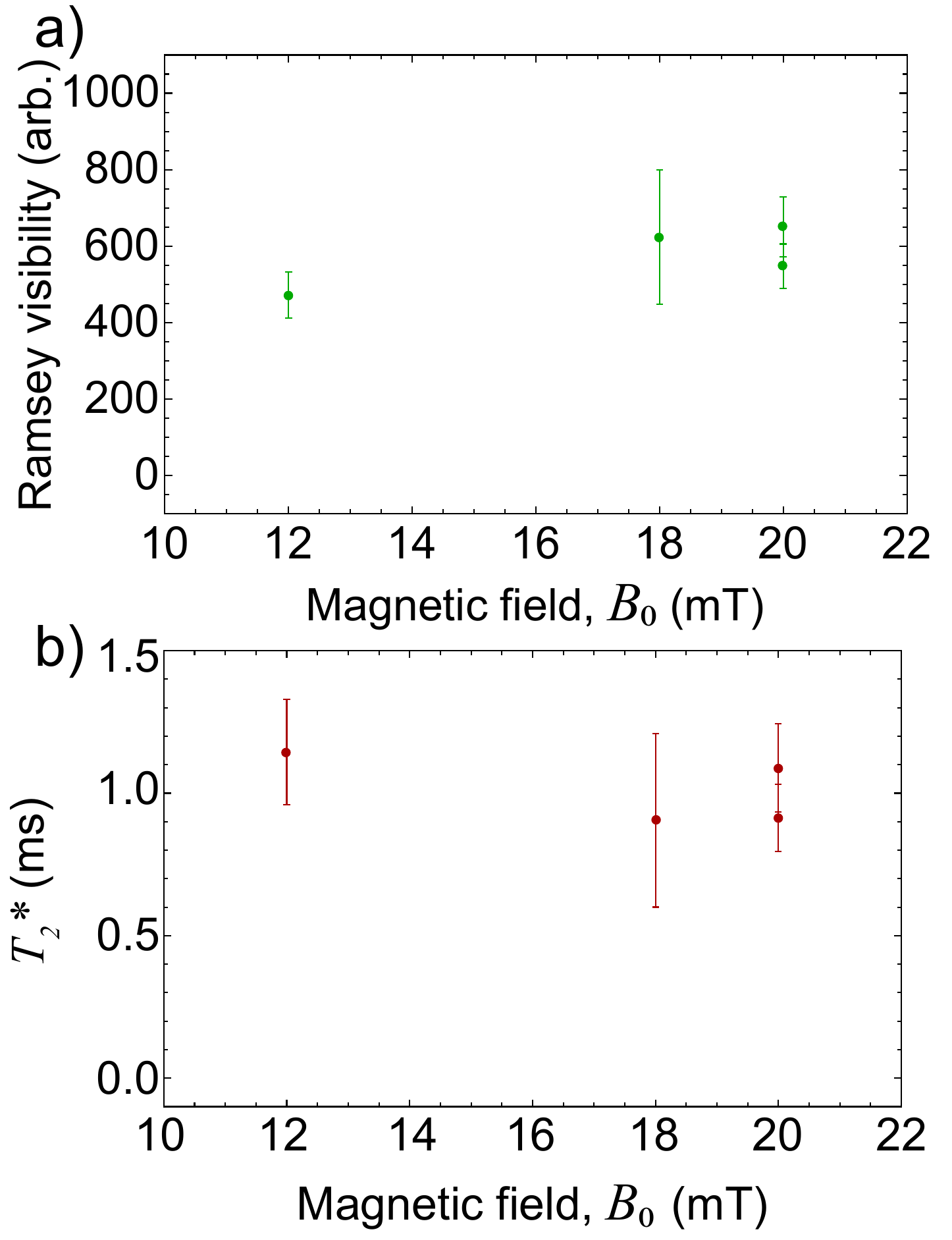}
    \caption{\textbf{ODNMR response above 12~mT.} a) Ramsey visibility versus magnetic field $B_0$ acquired using the alternative lock-in method during the detection phase. b) $T_2^*$ vs. $B_0$. In (a) and (b), error bars represent fit uncertainty.} 
    \label{fig:SI_higher_field_odnmr}
\end{figure}

We used the alternative lock-in amplifier detection mode (see~\ref{sec:SI_coilandNV}) to study the ODNMR behavior of diamond J5 from $B_0=12\mbox{--}20~{\rm mT}$. This technique involved a sinusoidal frequency modulation for the microwave drive field during the detection phase of the experiment instead of a triangle sweep. We also used a lock-in amplifier on the output of the photodetector to demodulate the signal at the sinusoidal modulation frequency. The laser was on for the entire detection sequence instead of being pulsed. Finally, we applied periodic, resonant RF $\pi$ pulses to the $^{13}$C nuclear spins to modulate their state at a rate that is slower than the sinusoidal frequency-modulation rate~\cite{SMI2025}. The resulting signal was a decaying square-wave oscillation with a period of twice the $\pi$ pulse repetition time.

For a given value of free precession time, $\tau$, we take the real part of the phased Fourier transform of the detection time-series data. The peak amplitude of this spectrum is the equivalent of the $\overline{\Delta F/F}$ metric from the main text. From there, the process is similar to that of the main text. We vary the $\tau$ delay and create a Ramsey interferogram and fit it to an exponentially-decaying sinusoidal function as in Eq.~\eqref{eqn:Ramsey}. The resulting Ramsey visibility, $|A|$, and dephasing time $T_2^{\ast}$ are shown as a function of magnetic field in Fig.~\ref{fig:SI_higher_field_odnmr}. These data provide further evidence that the ODNMR response is largely flat from $B_0\approx12\mbox{--}20~{\rm mT}$.

\clearpage

%

\end{document}